\documentclass[12pt]{article}
\usepackage{cite,amsmath,eucal,epsfig}
\newcommand{\ud}{\mathrm{d}}
\title{Microscopic model approaches to fragmentation of nuclei and phase 
transitions in nuclear matter}

\author{J.~Richert$^{a,1}$ and P. Wagner$^{b,2}$\\[0.5em]
$^a$ {\small Laboratoire de Physique Th\'eorique,}
{\small Universit\'e Louis Pasteur,}\\
{\small 3, rue de l'Universit\'e, 67084 Strasbourg Cedex (France)}\\[0.3em]
$^b$ {\small Institut de Recherches Subatomiques,}\\
{\small BP28, 67037 Strasbourg Cedex 2 (France)}\\[0.3em]}
 
\date{\today}

\begin{document}

\maketitle

\vfill


\vskip 1cm


\vskip 1cm

\noindent {\it e-mail:}\\
$^1$richert@lpt1.u-strasbg.fr\\
$^2$pierre.wagner@ires.in2p3.fr

\newpage

\begin{abstract}

  The properties of excited nuclear matter and the quest for a phase 
transition which is expected to exist in this system are the subject of 
intensive investigations. High energy nuclear collisions between finite 
nuclei which lead to matter fragmentation are used to investigate these 
properties. The present report covers effective work done on the subject 
over the two last decades. The analysis of experimental data is confronted 
with two major problems, the setting up of thermodynamic equilibrium
in a time-dependent fragmentation process and the finite size of nuclei. 
The present status concerning the first point is presented. Simple 
classical models of disordered systems are derived starting with the 
generic bond percolation approach. These lattice and cellular 
equilibrium models, like percolation approaches, describe successfully 
experimental fragment multiplicity distributions. They also show the 
properties of systems which undergo a thermodynamic phase transition. 
Physical observables which are devised to show the existence and to fix the 
order of critical behaviour are presented. Applications to the models are 
shown. Thermodynamic properties of finite systems undergoing critical 
behaviour are advantageously described in the framework of the 
microcanonical ensemble. Applications to the designed models and to 
experimental data are presented and analysed. Perspectives of further 
developments of the field are suggested.

\end{abstract}

\newpage

\tableofcontents

\newpage

\section{Introduction}
\indent

  Protons and neutrons are able to form bound nuclei with well defined
lifetimes. They are finite and contain a relatively small number of nucleons 
which varies from one to several hundreds. The nucleons experience a strong 
short range interaction which acts together with the long range Coulomb 
interaction between the charged protons. The interactions between nucleons 
determine the ground state properties and the low energy excitation spectrum
of nuclei which have been studied over several decades. Their description
by means of very different collective and microscopic, semi-classical and
quantum models has lead to a satisfactory understanding of their 
properties \cite{Bohr, Ring, Greiner}.

  With increasing excitation energy the detailed information coded in the 
spectrum and the wave functions of the system gets less and less 
meaningful and stringent for a satisfactory description of the system.
Statistical arguments get relevant. Random matrix theories, for instance, 
have proved to be a very efficient tool in this regime \cite{Mehta, Guhr}.
Under certain 
circumstances experiments reveal that the behaviour of nuclei can be 
described in the framework of thermodynamics. Excited nuclei behave like 
equilibrated compounds which may decay through emission of gamma rays, 
nucleons or light clusters at energies corresponding to unbound states of 
the system \cite{Bohr1}. The behaviour of nuclei in this regime and its 
description in terms of thermodynamic and statistical concepts leads to 
intriguing questions related to the general properties of thermodynamically 
equilibrated nuclear matter, either finite like nuclei or (quasi)infinite 
as it could be found in neutron stars, in the absence of the gravitational 
interaction. These questions are also triggered by the fact that nucleons 
are fermions and nuclear matter in its ground state, at zero temperature,
shows the properties of a Fermi liquid of strongly interacting
particles \cite{Landau, Abrikosov, Nozieres, Migdal}.
For high enough excitation energy it seems sensible to think
that such a quantum liquid may go over to a quantum gas by undergoing a 
phase transition which would be observable in a infinite system and for 
which one would find characteristic signals in finite nuclei.

  This question has become a major source of interest over the last two 
decades in the nuclear physics community. The only way to get an 
experimental answer to it goes through energetic collisions between nuclei 
or nucleons and nuclei
with the aim to generate excited finite nuclear matter. The analysis of the 
experiments must then be able to reveal its properties and eventually 
deliver signs for the existence of one or several phase transitions in the 
infinite system.

  Such ambitious objectives have led to intense experimental and 
theoretical investigations. Several reports have been written on the 
subject over the last ten years. In the present review we aim 
to discuss two aspects concerning excited nuclear matter. The first 
concerns the conditions under which it is experimentally generated. The 
second deals with specific microscopic models which we think allow a 
realistic study and description of experimental facts under the 
prerequisite that the 
physical conditions imposed by the experiment match those which are imposed 
by the theory.

  The content of the present review is the following. In section 2 we 
describe the status of the subject at the beginning of the early 80's. We 
present and discuss the attempts to interpret experimental results related 
to energetic heavy ion collisions and the theoretical models which were 
aimed to describe the outcome of such collisions, the fragmentation of the 
involved nuclei. We then turn over in section 3 to the two major points
which have to be taken in account in order to be able to compare experimental 
results with the type of models which will be introduced. These points 
concern the finiteness of the system and the problems related to its 
properties with respect to the concept of thermodynamic equilibrium. 
Section 4 is devoted to the outcome of very energetic nuclear 
collisions. Nuclei can break up into pieces. Fragments with different 
sizes, mass and charge numbers, are generated. With the presently available 
detection instruments they can be recorded. Their charge and, to some 
extent, their mass can be measured. Charge and mass distributions as well as 
related quantities are a priori non-trivial observables which contain 
interesting information about the fragmentation process and the behaviour of 
excited nuclear matter. We present and discuss the concepts and 
models which have been introduced in order to describe the fragmentation 
characteristics of nuclei, as well as the possible reasons for which rather
simple-minded models describe these observables
with a remarkable success. In section 5 we introduce microscopic
statistical models which may behave as generic approaches for the study
of the thermodynamic properties of excited systems composed of interacting 
nucleons. These models have either been borrowed from other fields of 
physics or consist of adapted derivatives of these models. They describe 
systems which show phase transitions and whose outcome can be confronted 
with experimental data. Since they are finite, tests which enable to 
characterize the phase transitions have been proposed. Some of these tests 
are introduced in section 6 and their application both on theoretical models 
as well as experimental results are presented and discussed in the 
perspective of the existence of a phase transition in excited nuclear 
matter. In section 7 we develop a summary, comment the present status of 
the field and suggest perspectives which may lead to further progress in the 
field.

  Nuclear fragmentation and the search of phase transitions is a very active
field of research. This is best seen by looking at the number of preprints 
and publications which appear regularly. We tried to cover the newest results 
concerning the aspects of the subject with which the present report is 
concerned, extending to summer 2000. We also tried to be as exhaustive as 
possible in the presentation and quotation of the work which has been done. 
We apologize for the contributions which could have been forgotten because 
of lack of information about their existence.

\newpage

\section{Properties of excited nuclei and nuclear matter : 
first experimental and theoretical attempts}
\indent

  Finite nuclei in their ground state and low-lying excited states
have been studied for over 50 years. They are still objects of interest,
in particular the so called exotic nuclei which contain a larger number of
protons or neutrons than those which lie in the so called valley of 
stability. 
These objects can be conceptually considered as finite 
representatives of infinite nuclear matter which may be realized in large 
objects of our universe like neutron stars where nucleons experience of 
course both the nuclear and gravitational interactions, the last one being 
absent from our present considerations.

  Extensive theoretical studies of the ground state properties of infinite 
nuclear matter have been pursued over several decades. They were essentially 
aimed to 
determine the non relativistic nucleon-nucleon interaction which originates 
from meson exchanges between nucleons. Its precise expression can be
fixed by means of nucleon-nucleon scattering amplitudes \cite{Lacombe,
Wiringa, Stoks, Machleidt, Engvik}. Different
expressions of this interaction have been introduced in order to construct 
effective interactions used in detailed spectroscopic studies of the ground 
state and low excited states of nuclei, see f.i. refs. \cite{Kahana, Brown}.

  At high enough excitation energy, nuclei which form liquid drops of 
fermions get unstable. It is tempting to trace a parallel with the 
behaviour of macroscopic liquids and to conjecture that the nuclear liquid 
may go over into vapour and that there may exist a transition from a liquid 
to a gas under precise thermodynamic conditions, in the limit of an 
infinite system. But this is not necessarily the case, the evolution
with increasing energy could well correspond to a smooth 
crossover from a bound to an unbound system of interacting particles. 
The question whether nuclear matter may exist in different phases is 
one of the important questions raised by the nuclear community and a central 
point of the present review. The first serious attempts to answer this 
question started about two decades ago. 

\subsection{Critical phenomena in nuclear matter : experimental facts and 
early interpretations}

\subsubsection{Fragmentation of nuclei}
\indent

  Experimental information about excited nuclear matter can only be gai\-ned 
through accelerator induced nuclear reactions by means of which excited 
nuclei can be generated and studied. Energetic beams of particles or heavy 
ions shot on nuclei lead to their fragmentation into pieces. The 
final multiplicity of fragments of different sizes which are collected by 
the detectors is correlated with the degree of violence of the collision.
To our knowledge, the first 
experiment which was concerned with the possible observation of signs 
related to a phase transition in nuclei was performed in 1982 by Minich et 
al. \cite{Minich}. Protons with energies between 80 and $350\,GeV$ were shot
on xenon and krypton targets. The resulting experimentally detected fragments 
with $A$ nucleons $(12 \leq A \leq 31)$ and $Z$ protons led to 
fragment yields $Y(A,Z)$ which could be parametrized in the form
\begin{equation} Y(A,Z) \propto A^{-\tau}f(A,Z,T) \label{Minc} \end{equation} 
where $\tau$ is a positive exponent and $f(A,Z,T)$ a Boltzmann factor which 
depends on a temperature $T$ and further constants which were fixed by a fit 
procedure to the experimental data. Expression (\ref{Minc}) was in fact 
inspired by Fisher's droplet model \cite{Fisher} which for the pupose of 
the analysis was generalized to two
types of particles, neutrons and protons. The value of the critical 
exponent extracted from the fit of the data was $\tau=2.64$ and 2.65 for
xenon and krypton respectively. It has to be compared to the value
$\tau=2.33$ obtained in mean field theory \cite{Jaqaman}. Fisher's 
condensation theory predicts that the transition from a liquid to a gas 
proceeds via the formation of droplets whose size distribution follows
a power law at the critical transition point. The authors do not
claim that these results would be a proof for the existence of a critical 
behaviour but that they are consistent with it. Following the droplet
model, the increase in excitation energy would generate instabilities in 
nuclear matter and would lead,
at some critical temperature $T_c$, to the disassembly 
of the system, which is here finite, into smaller pieces of all sizes which 
are experimentally detected.
 
  This pioneering experiment has been prolongated in further work which was 
pursued by Panagiotou et al. \cite{Panagiotou, Panagiotou1}. 
The authors introduced a 
systematic study of twelve different high energy reactions induced by 
protons and carbon ions on Ag, Kr, Xe and U targets. They measured fragment
yields corresponding to nuclei with $3 \leq Z \leq 22$. Using the 
same point of view as in ref. \cite{Fisher} the yields of fragments of size $A$
were parametrized in the explicit form
\begin{eqnarray}
P(A) & \propto & A^{-k} \exp\{[-a_s(T)A^{2/3}-a_v(T)A+\mu(T)A]/T\} \nonumber \\
& = & A^{-k}X^{A^{2/3}}Y^A \nonumber \end{eqnarray}
where \begin{equation} X = \exp[-a_s(T)/T] \nonumber \end{equation}
and \begin{equation} Y = \exp[-(a_v(T)-\mu(T))/T] \nonumber \end{equation}
Here $T$ is the temperature of the fragmenting system, $a_s, a_v, \mu$
are the surface, volume free energy per particle and the chemical potential 
respectively. $T$ was determined independently for the different systems by 
different means \cite{Panagiotou}.
Under the assumption that there exists a critical 
temperature $T_c$ at which $a_s(T)$ vanishes, the different quantities were 
further parametrized as
\begin{eqnarray}  a_s(T) & = & 18.4(1-T/T_c)^2 \nonumber \\
 a_v(T)-\mu(T) & = & b(1-T/T_c)^2 \nonumber \end{eqnarray} 
where $b$ is a coefficient which has to be fixed. For $T < T_c$ where the gas 
and liquid coexist
\begin{equation} a_v(T)-\mu(T) = 0 \nonumber \end{equation}
\begin{displaymath} \textrm{hence } Y=1 \end{displaymath}
at $T=T_c$. Furthermore
\begin{equation} a_s(T_c) = 0 \nonumber \end{equation}
\begin{displaymath} \textrm{hence } X=1 \end{displaymath}
For $T>T_c$ one assumes that $a_s(T) \simeq 0$.

  The expression $P(A)$ was fitted to the experimental data in order to fix 
$b$ and $T_c$, for different values of $k$ (=1.6 - 2.33) and 
in a temperature range $T'=T \pm 1$
where $T$ is the supposed
common temperature of all emitted fragments. The values of 
$T$ were chosen within an interval of $2\,MeV$ in order to take the 
uncertainty related to this quantity in account. If one fixes the critical 
temperature to the lowest value of $\tau(T)$ in a parametrization 
$P(A,T) \propto A^{-\tau(T)}$ where $\tau$ is a so called apparent 
exponent, then the fitted value of $T_c$ coincides for
$1.7 \leq k \leq 1.8$ with $T_c \simeq 12\,MeV$. Further analysis of the 
same type \cite{Panagiotou1} confirmed these results, i.e. a critical 
temperature which lies in the range 9.5 - 13.5 $MeV$ 
and an exponent $k$ smaller than 2.

\subsubsection{Thermodynamic interpretation of fragment yields : comments 
and discussions}
\indent

  The model which is aimed to reproduce the outcome of the fragmentation 
reactions presented above presupposes that the process can be described in 
the framework of thermodynamics and the grand canonical ensemble. It 
assumes that the fragmenting system is in thermodynamic equilibrium. This 
assumption is strong since it implies that there exists a relaxation time 
which is not longer than the time interval which separates the beginning of 
the process and the time at which fragments are formed. We shall come back 
to this point at different places below.

  The system undergoes a phase transition at a characteristic temperature 
$T_c$. For $T < T_c$ a gas and a liquid coexist. The liquid phase is made 
of particles and composite droplets (fragments) whose energy is composed of 
a volume and a surface contribution. At $T = T_c$ the surface contribution 
disappears, the droplets disassemble into independent nucleons, hence form 
an homogeneous gas phase which is reached for $T > T_c$.

  Taking the problem of thermal equilibrium apart, further thinking about 
this attractive picture leads to the following comments. First, the data 
analysis is restricted to a finite range in the size of fragments. Light 
and heavy species are not taken in account. This may introduce a bias which 
is not under control. It has to be noticed that the exponents $\tau$
and $k$ which were obtained in refs. \cite{Panagiotou}
and  \cite{Panagiotou1} do not coincide with the exponent expected from
the Fisher model 
which describes classical and infinite systems. Finite size effects 
should play a role in the determination of critical exponents, here they 
are not taken in account. Finally fragments and particles are free, i.e. 
they do not interact with each other. Even if the nuclear interaction is 
weak at the fragmentation stage because the system has already expanded, 
the Coulomb energy is present and certainly non negligible. Hence it 
should be taken into account.

  At this stage, even though there were encouraging signs for the 
existence of a critical 
behaviour of nuclear matter, it appeared that the observed phenomenon was 
not reducible to an ordinary liquid-gas phase transition as described by 
Fisher's droplet model. However, these first steps and the concepts they 
introduced paved the way to a huge amount of theoretical work
which we shall briefly describe below, as far as the existence of 
different phases is concerned.

\subsection{Microscopic interpretations}
\indent

  Here we restrict ourselves to those approaches which have been introduced 
as stationary and non-stationary descriptions of finite excited systems.
They were aimed to describe the response and fate of nuclei when energy is 
imparted to them.

\subsubsection{Temperature-dependent Hartree-Fock approaches}
\indent

\begin{figure}[p]

\centerline{{\large \bf FIGURES}} 

\centerline{\epsfig{figure=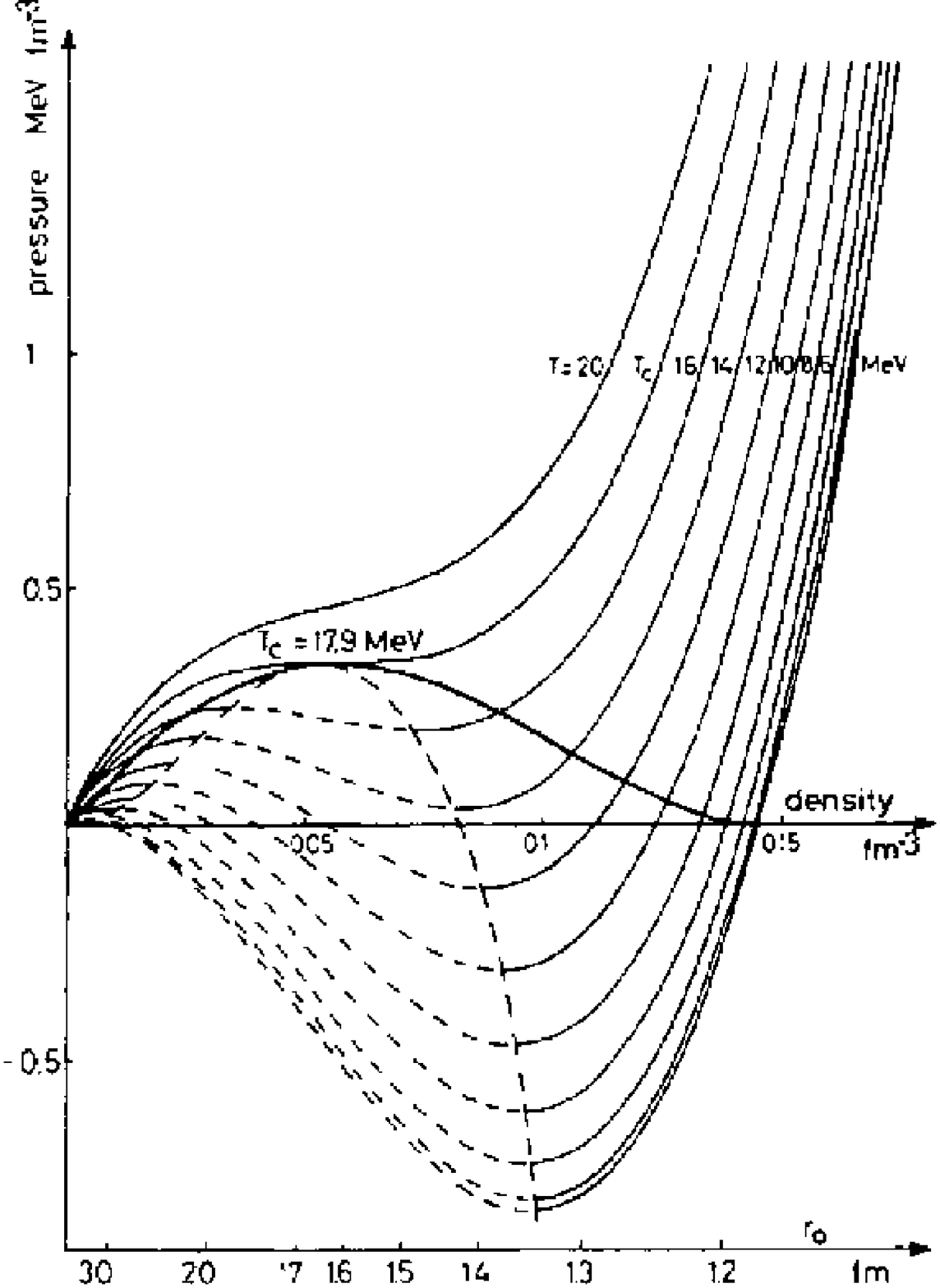,height=160mm}}
\caption{Equation of state relating the pressure to the density in nuclear 
matter \cite{Sauer}. The curves represent isotherms. The spinodal region 
is indicated by dashed curves.}
\label{fig1}
\end{figure}

  The droplet model \cite{Fisher}
is a classical and phenomenological description of an 
excited, inhomogeneous and unstable matter phase. The most common 
approximation introduced in the framework of the quantum many-body problem
is Hartree-Fock theory in which nucleons move in a self-consistent
mean field which can be described by
an effective density-dependent interaction. This framework has been used 
up to the present time for the study of static nuclear properties, a 
detailed review of its applications can be found f.i. in \cite{Quentin} and 
references therein. The theory has been extended to the study of systems at 
finite temperature in the framework of the canonical and grand canonical
ensembles \cite{Fetter, Sauer, Curtin, Schulz, Bonche, Bonche1}.
Using point contact and density-dependent effective 
interactions like Skyrme interactions it is easy to derive explicit 
expressions of different equations of state such as the expression of the 
pressure $P$ as a function of the density $\rho$ for different values of the 
temperature $T$. The typical behaviour of these quantities is presented 
in Fig.~\ref{fig1}. 
One observes the existence of an area limited by an envelope, the 
spinodal line, where the pressure decreases with increasing density. This 
is the signal for the presence of an instability of matter which 
corresponds to the formation of an inhomogeneous medium. It is 
interpreted as a zone where liquid and gas coexist and which is not 
thermodynamically accessible. The transition from an homogeneous to the 
unstable regime corresponds to a first order phase transition in the 
thermodynamic limit. The transition is of second order at the critical 
point corresponding to the critical temperature $T_c$, at the maximum of 
the spinodal line where $\ud^2P/\ud\rho^2=0$. This behaviour establishes a
connection with the droplet model physics. 
Condensation has been studied in this framework \cite{Jaqaman}. 
The calculations fix the critical density $\rho_c$ and the 
critical temperatures. The presence and importance of the Coulomb 
interaction was first taken into account by Levit and Bonche \cite{Levit}
who introduced a description in which the system is 
composed of a bound, excited nucleus surrounded by an external gas made of 
light particles in thermal equilibrium with the nucleus. The presence of 
the long range Coulomb force produces a sizable qualitative effect. It
results in the apparition of a so called limiting temperature, $T_{lim}$, 
which restricts the instability zone to temperatures which are lower than 
$T_c$ and hence introduces an important modification. Similar investigations 
were performed in refs. \cite{Jaqaman1, Jaqaman2}
with different Skyrme interactions and 
refinements concerning, in particular, the treatment of the vapour charge.

  The temperature-dependent Hartree-Fock approach with density-depen\-dent 
effective interactions is a simple and elegant approximation to the
many-body problem of finite excited systems. 
A priori it is possible to find the 
correspondence with central ingredients of the droplet model like the 
concept of surface in the liquid phase. It is however difficult to control 
the degree of realism of these microscopic quantum approaches. The fact 
that the process is described in the framework of the canonical or grand 
canonical ensemble when the fragmenting system is closed and has a fixed 
and small number of particles may lead to difficulties in the 
characterization of the transition. We shall come back to this point in the 
sequel. There are no means to obtain fragment yields which could be 
compared to the experiment since there are only two types of species, a 
bound nucleus and light particles which constitute respectively the liquid 
and the vapour phases. This is of course directly related to a major 
difficulty which comes from the mean field description. The phases are 
homogeneous. Many-body correlations induced by the neglected residual
two-body interactions are absent, but their effect is very important at the 
considered excitation energies. This may also explain the high values of the 
temperature $T_c \simeq 15 - 20\,MeV$ where the system becomes unstable. 
Indeed, the interpretation of experiments indicates that these temperatures 
could be much lower, of the order of $4 - 6\,MeV$.

  Hence this kind of approach is at most indicative for the determination 
of the properties of finite excited nuclear matter in thermodynamic 
equilibrium. It worked however as an incentive to develop more realistic 
approaches which take explicitly care of the many-body aspect of the 
problem. We describe some of them below.

\subsubsection{Time-dependent descriptions of nuclear fragmentation}
\indent

  The general microscopic framework describing energetic collisions leading
to the decay of the system into species of all sizes is the quantum
many-body scattering theory. In this fundamental approach the scattering
$S$-matrix elements $S_{fi}$ measure the overlap between any arbitrary
many-body initial scattering state $i$ with any arbitrary final state
$f$ \cite{Newton, Wu}. A full fledged microscopic approach of 
this type is however hopelessly out of scope because of its mathematical 
intricancy. Many pragmatic ways have been devised in order to circumvent 
this problem. They all rely on non relativistic formalisms in which non 
nucleonic degrees of freedom are parametrized by means of nucleon-nucleon 
potentials. 
All approaches which have been used in a time-dependent framework contain 
a minimum of classical ingredients. Indeed, all of them fix initial 
conditions at an initial time and follow the evolution of the system up to 
a final time in contradistinction with scattering theory which specifies 
both the initial and the final state of the system.

  It is not necessarily clear that classical concepts and assumptions should 
work in the present physical context, except for the fact that high energy 
processes are correlated with short wavelengths. One may however notice that 
processes like deep-inelastic or fission reactions which occur at lower 
collision energies can be reasonably and surprisingly well reproduced in 
terms of classical (or semi-classical) equations of
motion \cite{Weidenmuller, Frobrich}.

  An explicit time-dependent description generally called ``dynamic''
description shows a priori a further advantage, since it
allows to avoid the crucial point concerning thermodynamic equilibrium.
Indeed, the dynamical description of the evolution of the system may lead or
not to thermodynamic equilibrium which follows a transient
period of time in a collision process. 

  There exists essentially two classes of models of this type. The first 
one transcribes the many-body equations of motion into kinetic equations
of motion for the one-particle phase-space distribution function
$f(\vec r, \vec p, t)$ at point $\vec r$ with momentum $\vec p$. Each phase 
space cell $\{[\vec r,\vec r + \ud\vec r],[\vec p,\vec p + \ud\vec p]\}$
evolves in time through the transport equation
\begin{equation}
\frac{\partial f}{\partial t} + \frac{\vec p}{m} \cdot \frac{\partial f}
{\partial\vec r} + \frac{\partial V}{\partial \vec r} \cdot \frac{\partial f}
{\partial \vec p} = I[f] \label{transp} \end{equation}
where $V(\vec r)$ is a one-body potential and $I[f]$, the so called 
collision term contains in principle all the information relative to
$2-, 3-, n-$body correlations which are generated by the two-body (possibly
also $3-,\ldots, n-$body) potentials beyond the average field $V$. 
This quantity 
and $I[f]$ are in principle correlated and fixed in the framework of the 
initial many-body equation of motion in such a way as to satisfy 
conservation laws. In practice however they are 
determined through phenomenological considerations which preserve more or 
less the quantal aspects of the initial problem. These approximations are 
particularly tricky for $I[f]$ which has to be chosen in such a way that it 
can be expressed in terms of one-body distributions, in order to ensure the 
closed form of (\ref{transp}) \cite{Bonasera, Cugnon}. The mean field 
potential $V$ is generally fixed independently, from purely phenomenological 
formulations to self-consistent density-dependent expressions derived by
means of static Hartree-Fock calculations. As a consequence,
a large amount of models of the type of Eq.~(\ref{transp}) have 
been proposed in the litterature, see \cite{Bonasera} and refs. therein.
They have been used as descriptions of many-body systems,
among which high energy nuclear collisions.
It is not our aim to give a detailed report on the achievements of these 
descriptions. It has been done in different publications and reports, 
see the references above.
Here we want to restrict the subject to the discussion of some aspects
of nuclear fragmentation which have been addressed in this framework.
They are of interest for the understanding of the fragmentation process 
and problems related to thermodynamic equilibrium, the possible existence
and (or) the coexistence of phases in nuclear matter.

 In the formulation of Eq.~(\ref{transp})
the mean field $V$ plays an essential role 
and its choice is central for a realistic description of the evolution of the 
system at high energy. As we have seen in section 2.2.1, realistic mean field 
calculations in the framework of Hartree-Fock theory with effective
density-dependent potentials $V$ and the equation of state $P(\rho)$ 
lead to a 
zone of instability and possibly of metastability whose separation from pure 
``liquid'' and ``gas'' phases corresponds to the existence of a first order 
phase transition, see Fig.~\ref{fig1}. Transport equations may be 
considered as time-dependent extensions of mean field approaches which 
furthermore avoid the problem of the existence or not of thermal 
equilibrium in the expanding and decaying system. Much work has been done 
in this framework in order to identify the existence and consequences
of an instability zone 
where the initially homogeneous system gets inhomogeneous and appears as
formed of clusters which can be identified \cite{Bonasera1, Colonna}. This 
has indeed been achieved in a pure mean field approximation, in the absence 
of the collision term, when the mean field is supposed to possess a 
fluctuating contribution \cite{Bonasera2, Colonna1, Chomaz, Jacquot, 
Jacquot1}, and (or) when the collision term induces 
fluctuations in the one-body distribution function
$f(\vec r, \vec p, t)$ \cite{Colonna2, Colonna3}.
These fluctuations develop collective modes like zero sound, some of which
are unstable and break the system \cite{Colonna4, Ayik}.
The expansion of the system affects also the development 
of the instabilities. The presence of two types of particles, 
protons and neutrons has been taken into account \cite{Baran} and comparisons 
with experimental results have been made \cite{Rivet}. Quantum effects on 
the expansion dynamics have been investigated recently. It comes out that 
they are non negligible, especially when the energy of the collective modes 
which develop in the system are larger than the temperature \cite{Lacroix,
Wen}.

  This approach which describes the fragmentation of nuclei as a 
mechanical process shows however limitations due to 
the already mentioned fact that Eq.~(\ref{transp}) is an approximation to 
the many-body problem whose expression allows for many more or less 
controlled phenomenological ingredients, such as stochastic contributions 
in the mean field and (or) the collision term. These are supposed to
simulate the missing many-body features. Unfortunately it is 
difficult to control their degree of consistency, much physical information 
is hidden in numerics and may obscure the conclusions which can be drawn 
from comparisons with the experiment. On the other hand one can believe 
that transport models of the type 
given by Eq.~(\ref{transp}) are able to describe the 
time evolution of global observables like inclusive reaction cross
sections and collective properties like flow \cite{Bonasera}.

  The second approach concerns molecular dynamics. It is also an 
approximation to the time-dependent many-body problem.
Different formulations starting from the most 
classical one, Classical Molecular Dynamics (CMD) try to pick up quantal 
aspects at different levels, going from ``Quantum'' Molecular Dynamics 
(QMD) which tries to approximate the many-body wave function at the initial
time by means of coherent (gaussian) states and takes care of the Pauli 
principle \cite{Aichelin, Bonasera3} to the more recent Fermion Molecular
Dynamics (FMD) \cite{Feldmeier, Feldmeier1, Feldmeier2} and Antisymmetrized 
Molecular Dynamics (AMD) \cite{Horiuchi, Ono, Sugawa, Tosaka} which 
introduce the antisymmetrization of wave functions in a semi-classical way.

\begin{figure}[b!]
\centerline{\epsfig{figure=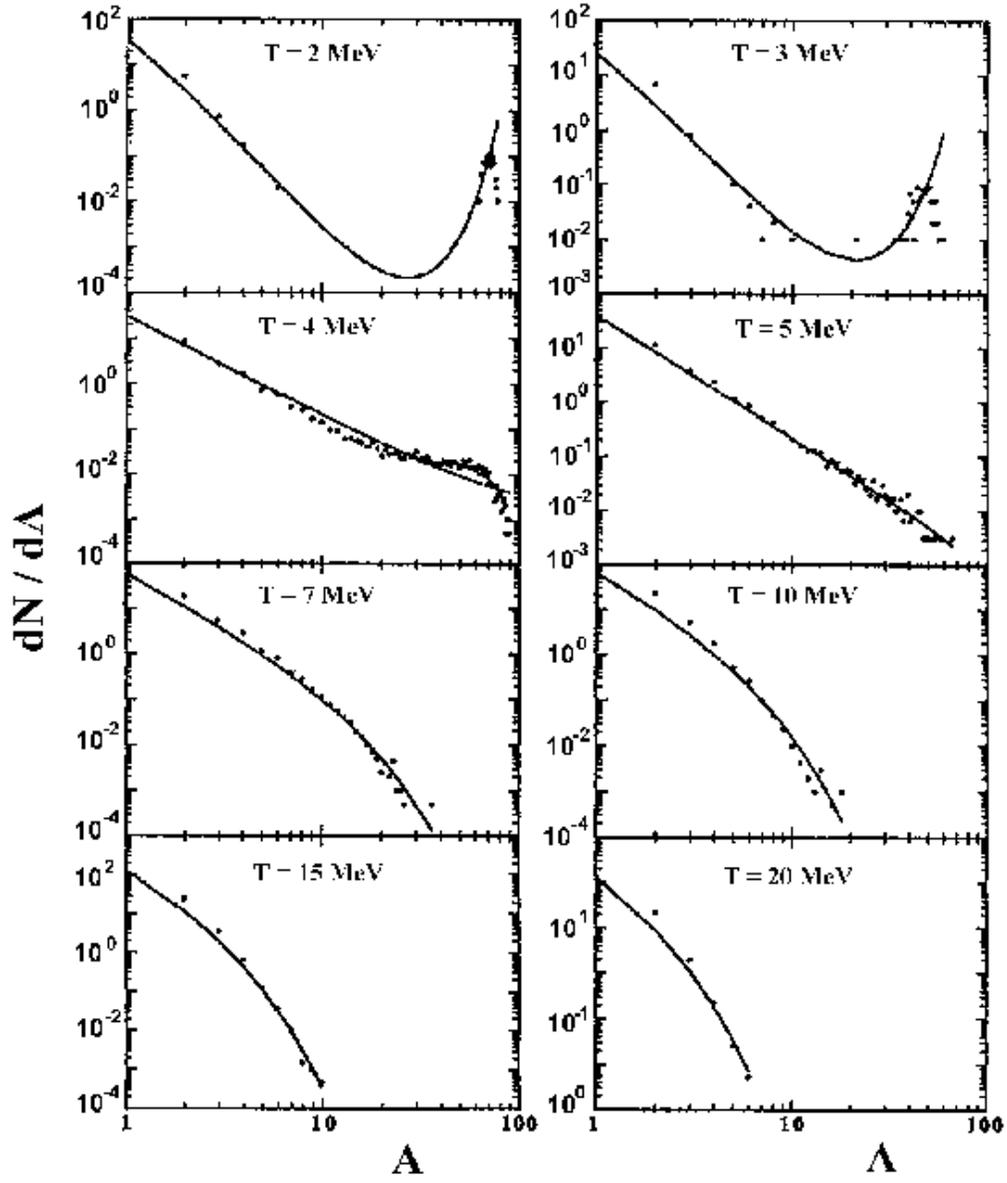,height=160mm}}
\caption{Mass distributions for an expanding system of $A = 100$ particles 
for different initial temperatures \cite{Belkacem}. The dots are obtained 
by means of CMD calculations, the lines are fits using Fisher's formula.}
\label{fig2}
\end{figure}

  In the simplest form (CMD) the time evolution of the system is 
described in terms of classical equations of motion for protons and 
neutrons which interact through a two-body potential representing the 
nuclear interaction and the Coulomb force which acts between protons. These 
interactions are implemented in such a way that the fundamental ground 
state properties of real nuclei are reproduced. The effects of 
the Pauli principle can be mimicked by means of a momentum-dependent 
potential \cite{Aichelin}. Applications of this formalism have been developed 
into different directions. We shall come back in the forthcoming section to 
the description of heavy ion collisions in which theoretically calculated 
and experimental determined observables are confronted in order to fix the 
possible relationship between fragmentation and phases in nuclear matter. 
Here we concentrate on applications in which Molecular Dynamics (MD)
has been implemented in order 
to study the properties of a thermalized system for which the velocity 
distributions of particles are maxwellians. From there on the system 
evolves in time by expansion for different initial values of the 
temperature and the density \cite{Latora, Belkacem, Finocchiaro}
One follows the evolution of the temperature,
the pressure and the fragment size distribution of the system 
with decreasing density. Fragments can be identified by means of different 
criteria which may be more or less realistic \cite{Belkacem, Dorso,
Strachan}. In some cases the system crosses the spinodal line and 
enters the instability zone. Different observables show a 
behaviour which can be interpreted in terms of Fisher's droplet
model, i.e. a fragment size distribution corresponding to a power 
law corrected for finite size effects and a corresponding critical exponent
$\tau \simeq$ 2.2 to 2.3, which goes over to other characteristic shapes 
for other choices of the temperature as expected in the droplet theory,
see Fig.~\ref{fig2}. These results along with the interpretation of other 
observables which will be defined below are interpreted as the outcome of a 
realistic simulation of the experimental situation and the sign for the 
existence of a critical behaviour of the system. The role and importance of 
the Coulomb interaction has been investigated \cite{Belkacem}. Molecular 
Dynamics has also been used in order to construct the equation of
state \cite{Finocchiaro}, i.e. the 
thermodynamic properties which link density, pressure and temperature. It 
is again shown that there exists a critical point for finite systems, and a 
power law distribution with $\tau \simeq 2.23$ which can be interpreted in 
the framework of Fisher's model.

  However these results and conclusions concerning criticality have been 
put in question by another study \cite{Pratt} which claims that the observed 
signs have not to be interpreted as being related to a critical behaviour 
but for the existence of an intermediate regime. This 
behaviour would correspond to a trajectory in the $(T,\rho)$ plane on which 
the evolving system does neither fall back to a drop nor develop into a gas.
A power law distribution can be extracted from the data. The power law 
exponent depends sensitively on the initial temperature and the size of the 
system. In ref. \cite{Pratt} it is claimed that the power law behaviour of 
mass yields expected for a system close to its critical point would be 
purely accidental.

  Time-dependent transport descriptions of nuclear fragmentation seem a 
priori to be the most realistic way to the description of high energy
nuclear collisions. They have been widely used in order to describe the 
possible out-of-equilibrium character of the fragmentation process which 
characterizes at least the first stage of the process \cite{Baldo}.
In the present section we concentrated on aspects related to a 
time-dependent description of thermodynamically equilibrated systems which 
expand and cool down, exploring different regions of the $(P,\rho)$ and 
$(T,\rho)$ planes. Calculations show that the results can be interpreted as 
the entrance of the system into a thermodynamic unstable regime. It seems 
however not clear how this can effectively be related to the existence of a 
phase transition which would characterize an equilibrated system heated up 
in a fixed volume. In an interpretation in which the system gets critical 
it is claimed that it is possible to extract critical exponents, at least 
the power law exponent $\tau$ \cite{Latora}. However this quantity is obtained 
from finite size systems and hence should be corrected for finite size 
effects. 

  The time coordinate which is introduced in these descriptions 
raises the question of time scales, i.e. the question concerning 
equilibration in energetic light particle reactions or heavy ion collisions.
This is the subject of section 3. Last but not necessarily least,
all these time-dependent approaches need a detailed description of the 
dynamics in terms of more or less sophisticated mean field and two-body 
potentials. It is always difficult to get a clear feeling for the sensivity 
of the outcome of numerical calculations and simulations to these details.
Paradoxically, it is at criticality that the information on microscopic 
systems is the less sensitive to the details of the Hamiltonians which 
govern them.

  Finally it is worthwhile to mention here that 
it has been shown recently \cite{Egolf} that even
far-from-equilibrium, spatially chaotic systems can show equilibrium 
properties such as ergodicity (equivalence between time and ensemble 
averages), detailed balance in microscopic processes, partition functions 
and renormalization group flow at coarse-grained lengths between 
microscopic and macroscopic scales. This suggests that it might be possible 
to describe the global behaviour of some far-from-equilibrium systems in the 
framework of equilibrium statistical mechanisms. Whether this result could be 
some clue to the actual question raised in the present framework is an open 
and highly interesting question.

\newpage

\section{Thermodynamic equilibration and phase space descriptions of nuclear 
fragmentation}
\indent

  If an isolated system of particles is such that there is no unbalanced 
force acting between its constituents and if it does not experience 
internal structure changes of its constituents it is said to be in 
mechanical and chemical equilibrium. Thermal equilibrium is realized if the 
relevant degrees of freedom which characterize the system share equally its 
macroscopic energy. One can then define a common temperature which 
characterizes the energy attributed to each degree of freedom.
When all three conditions are realized the system is said to be in 
thermodynamic equilibrium \cite{Zemanski} and can be described in terms of 
macroscopic coordinates, the thermodynamic coordinates like density, 
pressure and temperature which do not change with time and are linked 
together through equations of state. In the present section we aim to 
present and discuss the difficult and for a large part unsettled problem of 
thermodynamic equilibrium in small systems like nuclei excited by means of 
energetic collisions between particles or nuclei 
called heavy ions if their mass is larger than 4. In section 3.1 
we discuss questions related to the interval of time over which equilibrium 
can be reached in terms of model estimates. A confrontation between 
experimental facts and phase space equilibrium models is presented in 
section 3.2. 
  
\subsection{Thermodynamic equilibration in excited nuclear systems 
generated by nuclear collisions}
\indent

  What is the scenario which governs energetic collisions? Data analysis 
and their interpretation show that several scenarii are at hand. Typically 
one expects that part of the bombarding energy is transferred to the 
internal degrees of freedom of the system, the system may fall into pieces, 
particles and fragments, and expand in time. The remnants are then collected
at asymptotic distances in detectors which identify their charges and 
eventually their masses, along with their kinetic energies and 
distributions in space. The setting up of thermodynamic equilibrium over 
the whole or part of the interacting system of particles is a central but 
open question. Under favorable circumstances one may figure out that after 
some transient interval of time which starts at the beginning of the 
collision the system reaches at least partial equilibrium at some stage and 
then expands to infinity.

\subsubsection{Low energy time scales}
\indent

   At very low projectile energies the excitation of nuclei by means of 
light particles can lead to the formation of an excited complex whose 
behaviour can be interpreted within the framework of thermodynamics as a 
system in thermodynamic equilibrium and which can loose part of its 
excitation energy through the emission of particles, photons or through 
symmetric or asymmetric fission. The complex is the so called compound 
nucleus whose formation and decay properties were already studied in the 
30's by N. Bohr \cite{Bohr, Bohr2} who conjectured that, in the framework
of this scenario, the process can be explained as a two step mechanism in 
which the decay of the equilibrated system would be totally uncorrelated 
from the way it was generated. From a theoretical point of view compound 
nucleus theory is a phenomenological approach. It has been largely 
investigated and worked out in terms of microscopic models, among others 
random matrix theory. There exists an extended litterature on this subject 
which covers several decades of intensive work \cite{Mahaux}. Indeed, the 
consequences of the model have been abundantly verified by means of 
physical observables like excitation functions and angular distributions of 
emitted particles \cite{Mahaux, Hauser}. The lifetime $\tau$ of the 
compound system can in principle be read from the compound nucleus decay 
width $\Delta E$ through $\Delta E \cdot \tau \geq \hbar$ leading to 
$\tau \simeq 10^{-20} - 10^{-17} s$ depending in particular on the size of 
the system. These times are substantially larger than 
$\tau_0 \sim \ell/3 \times 10^{-23} s$ corresponding to the time necessary 
to light to cross a nucleus of diameter $\ell\,fm$.

  At higher energies, above the Coulomb barrier which corresponds to 
bombarding energies per particle 
$E/A \simeq 1 - 2\,MeV$, reactions involving heavy ions lead to a new 
mechanism the so called deep inelastic collision (DIC) process. Typical 
contact times during which the projectile and the target interact through 
the short range nuclear potential are much shorter than compound nucleus 
decay times, of the order of $10^{-21} s$, not much larger than the 
contact time in so called direct reactions in which the ions strike each 
other peripherally in typical time intervals of $10^{-22} s$. The 
interpretation of physical observables measured in these DICs shows that 
although the characteristic reaction times are quite short, the complex which 
is formed over the contact time of the two ions is sufficient to transfer 
sizable amounts of energy from the relative motion between ions
into the excitation of the 
nucleonic degrees of freedom and the characteristic features of the system 
can be interpreted in terms of an equilibrated system whose equilibration 
is reached in as short as some units of $10^{-22} s$ \cite{Weidenmuller,
Boose}.

\subsubsection{Approaches to the equilibration problem at high energies}
\indent

\begin{figure}[b!]
\centerline{\epsfig{figure=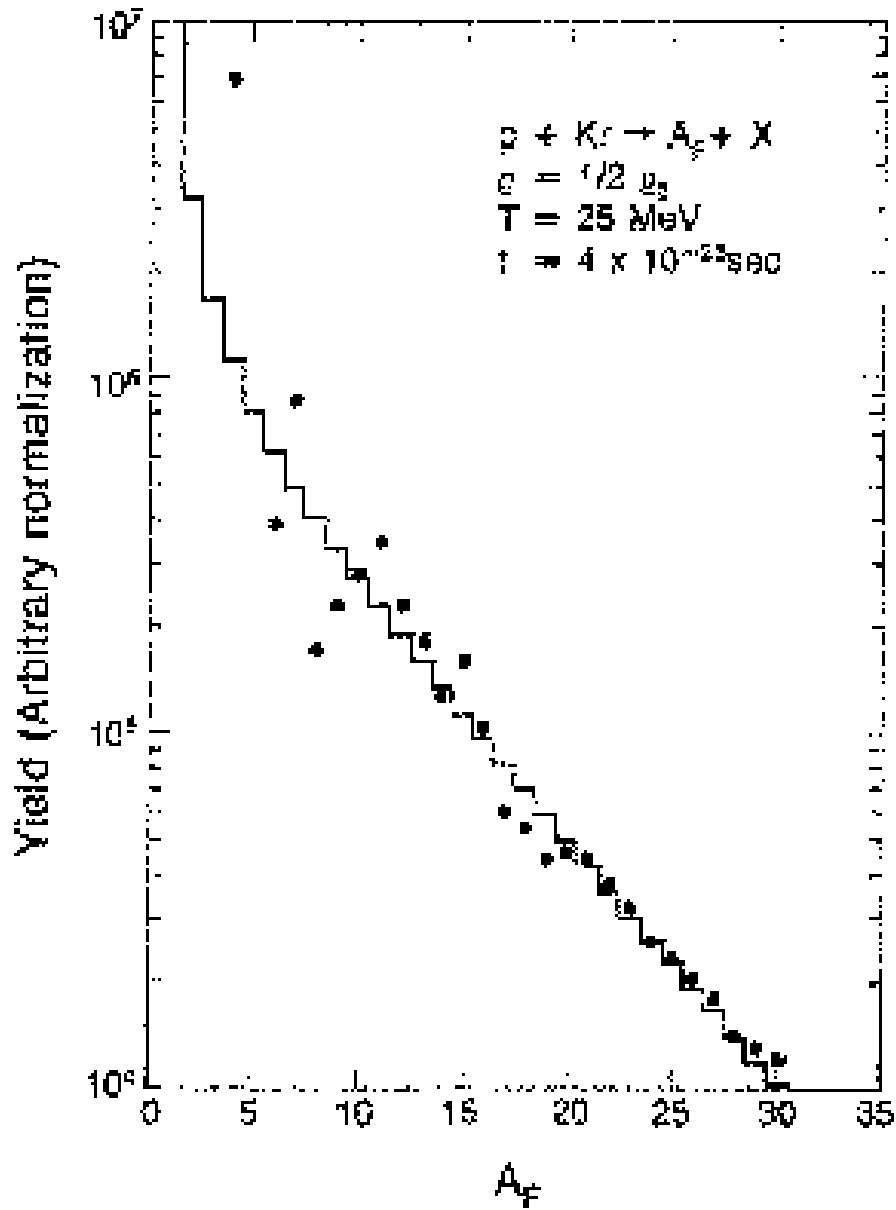,height=160mm}}
\caption{Histogram of mass yields obtained by means of the rate equations
(\ref{rate}) after $ t = 4 \times 10^{-23} s$ \cite{Boal}. The dots are 
experimental results of the reaction $p + Kr$ at $E = 80 - 350 \, GeV$.}
\label{fig3}
\end{figure}

  How does this behaviour extrapolate to high energy events which can lead 
to the fragmentation of the system into many pieces of different sizes? The 
problem is intricate since it is difficult to guess the scenario (or the 
different scenarii) as already mentioned above. Hence one may proceed in 
steps, starting with simple-minded models. To our knowledge, the first 
attempt was made by Boal \cite{Boal}, at the time where the experiments, 
which led to the belief that signs for the existence of a phase transition 
had been seen, were performed \cite{Panagiotou, Panagiotou1}. 
In fact the question
which was raised did not concern the equilibration but the fragment 
formation time. Former experiments \cite{Anderson} had shown that the ratio 
between inelastic proton reactions $\sigma(p,p')$ and proton-neutron 
exchange reactions $\sigma(p,n)$, $\sigma(p,p')/\sigma(p,n)$, is of the 
order of 2 at $100\,MeV$ incident energy after correction for the $Z/N$ 
ratio $(Z, N = $ 
proton and neutron number). This is evidence for the fact that chemical
equilibrium is not reached, hence strictly speaking thermodynamic 
equilibrium  cannot be either, see above. The author of ref. \cite{Boal}
postulates that a hot system generated at high temperature cools down to a 
fraction of this temperature and, from thereon, experiences particle 
coalescence leading to the formation of clusters which come out as 
fragments. Defining $N_i(t)$ as the number of clusters of size $i$ which 
exist at time $t$ normalized to the total number of particles $A_T$, the 
initial conditions are supposed to be given by
\begin{eqnarray} N_1(t=0) &=& \rho  \nonumber \\
N_i(t=0) &=& 0  \qquad 2 \leq i \leq A_T \nonumber \end{eqnarray}
where $\rho$ is the matter density. The system starts at temperature $T$ 
from an homogeneous distribution of particles and its evolution is given by 
the postulated rate equations
\begin{equation} \frac{\ud N_k(t)}{\ud t} = \sum_{i,j}
\frac{N_i(t)N_j(t)}{1+\delta_{ij}} \overline{\sigma_{ij}} \delta_{i+j,k} -
\sum_iN_i(t)N_k(t) \overline{\sigma_{ik}} \label{rate} \end{equation}
where the first term on the r.h.s. of (\ref{rate}) corresponds to the 
coalescence of species of size $i$ and $j$ to species of size $k$ and the 
second term to the decay of species of size $k$ to species of size $i$. The 
coalescence and break-up rates are given by
\begin{equation} \overline{\sigma}=4 \pi \bigg(\frac{\mu}{2 \pi T}\bigg)^{3/2}
\int \ud v v^3 \sigma(v) \exp(-\mu v^2/2T) \nonumber \end{equation}
where $\mu$ is the reduced mass of the colliding species, $v$ their 
relative velocity and $\sigma(v)$ the corresponding cross section. The 
solutions of the system of coupled equations (\ref{rate}) reaches the 
asymptotic regime after $4 \times 10^{-23} s$. The results are shown in 
Fig.~\ref{fig3} by the histogram for the case of a reaction $p + Kr$ at 
energies between $80$ and $350\,GeV$. Their global trend agrees quite nicely
with the experimental points. The asymptotic time is in agreement with the
$\sigma(p,p')/\sigma(p,n)$ ratio and the estimated rate of cooling of the 
excited source, $\sim 10^{-23} s$. One should mention here that the energy 
is very high and that thermal equilibrium is postulated since one defines a 
temperature $T$ which is furthermore kept constant over the duration time 
of the process. The expansion of the system in space is not taken into 
account. In a further work \cite{Boal1} Boal and Goodman abandoned the 
coalescence scenario of an initially homogeneous and expanded system. They 
proposed a break-up mechanism related to the entrance of the system into an 
instability region of the equation of state as already discussed in section 
2. The collision of energetic protons $(E_p = 300\,MeV)$ with nuclei was 
simulated by means of a simplified transport equation of the 
Boltzmann-Uehling-Uhlenbeck (BUU) type. The authors emphasize the fact that 
thermal equilibrium, requested by a classical collision time
between nucleons which is short compared to the expansion time of the 
nucleon gas, may not be established at a time where the system is already 
dilute, a regime which is reached after a very short characteristic time 
which is smaller than $10^{-22} s$. This goes in the same direction as the 
non equilibrium $\sigma(p,p')/\sigma(p,n)$ which was experimentally found. 
Notice however that the energies involved here are very large, non 
nucleonic degrees of freedom are not taken into account and the transport 
description is somewhat schematic.
The different models are used in order to explain the increase of entropy 
of the system when it crosses the spinodal line and enters the instability 
zone. This entropy agrees with the quantity extracted from the experiment. 

  Some aspects of fragmentation related to time scales have been discussed by
Gross and coll. \cite{Gross}. The reaction process goes through a transient 
out-of-equilibrium phase which may ultimately lead to the formation of an 
equilibrated excited system. In order to estimate the transient time the 
authors use a BUU-type equation \cite{Bauer, Bao, Bao1, Gross1} which
is a mean field description without collision term as described in section~2.
They argue that this description may be valid at the beginning of the 
collision process, up to and not further than the time when the system may 
break into pieces. At energies of 50 to $60\,MeV\!\cdot\!A$ this happens 
after $50 - 60\,fm/c$ for a reaction $^{96}Mo$ on $^{96}Mo$. Then the system 
expands and apparently reaches equilibrium at about $200\,fm/c$ due to the 
dissipation of energy in the nucleonic degrees of freedom.
How far the equilibrium is reached is however not firmly established.
Once the distance between the different 
species gets larger than $3\,fm$, they do no longer interact by means of the 
nuclear potential. The system has then reached the so called freeze-out 
stage in which it is in chemical equilibrium. A phase space description of 
fragments and particles which move under their mutual Coulomb interaction 
can then be introduced. The early time behaviour is confirmed by 
experimental results \cite{Gelderloos}.

  Recently relativistic transport equations \cite{Fuchs} have been used 
in order to simulate the heavy ion reaction $Au$ on $Au$ which has been 
studied experimentally \cite{Gaitanos, Gaitanos1}. The calculations show 
that after the collision the excited system separates into a participant 
region of highly excited matter and a spectator which reaches thermal 
equilibrium, approximately after $2 \cdot 10^{-22} s$. Indeed, a local 
temperature can be defined through the knowledge of fragment kinetic 
energies. It characterizes a system which is mechanically unstable and 
decays into fragments. The values of the temperatures obtained for 
different beam energies compare well with those which are extracted from 
the experiments \cite{Lisa, Reisdorf}.

  Other decay scenarii borrowed from low energy reaction mechanisms have 
been introduced \cite{Moretto, Friedman, Friedman1, Barbagallo, Richert, 
Durand}. Fragments are formed sequentially in 
time, by means of the successive binary decay of the excited species which 
already exist at a given time step. The process starts from an equilibrated 
excited system. In ref. \cite{Friedman1} this mechanism 
is supplemented by a boost in time due to the early compression of the 
system at the beginning of the process. Most of these approaches rely on 
the Weisskopf formulation of the break-up rates \cite{Weisskopf} of an 
excited bound cluster of particles. In
ref. \cite{Richert} the binary decay is interpreted in terms of fission 
barriers, using a formulation introduced by Swiatecki \cite{Swiatecki}. The
use of the Weisskopf expressions raises problems which shed
doubt about their physical relevance at high energies. First, they 
get unrealistically small at energies corresponding to temperatures above 
$4 - 5\,MeV$ \cite{Richert1}. Second \cite{Gross}, each decay event in this 
formulation should be independent of the preceding one. This is only 
possible if the characteristic emission time exceeds the characteristic 
time over which the Coulomb emission barrier changes for the following 
emission. This time is of the order of $10^{-21} s$, much longer than the 
Weisskopf emission time and not compatible with characteristic fragment 
formation times as estimated from the experiments \cite{Abouffirassi}. 
Hence, although confrontation of mass yields obtained in this framework 
with experiments may seem satisfactory it is difficult to believe this type 
of description \cite{Wagner} for highly excited systems.
It is of course sensible to believe that 
fragmentation of nuclei proceeds through a sequence of break-ups, but it is 
difficult to guess a priori or to derive from some microscopic model the 
realistic expressions of the rates which govern them when the excitation 
energy is as high as several $MeV/A$.  

  The evolution of the fragment emission time has been studied recently by 
means of two-fragment correlation measurements over a large range of 
excitation energies in reactions $\pi^-$ and $p$ on $^{197}Au$ 
\cite{Beaulieu}. The fragments are emitted from a unique source which is 
considered to be in thermal equilibrium. Confrontation of the data with 
results obtained by means of classical trajectory calculations allows to 
follow the emission time which decreases with increasing excitation energy. 
The result is interpreted as evidence for a cross-over from surface 
emission over long times as described by sequential decay and fast bulk 
emission which is characteristic of a break-up process due to mechanical 
instability.

  It has also been looked for further direct insight into the time scale 
problem. In ref. \cite{Cebra} central-impact-parameter events from the 
reaction $^{40}Ar+^{51}V$ with incident energies between 35 and 
$85\,MeV\!\cdot\!A$ were analysed for the shape of the emitted particle and 
fragment events in momentum space. It is expected that fragments which are 
emitted quasi-simultaneously show a close to isotropic emission in the centre 
of mass system, hence are spherical in shape, whereas emission times which 
are long as it may be in the case of sequential emission would correspond to 
the existence of an elongated emission axis, hence an ellipsoidal shape.
Experimental events were confronted with numerical simulations. It comes out 
that for energies below $35\,MeV\!\cdot\!A$ the fragmentation process is 
slow. For higher energies, the experimental momentum distributions lie in 
between the predictions corresponding to a slow and a quasi simultaneous
break-up process. Even though the result does not give any quantitative 
clue to the problem, it shows at least that the characteristic 
fragmentation time decreases with increasing energy. A fragmentation 
experiment induced by $^{40}Ca + ^{40}Ca$ at 
$35\,MeV\!\cdot\!A$ \cite{Hagel} confronted with different
models related to quick and slow processes 
confirms the fact that ordinary binary sequential decay may not be the 
appropriate fragmentation mechanism.

\begin{figure}[b!]
\centerline{\epsfig{figure=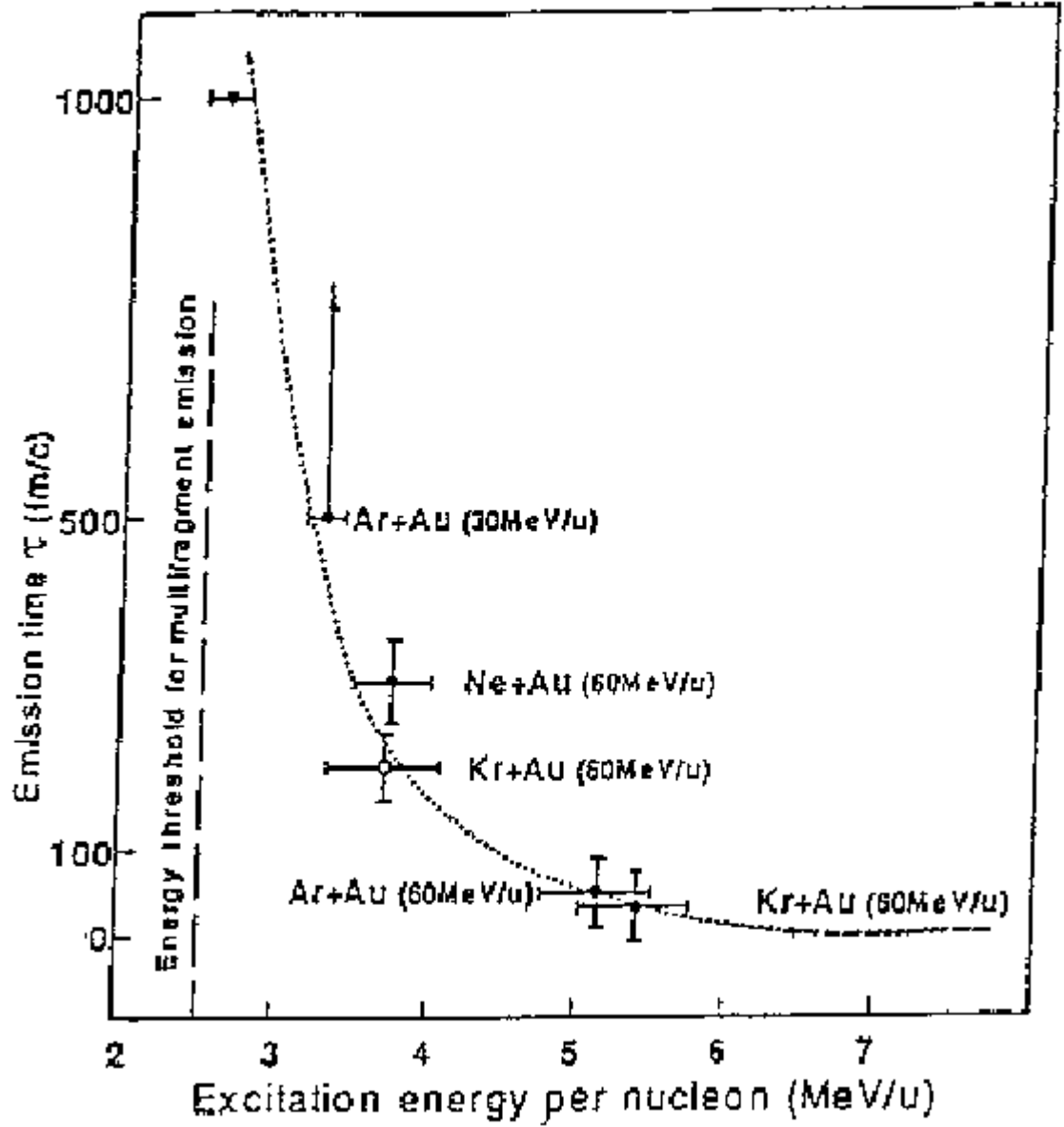,height=160mm}}
\caption{Evolution of the average length of the interval of time between 
two successive break-ups in the fragmentation process with increasing 
excitation energy \cite{Tamain}.}
\label{fig4}
\end{figure}

  Other characteristic time scales can be introduced such as the time for 
the occurence of the first break-up of the system $\tau_0$, and the average 
length of the interval of time between successive break-ups 
$\tau_{FF}$ \cite{Tamain}.
These times can be estimated by means of relative angle correlation 
measurements between breaking species. If the break-up time is slow, 
Coulomb repulsion will hinder the emission at small relative angle and low 
relative velocity. Simulations of the correlation functions indicate that
$\tau_{FF}$ corresponds to prompt decay for excitation energies exceeding
$4\,MeV/A$. Estimates of $\tau_{FF}$ are shown in Fig.~\ref{fig4}. 
A long time interval
$\tau_0$ has been observed for excitation energies lower than $3\,MeV/A$ in
peripheral collision events. Even for high excitation energies of the order 
of $5\,MeV/A \ $ $\tau_0 \sim 100 - 150\,fm/c$. This leads to characteristic 
times of $300 - 400\,fm/c$ for complete break-up which are rather large for
heavy nuclei, close to the Fermi energy. Recently further emission times 
have been deduced from IMF-IMF correlation functions obtained from hadron 
induced multifragmentation events \cite{Beaulieu}. One observes an 
evolution from typical emission times $\tau \simeq 500 \, fm/c$ at excitation 
energies $E^*/A = 2\,MeV$ to $\tau \simeq 20 - 50 \, fm/c$ for $E^*/A = 
5\,MeV$ and above. The observation of the transition is interpreted as a 
transition from surface to bulk-dominated emission. The characteristic 
times are somewhat shorter than those obtained in ref. \cite{Tamain, 
Durand1} for excitation energies lower than $6\,MeV/A$.

  Albeit the numerous and different types of investigations which have been 
made up to now the arguments concerning a quick thermodynamic equilibration of 
the system and relying explicitly on time remain the subject of controversies 
since time is not directly under experimental control. These controversies are 
in particular raised by time-dependent descriptions of the collision process, 
in the framework of the Quantum Molecular Dynamics (QMD)
models \cite{Aichelin1, Peilert, Aichelin}. These are classical molecular 
dynamics models which, to some extent, take care of the quantum nature of 
nucleons. In recent applications of this type of models to specific reactions 
which have been investigated experimentally in some details like 
$^{129}Xe$ on $Sn$ at $50\,MeV\!\cdot\!A$ incident energy,
the confrontation between calculations 
and experimental results shows good quantitative agreement, except for the 
mass yields corresponding to fragments with more than 30 nucleons and too 
small kinetic energies for fragments and particles \cite{Nebauer}. It is 
argued that the system of interacting particles is far from equilibrium, 
there is no sign for a thermally equilibrated behaviour of the species 
which are generated during the early stages of the process. In a further 
study \cite{Nebauer1} QMD calculations are confronted with 
$50\,MeV\!\cdot\!A$
$Xe + Sn$ data which were also analysed in the framework of the 
Statistical Multifragmentation Model (SMM, see below) \cite{Bondorf}. 
The approach assumes 
thermodynamic equilibrium at the freeze-out. Experimental data related to 
the multiplicities of light charged particles (LCP), the average kinetic
energy of LCPs and intermediate mass fragments (IMF)
are very close to those obtained by means of QMD and SMM simulations.
This result concerning the average transverse kinetic energy of 
IMFs and the total kinetic energy of LCPs contradicts two arguments
which were supposed to help to distinguish between a fragmentation
in an equilibrated and non equilibrated system \cite{Tocke}.

  The preceding discussion shows once more 
that the experimental data related to
fragmentation processes are ambiguous as far as the production mechanism is 
concerned. It raises the question why the outcome of different descriptions 
should be so close to each other when the process is over, i.e. at the 
freeze-out. At the present stage where different models lead to similar 
results which agree more or less satisfactorily with the experiment it is 
tempting to conclude that all of them contain part of the truth, at least 
to a sufficient degree as to agree with the measured quantities. Since 
there exists no trustful method to settle the point, one has to collect a 
maximum amount of informations, as exclusive as possible, and confront them 
directly with theoretical descriptions. In the next subsection we intend to 
show that in fact most of the data can be understood in terms of systems 
which reach thermodynamic equilibrium at the freeze-out. 

  As a final remark, one may wonder why the occupation of phase space 
leading to equilibrium goes so fast, seemingly 
even faster at high than at low excitation energy. In a quantum mechanical 
microscopic description, the decay width of a state $\vert f \! >$ into a 
state $\vert i \! >$ at the energy $E$ is given by the golden rule 
expression
\begin{equation} \Gamma(E) \propto 2 \pi \vert V_{fi} \vert^2 \rho_f(E)
\nonumber \end{equation}
where $V_{fi}$ is the interaction matrix element between the initial and 
final state and $\rho_f$ the density of states at the energy of the final 
state. For increasing energy $E$, $\rho_f(E)$ will increase exponentially 
and if $V_{fi}$ does not decrease as fast as or faster than $\rho_f$ which 
is certainly the case, $\Gamma(E)$ increases with $E$. If we associate a 
transition time $\tau (E) \propto \hbar/\Gamma(E)$ to this process, one 
sees that the transition time between different states of the system can 
decrease very fast and hence drastically accelerate the path to 
equilibration.

\subsection{Thermodynamic equilibrium : confrontation of experimental 
data with phase space models}
\indent

  Excited nuclei in thermodynamic equilibrium are described by means of so
called statistical models. They generally describe a situation where 
the system is at the freeze-out, the stage at which particles and fragments 
present in the system are located at relative distances $d$ from each other 
which are
larger than $2 - 3\,fm$, so that the nuclear interaction between them is 
negligible. Before we show applications of this type of models to the 
analysis of experimental results we present and discuss some of their 
aspects.

\subsubsection{The first phase space models}
\indent

  To our knowledge, the first attempts to describe strongly excited systems 
of particles and bound clusters were made in the beginning of the 80's. The 
concept of statistical multifragmentation was first introduced in ref. 
\cite{DGross}. In 1981, Randrup and Koonin proposed a thermodynamic phase 
space model \cite{Randrup}.
They defined the partition function $\mathcal{Z}$ of a classical system in 
the grand canonical ensemble in which the number of particles $A$, $N$ 
neutrons and $Z$ protons, as well as the energy, is fixed in the average.
If $A_0$ is the average number of particles, one defines 
\begin{displaymath} \omega = \ln \mathcal{Z}/A_0 = \sum_{AT} 
\omega_{\scriptscriptstyle AT}
\end{displaymath} where $T=1/2(N-Z)$. The multiplicity of fragments 
characterized by $A$ and $T$ is given by 
\begin{equation} \omega_{\scriptscriptstyle AT} = \bigg( \frac{4\pi}{3} 
r_0^3 \chi \bigg) \cdot \bigg( \frac{2\pi m A}{\beta \hbar^2} \bigg)^{3/2} 
\cdot z_{\scriptscriptstyle AT} \cdot \exp[-\beta (V_{\scriptscriptstyle AT} 
- \mu A - \nu T)] \nonumber \end{equation}
In the first term which is an effective volume, $\chi$ is a parameter of 
order unity which can be fixed by comparison with experiment, $r_0 (= 1.15\,
fm)$ the nuclear radius constant $(R_{syst}=r_0A^{1/3})$. In the 
second term which is related to the kinetic energy contribution $m$ is the 
nucleon mass and $\beta$ the inverse temperature. The third factor takes 
care of the internal energy of the fragments
\begin{equation} z_{\scriptscriptstyle AT} =
\sum_i g_{AT}^{(i)} \exp(-\beta \epsilon_{AT}^{(i)}) \nonumber \end{equation}
where $g_{AT}^{(i)} = 2j_{AT}^{(i)}+1$ is the degeneracy of the excited 
level $i$ and $\epsilon_{AT}^{(i)}$ the corresponding energy. In the fourth 
term $V_{AT}$ is the gound state mass excess which is taken from a liquid 
drop formula \cite{Myers}, $\mu$ and $\nu$ are Lagrange multipliers which 
are fixed in such a way that the total energy $E_0$, number of particles 
$A_0$ and isotopic composition are fixed,
\begin{eqnarray}
E_0 &=& -\frac{\partial}{\partial\beta} \ln \mathcal{Z} \nonumber \\
A_0 &=& \frac{1}{\beta} \frac{\partial}{\partial\mu} \ln \mathcal{Z} 
\nonumber \\
T_0 &=& \frac{1}{\beta} \frac{\partial}{\partial\nu} \ln \mathcal{Z}
\nonumber \\
\textrm{where } \mathcal{Z} &=& \sum_f \exp[-\beta (E_f - \mu A_f - \nu T_f)]
\nonumber \end{eqnarray}
and the sum over $f$ extends over all final states which are characterized 
by the number of fragments, their mass numbers, isospin projections, 
internal excitations, positions and momenta. In this description, particles 
and fragments do not interact, hence their spatial location is not relevant.

  The first confrontation with experiment was made with an extension of 
this model in which unstable species decaying by particle emission were 
included \cite{Fai}. The fragmentation was supposed to consist of a quick 
explosion followed by a slow evaporation from unstable excited fragments.
The light particle generation ratios $d/p$, $\alpha/p$, $^3He/p$, $t/p$ as 
well as the pion rates $\pi^-/p$ and $\pi^+/p$ were worked out and showed 
good agreement with experiment.

  Nuclei are finite objects and closed systems, with a fixed number of 
particles and a fixed energy. An approximate microcanonical description was 
introduced later \cite{Fai1} along with an extension which allows for the 
existence of different fragmentation sources generated in the 
collision process, which either ``explode'' for a large enough energy 
content or decay sequentially for low excitation energy. In this form the 
model has been used for explicit numerical simulations \cite{Fai2}. 

  In a further step Koonin and Randrup \cite{Koonin} implemented the model 
with a more rigorous canonical and microcanonical description in which the 
interaction between fragments and particles is taken into account. The 
essential quantity in the microcanonical framework is the density of states 
for a system of $A_F$ particles with energy $E_F$ and located in a volume 
$\Omega$
\begin{equation} \rho(\Omega,A,E) = \sum_F \delta(A_F-A) \delta(E_F-E)
\nonumber \end{equation}
where $F$ is the set of fragmentation configurations defined by a set of 
variables
$\{ A_n, \vec r_n, \vec p_n, \epsilon_n\} \ (n=1, \cdots ,N_F)$ which
specifies the mass, charge, position, momentum and internal energy of the 
species labelled by $n$. The total energy is written as
\begin{equation} E_F = \sum_{n=1}^F \bigg[ \frac{\vec p_n^{\ 2}}{2 m A_n} 
- B_n + \epsilon_n +\frac{1}{2} \sum_{n' \neq n} V_{nn'} \bigg]
\nonumber \end{equation}
where $B_n$ is the ground state binding energy and
\begin{equation} V_{nn'} = \frac{e^2 Z_{n} Z_{n'}}
{\vert \vec r_{n} - \vec r_{n'} \vert} + V_{nuc}(\vert \vec r_{n} - 
\vec r_{n'} \vert) \nonumber \end{equation}
the potentials which act between the fragments and the particles.
Applications of the model which includes metastable states showed the 
effect of the finiteness of the system which is contained in a fixed volume 
and the importance of the interaction energy between the constituents.

\subsubsection{The Berlin and Copenhagen phase space models}
\indent

  The first statistical theory on fragmentation of finite nuclei which 
included an exact treatment of the Coulomb energy between charged particles 
and fragments was developed in ref. \cite{DHEGross}. The
most popular and successful models are the Berlin model called MMMC 
(Microcanonical Metropolis Monte Carlo) \cite{Xiao, Gross2, Gross3, Gross} 
and the already quoted SMM (Statistical Multifragmentation Model)
\cite{Bondorf, Bondorf1, Bondorf2, Barz}. A description closely related to 
MMMC in its spirit was developed in Beijing \cite{Hao}.

  Except for technical though important points concerning Monte Carlo event
samplings which 
are used in the practical implementation of these models, the main 
difference between MMMC and the Koonin-Randrup model concerns the fact that 
the system which is at the freeze-out density does not experience 
evaporation of light particles in its final stage. The volume of the system 
is kept fixed and spherical, the Coulomb interaction between the 
constituents is treated rigorously. The model has been extensively 
used in order to confront calculated 
and experimental observables such as fragment correlation functions at high 
energy \cite{Li}  and also low energy data \cite{Gross4}. It has been 
tested for the possible existence of a first order phase transition, see 
below, section 5. Recently it has been extended to non spherical shapes of 
the volume and the consequences of this generalization have been 
studied \cite{Lefevre}.

  The Copenhagen approach has been initially worked out in the framework 
of the canonical ensemble \cite{Bondorf1,Bondorf2} and later on a 
microcanonical formulation was developed \cite{Bondorf}. The approach is 
rather close to MMMC,
see ref. \cite{Gross5}. The main differences concern the fact that 
particles can evaporate from excited fragments, the Coulomb 
interaction is treated in the Wigner-Seitz approximation and the volume in 
which the system is enclosed at freeze-out is not kept fixed. The problem 
of equilibration cannot be raised in the framework of the model. The 
question has been investigated by means of simple-minded transport 
equations \cite{Bondorf}. The authors rely on the observation of collective 
flow of matter which is generated and whose transverse energy carried 
perpendicularly to the incident beam axis by the particles and the 
fragments can be measured. In this framework the confrontation with 
experiment shows good agreement.
The interpretation of the data indicates that fragments are 
generated very early in the process, over time intervals as small as
$50\,fm/c$. This shows that fragment formation goes as a break-up and not a 
condensation process if the description is realistic. The instability 
growth is accelerated by the collective expansion of matter when the system 
enters the spinodal region. It is argued that the energy transferred to the 
system is converted into heat, generating a chaotic system which 
thermalizes somewhat before break-up, hence after a short interval of time.
  
\subsubsection{Thermal equilibrium : confrontation of models with 
experimental facts}
\indent

  Since there exists no direct way to measure the evolution and to find out 
experimentally whether thermodynamic equilibrium is reached or not, one has 
to rely on the confrontation between measured and model-dependent physical 
observables extracted from numerical simulations. This procedure has been 
followed a number of times in the recent past \cite{Moretto1, Tso, 
Donangelo, Beaulieu1, Phair, DAgostino, Phair1}.
We give here some examples which show how far equilibration is at least 
compatible with the experimental outcome even if it is not possible to 
prove it rigorously.

  In an experiment $^{36}Ar + ^{58}Ni$ at different energies (52, 74, 84 
and $95\,MeV\!\cdot\!A$) Borderie et al. \cite{Borderie, Borderie1} 
analysed the outcome of fragmentation events in which 90 \% of the nucleons 
are identified with the detector INDRA. They showed that these events 
correspond essentially to the vaporization of the system into light 
particles and clusters. The forward and backward spectra of particles are 
superimposable, which at low energy would be interpreted as a sign for the 
formation of a compound system.
The energy distribution functions of the different light 
species show experimental tails whose slopes are comparable to within 30 \%.
The authors introduce simple thermodynamic models like the so called 
Quantum Statistical Model (QSM) \cite{Mekjian, DasGupta, Gulminelli} which
works in the framework of the grand canonical ensemble, in an attempt to 
reproduce the measured observables. A detailed comparison shows that all 
measured quantities are well reproduced, in particular light particle 
yields, their average kinetic energies and the variances of multiplicity 
distributions.

\begin{figure}[b!]
\centerline{\epsfig{figure=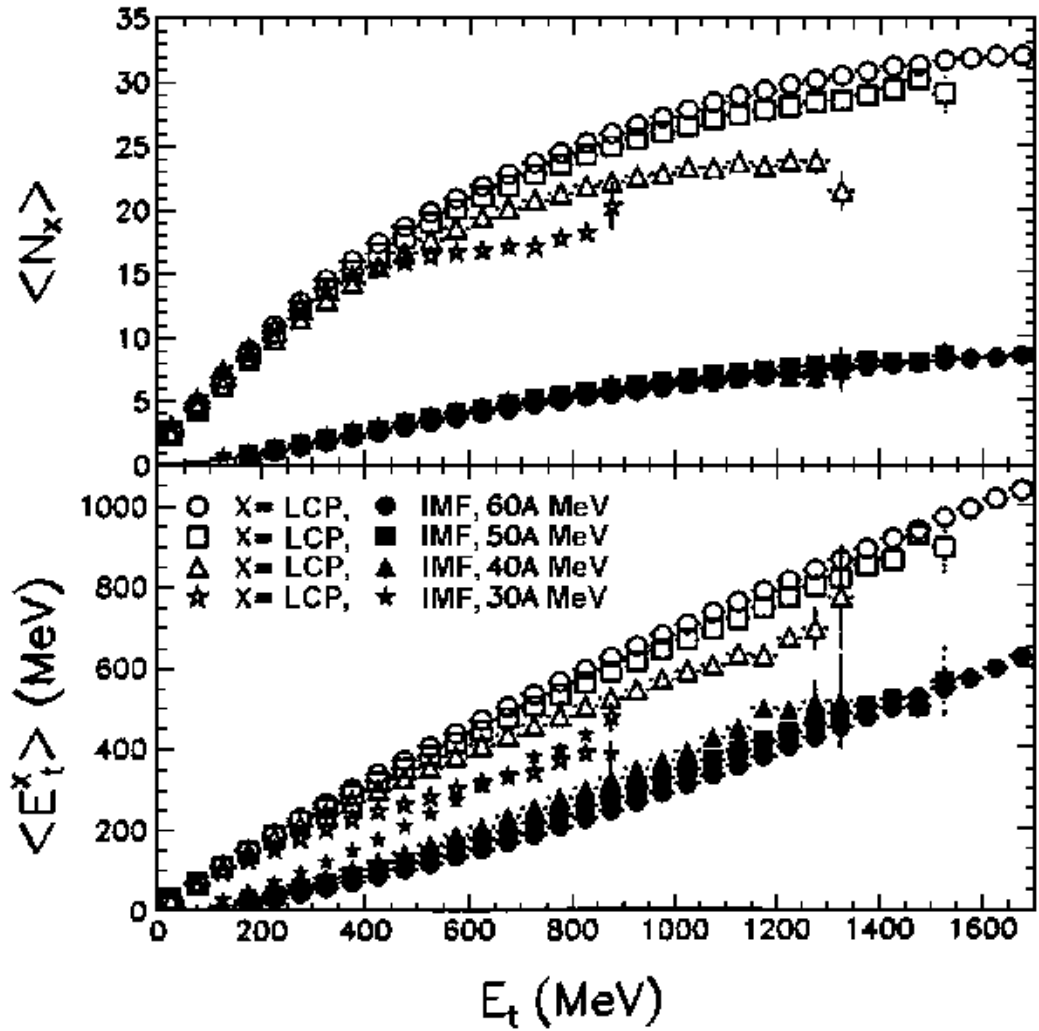,width=140mm}}
\caption{Top panel : average number of IMFs (solid symbols) and LCPs
(open symbols) as a function of the total transverse energy $E_{t}$.
Bottom panel~: corresponding average transverse energies as a function of 
$E_{t}$. The different symbols correspond to different bombarding energies 
indicated in the bottom panel \cite{Phair2}. See text.}
\label{fig5}
\end{figure}

  The fragmentation events observed in the reaction $^{129}Xe +^{197}Au$
at 30, 40, 50 and $60\,MeV\!\cdot\!A$ projectile energy have been 
analysed in some detail \cite{Phair2}. The average transverse energy
$(E_{t} = \sum_i E_i \sin^2 \theta_i \,$, where $E_i \sin^2 \theta_i$ is 
the projection of the energy associated with species $i$ in the direction 
perpendicular to the incident beam) measures the energy transferred from the 
relative motion of the ions to the excitation energy of the system. The 
analysis shows that the average number of light charged particles 
$< N_{LCP} >$ and the average number of IMFs and their corresponding 
transverse energies increase monotonously with increasing $E_t$ as shown in 
Fig.~\ref{fig5}. This contradicts the results of \cite{Toke} where the 
transverse LCP energy saturates. This saturation can in fact be understood 
as being due to instrumental problems. The features observed in 
Fig.~\ref{fig5} can be interpreted and understood in the framework of a 
statistical decay mechanism \cite{Beaulieu1}. The results can also be 
reproduced in the framework of phase space models like SMM \cite{Bondorf}.

\begin{figure}[b!]
\centerline{\epsfig{figure=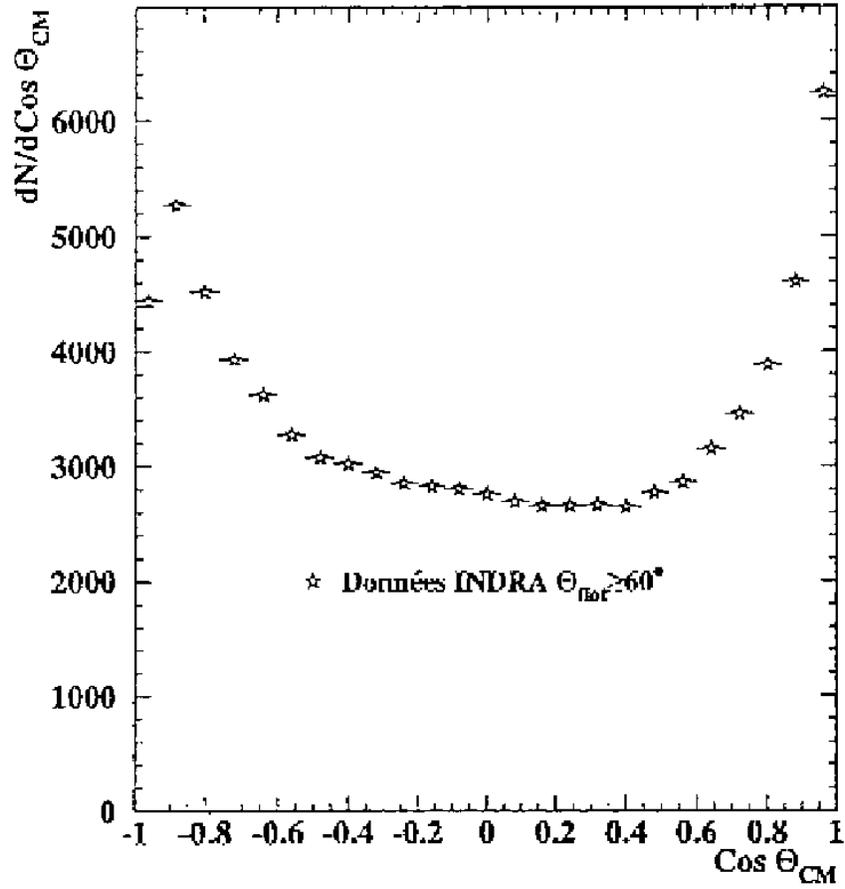,height=120mm}}
\caption{Angular distribution of light particles $(Z=1,2)$ in the flow 
angle $\theta \geq 60 ^\circ$ corresponding to events generated by a unique 
source in the reaction $Xe + Sn$ at $50\,MeV\!\cdot\!A$ \cite{LeNeindre}.}
\label{fig6}
\end{figure}

  Further analysis of the reaction $^{129}Xe + Sn$ have been presented and 
discussed in the recent past \cite{Marie, Marie1, LeNeindre}. In
ref. \cite{Marie1} correlation techniques were used in order to determine 
the multiplicities of $H$ and $He$ isotopes emitted by the excited 
fragments produced at the disassembly stage of an equilibrated hot source.
The determination of LCP multiplicities and kinetic energies led to the 
fragment excitation energies. These energies per particle are the same for 
all IMFs, which is interpreted as a sign for the fact that thermodynamic 
equilibrium is established at the stage where the source decays. A very 
detailed analysis of the same reaction was presented and discussed in
ref. \cite{LeNeindre}. It is shown there that the experimental fragmentation 
data can be interpreted in terms of unique source whose existence could be 
ascertained by means
of kinematical arguments related to global observables 
like the collective flow angle. For events which are selected under 
specific criteria the angular distributions of light particles and 
intermediate mass fragments show the characteristic behaviour of 
statistical evaporation, i.e. approximate symmetry with respect to 
$90^\circ$, see Fig.~\ref{fig6}, and excitation functions decrease 
exponentially as a function of energy. The experimental results were further 
confronted with simulations by means of the SMM model in its microcanonical 
version. This confrontation confirmed that the assumption of thermodynamic 
equilibrium is compatible with information extracted from experimental data.
The analysis relies on the so called backtracing method developed in
ref. \cite{Bondorf} which solves the inverse problem by tracing back the 
issue of the process to the characteristics of its origin, i.e. the 
characteristics of an excited equilibrated source. This method has also 
been applied to experiments performed by the ALADIN collaboration at the 
GSI for reactions of $^{197}Au$ on $C,\ Al,\ Cu$ and $Pb$ targets at
$600\,MeV\!\cdot\!A$ energy \cite{Hubele, Ogilvie, Hubele1}.

 There exist some indications for equilibration from purely experimental 
origin \cite{Trautmann}, corresponding to so called spectator reactions,
i.e. more or less peripheral heavy ion collisions, in particular 
$^{197}Au$ on $^{197}Au$ at high bombarding energies, 0.6 and
$1\,GeV\!\cdot\!A$. In this type of reactions it is assumed that part of 
the nucleons, the spectators, form a highly excited system which experiences 
only disordered motion and from which any coherent collective motion is 
practically absent. Several arguments lead to the conclusion that these 
systems could be in thermodynamic equilibrium. Recently the relative 
velocity correlation between LCPs and fragments has been used in order to 
reconstruct the size and excitation energies of the so called primary 
fragments which are produced at the early stage of the collision of $Xe$ on 
$Sn$ at 32 and $50 MeV\!\cdot\!A$ \cite{Hudan}. The constancy of the 
excitation energy imparted to the system with increasing bombarding energy 
suggests that thermodynamic equilibrium may be reached at freeze-out.

  The fragment distributions are invariant with respect to the formation 
process, i.e. the entrance channel of the reaction. As we shall see in 
section 5, there has been an attempt to fix a temperature to the spectator 
system by means of the measurement of the isotope ratios of light nuclei
$(He, Li)$ which are generated through fragmentation \cite{Xi}. These 
ratios are invariant with respect to the bombarding energy. The result is 
however, at first sight, not compatible with other observables from which 
it is in principle possible to extract a temperature. This is so, in 
particular, for the slopes of the energy distributions of fragments which, 
if they are interpreted as Maxwellian distributions, lead to much higher 
temperatures. Several arguments which try to explain this fact have been 
proposed \cite{Trautmann}, among them the role played by the Fermi 
motion \cite{Goldhaber} of nucleons related to a fast fragmentation process.

  The equilibration problem has also been raised by the FOPI collaboration 
in the study of reactions involving more central heavy ion 
collisions \cite{Reisdorf}. In this type of events, one observes a 
collective flow of particles. This fact complicates the experimental 
analysis and does not allow for a clear-cut conclusion about equilibration. 
Indeed, different models, both time-independent and time-dependent seem to 
be able to reproduce the data \cite{deSchauenburg}. Recently isospin 
arguments have been used in order to test equilibration \cite{Rami}. They 
indicate that equilibration is not reached as far as isospin degrees of 
freedom are concerned.

\newpage

\section{Percolation models and fragment size distributions}
\indent

  The detection and identification of the charge and the mass of fragments 
generated through violent collisions is of fundamental importance in the 
study of excited nuclei. Many detectors like ALADIN, FOPI, INDRA, the
EOS setup, the MINIBALL, LASSA, CHIMERA are able to detect and identify 
particles and fragments event by event and to determine their asymptotic 
kinematical properties which constitute the information which can be 
experimentally collected.  

  As already seen above, the information contained in the outcome of 
processes like fragmentation are of extreme complexity and the formation 
mechanism largely unknown, certainly different under different physical 
conditions such as the energy range and the impact parameter. On the other 
hand, the use of the concepts of statistical mechanisms has shown that 
paradoxically the most complex systems can be described in terms of minimum 
information theories which bypass the detailed knowledge of the underlying 
microscopic systems. Here it may be tempting to follow this philosophy and 
to try to describe the outcome of nuclear fragmentation processes within 
the simplest possible physical framework. This has been done starting in 
the early 80's by means of minimum information approaches like percolation 
models. These models ground on purely topological and statistical concepts, 
they avoid the explicit introduction of Hamiltonians. If they work, they 
raise of course the question why and how this is the case.

  In percolation theories, space contains empty regions and parts which are 
occupied by individual objects which can be considered as bound to form 
larger entities defined by means of different criteria. The essential 
concepts are connectivity, localization and percolation which characterizes 
the properties of clusters (animals, fragments). Percolation itself is 
established when, in an infinite space, at least one cluster of infinite 
size is present.

  Percolation models have been used in many fields which deal with geometry,
disorder, statistics and physics such as aggregation, localization, 
diffusion, conductivity \ldots  Before going over to its application in the 
field of nuclear fragmentation it may be worthwhile to recall general 
concepts, definitions and properties which appear in the framework of this 
type of descriptions.

\subsection{Random-cluster models : general concepts and properties}
\indent

  The first percolation model was introduced by Broadbent and 
Hammersley \cite{Broadbent} in order to describe the spread of matter 
through a medium such as a liquid through a porous medium. It was then soon
recognized that this type of model is acquainted to already known models 
like the Ising model \cite{Hammersley} and other statistical spin models 
like those developed by Ashkin and Teller \cite{Ashkin} and 
Potts \cite{Potts}.

  All these descriptions can be studied in a general framework called
random-cluster model. This has been worked out by Fortuin and
Kasteleyn \cite{Kasteleyn,Fortuin}. In the first paper \cite{Kasteleyn} the 
authors showed the close connection between the connectivity problem of 
different systems with uncorrelated bonds between sites which are occupied 
by single objects (particles) and the
well known Ising model. Since Ising systems exhibit phase transitions
in $2d$ and $3d$ spaces it is not surprising that the random-cluster model 
shows itself a phase transition in the infinite space limit. In later 
developments \cite{Fortuin} the authors studied the general topological 
properties of the aforementioned models by means of graph theory, in 
particular those of the random-cluster model. They showed why the models 
quoted in refs. \cite{Hammersley,Ashkin,Potts} as well as linear 
resistance networks 
and percolation models can be studied in a unique framework by means of 
graphs defined in terms of vertices (sites), edges (bonds), connections 
between edges and vertices, and probabilities $p$ for edges to exist or not. 
In practice they introduced a generalized partition function (cluster 
generation function) and ``free energy'' which enables to work out physical 
observables like thermodynamic quantities in the case of the Ising, 
Ashkin-Teller and Potts models and correlation functions.

  Further analytical developments on percolation models were pursued by 
Kunz and Souillard \cite{Kunz,Kunz1} related to previous work by
E.H. Lieb. In the case where sites do not interact, i.e. are not 
correlated, they introduced the free energy
\begin{equation} f_p(h) = \sum_n P_n \frac{e^{-hn}}{n} = \sum_{C} e^{-h|C|} 
\cdot \frac{p^{|C|}q^{\partial|C|}}{|C|} \nonumber \end{equation} 
where $p$ is the bond probability, $q=1-p$, $P_n$ the 
probability that a given site belongs to a cluster with exactly $n$ sites,
$C$ are all finite clusters containing the origin of space $0$, $|C|$ 
is the size of the clusters and $\partial|C|$ the size of the boundary of 
clusters, i.e. their perimeters (surfaces). The study of the properties of 
$f_p(h)$ shows that this function is analytic for $h=0$ and $p < p_c$, and 
develops a singularity for $p \geq p_c$ if the system is infinite, 
indicating a phase transition of second order at $p = p_c$. For 
$p \geq p_c$ the system percolates in the sense defined above, i.e. at 
least one cluster of infinite size appears.

\subsection{Percolation models : definitions and general properties}
\indent

  Before discussing the applications of percolation concepts to nuclear 
fragmentation we present here some essential properties of percolation 
models we shall need in the sequel.

 Percolation models  consist of a discrete number of sites which cover a 
finite or infinite part of a $d$-dimensional space \cite{Stauffer, Essam,
Stauffer1}. Site percolation corresponds to the case where each site is 
either occupied by one ``particle'' or is empty. Bond percolation corresponds 
to the case where each site is occupied, neighbouring sites being bound or 
not to each other. The number of neighbouring sites is fixed by the 
connectivity which depends on the  geometric structure of space occupation 
(f.i. a triangular lattice for $d=2$, a cubic lattice for $d=3$, \ldots).
The combination of site occupation and bond formation leads to hybrid
site-bond models.

  Except for the results presented in section 4.1 above \cite{Kasteleyn, 
Fortuin, Kunz, Kunz1}, very little is known analytically about these models. 
Most information has been gained by means of numerical simulations. 

  In the case of site percolation, one fixes a probability $p \in [ 0, 1]$.
For each site, one draws a random number $\eta$ taken from a uniform 
distribution in the interval $[ 0, 1]$. If $ \eta \leq p$ the site is said 
to be occupied, if $ \eta > p$ it is empty. Hopping over all sites, one 
generates an ensemble of occupied and empty sites. If the number $N$ of 
sites gets very large, one finds $Np$ occupied and $N(1-p)$ empty sites. If 
neighbouring sites are occupied they belong to the same cluster. Space is 
finally covered with clusters and empty space. 

  A similar procedure works in the case of bond percolation. For a fixed 
bond probability $p$ consider a pair of neighbouring sites and draw a 
random number $\eta \in [ 0, 1]$. If $ \eta \leq p$ the sites are linked by 
a bond, if $ \eta > p$ they are not. Testing this way all possible bonds 
between neighbouring sites generates linked clusters of bound sites, each 
cluster being made up of an ensemble of connected particles. The 
determination of the whole cluster network will be called event or 
realization. In each realization there are clusters of different sizes. 
They appear with a given multiplicity which depends of course on $p$. 
Averaging over many realizations leads to a characteristic cluster size 
distribution in the infinite system.

  Clusters with a fixed size can have different shapes for $d \geq 2$. The 
average number of clusters of size $r$ per lattice site taken over a set of 
realizations is given by
\begin{equation} n_r = \sum_s g_{rs}p^r(1-p)^s \label{def-nr}
\nonumber \end{equation} 
where $g_{rs}$ is the number of clusters with size $r$ and perimeter $s$, 
the perimeter being the number of empty (non connected) sites which 
surround the cluster. Generally $g_{rs}$ is not analytically accessible. We 
shall come back to this point below.  

  Individual clusters of size $s$ can also be characterized by a 
``gyration'' radius $R_D$, defined by
\begin{eqnarray} R_D^2 & = & \sum_{i=1}^s \frac{|\vec r_i - \vec r_0|^2}{s} 
\nonumber \\
\textrm{where} \quad  \vec r_0 & = & \sum_{i=1}^s \frac{\vec r_i}{s}
\nonumber \end{eqnarray} 
is the centre of mass of the cluster.

  Then the average squared distance between two cluster sites is
\begin{equation} 2R_s^2 = \sum_{i \neq j} \frac{|\vec r_i - \vec r_j|^2}{s^2}
\nonumber \end{equation}
  If $g(r)$ is the probability that a site which is a distance $r$ apart 
from an occupied site belongs to the same cluster, one can define a 
correlation length $\xi$ such that \cite{Stauffer1}
\begin{equation} \xi^2 = \frac{\sum_r r^2g(r)}{\sum_r g(r)} \label{def-xi}
\end{equation} 
$2R_s^2$ is the average distance between two sites of a cluster and $sn_s$ 
the probability that a site belongs to a cluster of size $s$. Since such a 
site is connected to $s$ other sites the correlation distance $\xi$ can be 
rewritten as
\begin{equation} \xi^2 = \frac{2\sum_s R_s^2s^2n_s}{\sum_s s^2n_s} 
\label{xi2b} \end{equation}
  As it has been mentioned above \cite{Kasteleyn, Fortuin, Kunz, Kunz1} 
there appears a percolation (second order) phase transition for $d \geq 2$ 
and some value $p = p_c < 1$. For $p > p_c$ one or several infinite 
clusters are present. The system is said to percolate. The value of $p_c$ 
depends on the dimensionality of the system, the type of percolation (site 
or bond) and its connectivity (the number of maximum occupied sites in the 
next neighbourhood of a site or possible bonds of a site to the nearest 
neighbours) \cite{Stauffer1}.

  The central observable in percolation physics is of course the cluster 
size distribution. In practice, nothing is known analytically about it, 
except for conjectures which agree very precisely with numerical 
simulations. For $p < p_c$ and clusters of large size $(s \to \infty) \
\ln n_s \propto -s$, i.e. one observes an exponential tail when 
space occupation (site percolation) or the number of bonds ( bond 
percolation) is sparse. For $p > p_c$ and  $s \to \infty \
\ln n_s \propto -s^{1-1/d}$. As one can see the space dimension enters in 
the case where space occupation or the number of bonds is large. This is 
not the case in the non percolative regime. Finally for $p \simeq p_c$ and 
large clusters it has been conjectured \cite{Stauffer1} that
\begin{equation} n_s(p) \simeq s^{-\tau} f\big( (p-p_c) s^\sigma \big)     
\label{def-n} \end{equation} 
which is true for all percolation models, $\tau$ and $\sigma$ being 
universal exponents and $f$ a model dependent function. For $p = p_c, \
f(0) = 1$, leading to a power law behaviour which intriguinly reminds 
Fisher's expression (\ref{Minc}). As stated above, these relations are very 
accurately verified by numerical tests.

\subsection{Moments of the cluster size distribution and critical exponents}
\indent

  For infinite systems it is possible to work out the expression of the 
cluster size distribution $n_s(p)$ in the vicinity of the critical point
$p = p_c$.

  By definition the $k$th moment is
\begin{equation} m_k(p) = \sum_s s^k \cdot n_s(p) 
\nonumber \end{equation} 
Going over to the continuum limit
\begin{equation} m_k(p) = \int_0^{\infty}\textrm{d}s \,s^k n_s(p)
\nonumber \end{equation} 
and introducing the analytical expression (\ref{def-n}) with
\[f\big( (p-p_c) s^\sigma \big)=\exp(-|p-p_c|^{1/\sigma}s)\] \noindent
for $p$ in the vicinity of $p \sim p_c$ leads to 
\begin{eqnarray}m_k(p) & \sim & |p-p_c|^{\mu_k} \label{momk} \\
\textrm{where} \quad \mu_k & = & \frac{\tau - k - 1} {\sigma} \label{muk}
\end{eqnarray} 
For $k > \tau - 1$ one sees that the moments diverge at $p = p_c$, a sign 
for the existence of a phase transition. Hence, if $\tau < 3$ which is the 
case for $d = 2, 3$
\begin{eqnarray} m_2(p) & \sim & |p-p_c|^{-\gamma} \nonumber \\
\textrm{with} \quad \gamma & = & (3 - \tau) / \sigma      
\label{def-gam} \end{eqnarray} 
For $d = 3$ in a cubic lattice $\gamma = 1.74$.

  Critical exponents characterize other quantities of interest. If in the 
infinite system $P(p)$ is the probability that a given site belongs to the 
infinite cluster for $p > p_c$, then probability conservation implies
\begin{equation} P(p) + (1-p) +\sum_s s n_s(p) = 1 
\nonumber \end{equation} 
The second term corresponds to the probability that the site is empty and 
the third that it belongs to a finite cluster of size $s$.
\begin{eqnarray} \textrm{At} \quad p = p_c \quad & P(p_c) = & 0 \nonumber \\
\textrm{and} \quad & \sum_s s n_s(p_c)  = & p_c \nonumber \end{eqnarray}
Hence in the vicinity of $p_c$, $P$ can be written \cite{Stauffer1}
\begin{equation} P = \sum_s [n_s(p_c) - n_s(p)]s + \mathcal{O}(p-p_c)        
\label{def-P} \end{equation} 
If one introduces the parametrization $n_s(p) \propto s^{-\tau} \exp(-cs)$
with $c = 0$ at $ p = p_c$ and goes over to the continuum 
limit with (\ref{def-P}) it comes 
\begin{eqnarray} P & \propto & (p - p_c) ^\beta \nonumber \\
\textrm{where} \quad \beta & = & (\tau - 2)/\sigma
\label{def-bet} \end{eqnarray} 
which consistently tends to zero when $\tau > 2$ at $p = p_c$.

  Since percolation systems show a continuous transition at $p = p_c$ one 
expects that the correlation length $\xi$ defined in (\ref{def-xi}) shows a 
singular behaviour at the threshold where an infinitely large cluster 
merges. This is indeed the case 
\begin{equation}\xi \propto |p - p_c|^{-\nu} 
\label{def-nu} \end{equation} 
and the positive exponent $\nu$ can be determined numerically.

  Percolation clusters are generally fractal objects. The radius of gyration 
$R_D$ is related to the volume $s$ through
\begin{equation} s \propto R_D^{d_f} \nonumber \end{equation} 
where $d_f$ is an effective, not necessarily integer space dimension. From 
the expression (\ref{xi2b}) it follows that $\xi^2$ behaves like 
$s^{2 + 2/d_f}$. The same type of derivation as the one used for the 
moments $m_k(p)$ leads to a power law expression of $\xi^2$ whose numerator 
diverges with an exponent $(3 - \tau + 2/d_f)/ \sigma$. Hence $\xi^2$ 
itself diverges like $2/d_f \sigma$ and an identification with 
(\ref{def-nu}) leads to
\begin{equation}\sigma \nu = 1/d_f \label{sig-nu} \end{equation} 
  At $p = p_c \quad R_s \propto s^{1/d_f}$, hence the size of the largest 
cluster increases like $L \propto s^{1/d_f}$. If $P$ is the strength of the 
infinite cluster as defined above, $PL^d$ is the volume occupied by the 
largest cluster. If the linear extension of this cluster is of the order of 
the correlation length $\xi$
\begin{eqnarray} PL^d & \propto & L^{d_f} \nonumber \\
L & \propto & \xi \propto |p - p_c|^{-\nu} \nonumber \end{eqnarray}
  Recalling the behaviour of $P$ in the neighbourhood of $p_c$ one gets
\begin{equation}d_f = d - \beta/\nu \label{df-bet} \end{equation}
  In summary, general considerations lead to six critical exponents $\tau, 
\sigma, \beta, \nu, d_f, \textrm{and } \gamma$ which are linked by four 
relations (\ref{def-gam}), (\ref{def-bet}), (\ref{sig-nu}) and (\ref{df-bet}).

  Up to now we considered essentially the properties of infinite systems 
and concentrated on their behaviour in the vicinity of the percolation 
threshold which allows to determine critical exponents. These exponents are 
very useful quantities since they fix the universality class to which a 
critical system belongs. If excited nuclei are the finite counterparts of 
critical nuclear matter, the singular behaviour of observables at the phase 
transition is quantitatively characterized by critical exponents. We shall 
try to see and to understand below if, how and why percolation concepts 
could be related to nuclear fragmentation. But nuclei are finite systems. 
Before we present and discuss the models which have been proposed, we 
present a study of the effects generated by the finiteness of percolation 
systems.

\subsection{Finite size constraints on random-cluster systems}
\indent
 
  Percolation deals with the critical behaviour of systems in the limit of 
infinitely extended space. Microscopic and mesoscopic systems like nuclei, 
atoms, molecules, aggregates are finite. As far as percolation concepts can 
be applied to their description and directly compared to experiment as we 
shall do below, it is necessary to get some insight into the behaviour of 
percolation models in finite space. Finite size effects enter because of 
the presence of a surface which induces constraints on the system, 
especially when the space occupation is large. We concentrate here on site 
percolation.

  The aim of the study concerns the determination of the fragment size 
distribution of a system with $V$ sites, with an occupation number equal 
to 0 or 1 for each site. The number of occupied sites $A$ is fixed, the 
density is $\rho = A/V$. Following percolation concepts one qualitatively 
expects a set of small clusters when $\rho$ is small, hence a cluster size 
distribution which decays monotonously as a function of the cluster size 
and for large values of $\rho$
one or several large clusters together with small clusters showing a 
kind of $U$-shape. For intermediate values of $\rho$ small and intermediate 
size clusters may coexist. These shapes are in fact observed in 
the experiment. The crucial point concerns the behaviour of the 
distribution for these intermediate values of $\rho$. 

  The behaviour of finite percolation systems has been investigated 
in \cite{Biro}. One considers a finite volume $V$ in a $d$-dimensional 
space. Space occupation by $A$ particles is defined by $O=(i_1, \ldots, i_A)$
where $i_1, \ldots, i_A$ label $A$ occupied sites. There are 
$ \binom{V}{A} \equiv V!/A!(V-A)! $ possible arrangements. To each occupation
can be attributed a canonical weight, $\exp[-\beta H(O)]$ where $H$ is the 
Hamiltonian of the system and $\beta$ can be interpreted as an inverse 
temperature. Define
\begin{eqnarray} H(O) & = & \frac{1}{2} \sum_{i \neq j} H_{ij} \nonumber \\
\textrm{with} \quad H_{ij} & = & H_{ji} \nonumber \end{eqnarray} 
A simple choice would be
\begin{equation}   H_{ij} = -eL_{ij} \nonumber \end{equation} 
where $e$ is a positive constant and $L_{ij} \ (i \neq j =1,\ldots,V)$
taking values 0 and 1 defines a link structure which decides whether two 
sites are connected or not.

  The link structure introduces the concept of clusters as sets $c$ of 
connected occupied sites $c = (i_1, \ldots, i_a)$, with size $a(c)$ and a 
surface (called perimeter in section 4.1) made of those unoccupied sites 
which are the nearest neighbours of occupied ones.

  Given a cluster of size $a$ and surface $s$, we call $f(c)=V-a-s$ the 
free complement of $c$ which is such that
\begin{equation}   c \cup s \cup f = V \nonumber \end{equation}
  The central observable is the average multiplicity of clusters of a given 
size in the canonical ensemble, allowing for all possible occupations with 
fixed $a$, weighted by the Boltzmann factor $\exp(-\beta H)$.

  The average of an observable $\Omega$ is defined by
\begin{equation} <\Omega>=\frac{1}{\mathcal{Z}} \sum_{O \subseteq V}\Omega(O) 
 \exp[-\beta H(O)] \, \delta \big( a(O)-A \big) 
\nonumber \end{equation} 
where obviously 
\begin{equation} \mathcal{Z} = \mathcal{Z}(V,A,\beta) = \sum_{O \subseteq V}
 \exp[-\beta H(O)] \, \delta \big( a(O)-A \big) 
\nonumber \end{equation} 
is the canonical partition function. In particular, the energy $U$ and the 
entropy $S$ read
\begin{eqnarray} U & = & <H> \\ S & = & \ln \mathcal{Z} + \beta U
\nonumber \end{eqnarray} 
Notice that, as generally admitted in random-cluster models, there exists no 
cluster-cluster interaction. This is due to the fact that the interaction 
is restricted to nearest neighbour occupied sites.

  The probability to find a cluster $c$ of size $a$ in a configuration is 
given by
\begin{equation}  P(c)=e^{-\beta H(c)}\mathcal{Z}(f,A-a,\beta)
  / \mathcal{Z}(V,A,\beta) \nonumber \end{equation} 
Then
\begin{equation}  <\Omega>=\sum_c P(c) \Omega(c) 
\nonumber \end{equation} 
for any observable $\Omega$.

  In particular, the multiplicity of clusters of size $a$ is given by
\begin{equation}  m(a)= \sum_c P(c) \delta \big( a-a(s) \big)
\nonumber \end{equation} 
The essential constraint on $m(a)$ is fixed by the fixed number of 
particles
\begin{equation}  \sum_{a=1}^A a \, m(a)= A  \label{mass} \end{equation}
  Monte Carlo simulations \cite{Biro} reproduce the expected cluster size 
distributions which were predicted above, in particular the $U$-shape at 
large density. This is true both in the case when the coupling constant
$e$ is zero or finite except that the transition
from the falling distribution for small $\rho$ to the $U$-shape 
is located at different values of $\rho$ for different values of $e$. Hence 
in order to study the consequences of the constraint imposed by (\ref{mass})
one may restrict this study to the case $e=0$ which allows for the use of 
simple analytical arguments. Then
\begin{eqnarray}  \mathcal{Z}(V,A) & = & \binom{V}{A} \nonumber \\
  m(a) & = & \sum_s g(a,s) \binom{V-a-s}{A-a} \Big/ \binom{V}{A}
\nonumber \end{eqnarray} 
where s is the surface defined above and $g(a,s)$ the number of clusters of 
size $a$ and surface $s$.

  In the case where the system is dilute $(\rho=A/V \ll 1)$
\begin{equation} m(a) = \sum_s g(a,s) \rho^a (1-\rho)^s 
\end{equation} \nonumber 
The degeneracy $g$ is a complicated function of the link structure $L$. 
There exists some knowledge about it \cite{Stauffer, Essam, Samaddar}. In 
fact $g(a,s)$ differs from zero only in a narrow range of surfaces.

  It is convenient to introduce
\begin{equation} g(a,s) = \gamma(a) P_a(s) \nonumber \end{equation} 
where $P_a(s)$ is the distribution of surface sizes for fixed $a$. The 
surface $s$ varies between a minimum and a maximum limit for fixed $a$. The 
minimum is given for the more compact forms
\begin{eqnarray} s_{min}(a) & \propto & a^{1-1/d} 
\quad\quad\quad\quad \textrm{for} \quad a < \frac{V}{2} \nonumber \\
    & \propto & (V-a)^{1-1/d} \quad \textrm{for} \quad a > \frac{V}{2}
\nonumber \end{eqnarray} 
The maximum corresponds to chain-like clusters, hence
\begin{equation} s_{max}(a) \propto 2+2(d-1)a \label{smax} 
\end{equation} \nonumber 
in a $d$-dimensional hypercubic lattice. Of course $s \leq V-a$.

\begin{figure}[b!]
\centerline{\epsfig{figure=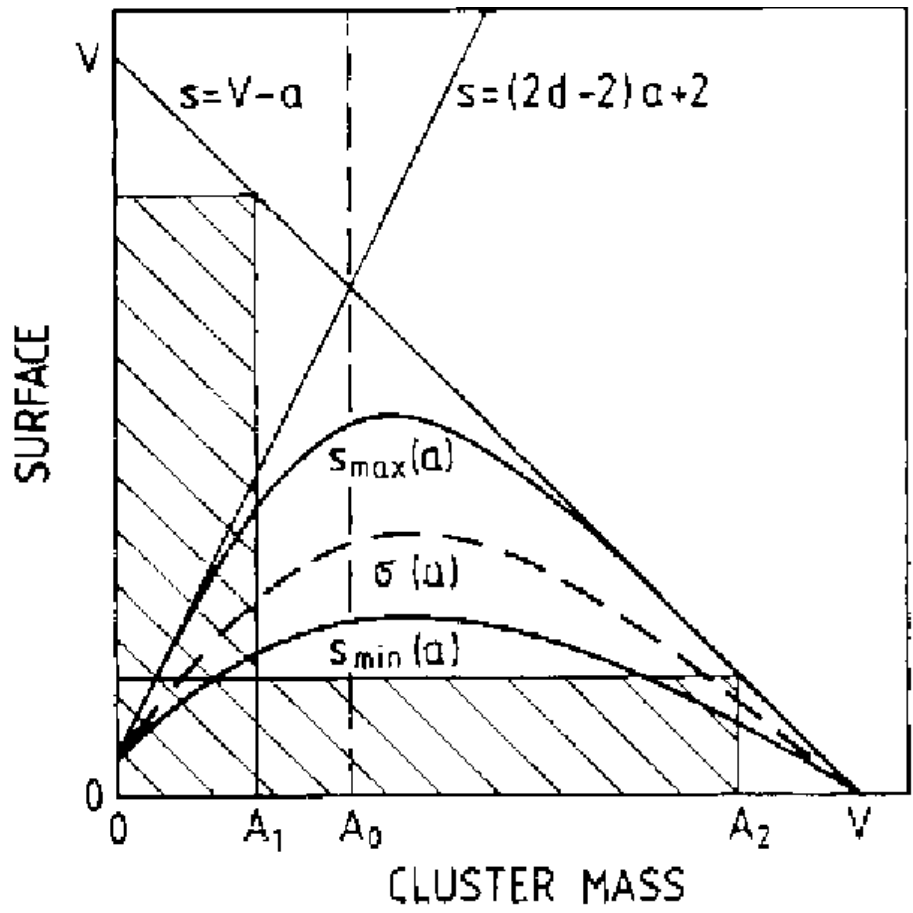,height=120mm}}
\caption{Plot of the $(a,s)$ plane where the degeneracy $g(a,s)$ is non 
zero. For fixed $A$ $(= A_1$ or $A_2)$ only the hatched areas contribute to 
the mass spectrum \cite{Biro}.}
\label{fig7}
\end{figure}

  The overall behaviour of the surface as a function of the cluster size is 
shown in Fig.~\ref{fig7}. If one defines an average surface
\begin{equation}  \sigma(a) = \sum_s s P_a(s) 
\nonumber \end{equation} 
then for the small mass sector one may parametrize $\sigma(a)$ in the 
following form
\begin{equation}
  \sigma(a) = \lambda a + \sigma_0 \quad a \ll A_0 \, , \quad \lambda \le 2d-2
\nonumber \end{equation} 
and for the large mass sector
\begin{equation}  \sigma(a) = K(V-a) \quad a \gg A_0 \, , \quad K\le 1
\label{sigma} \end{equation} 
where $A_0$ is the solution of $s_{max}(A_0) = V-A_0$, see (\ref{smax}).
Hence
\begin{equation} 2(d-1)A_0 + A_0 = V-2 \nonumber \end{equation} 
and
\begin{equation} A_0 = (V-2)/(2d-1) \nonumber \end{equation} 
Consider first the small mass sector. Since the distribution $P_a(s)$ is 
not known, one might investigate different types of distribution. Analytic 
results can be obtained for a Poisson or a sharp distribution. In the 
simplest case
\begin{equation} P_a(s) = \delta \big( \sigma(a)-s \big) \label{distrib}
\end{equation} 
The distribution may indeed be rather sharp when $a$ does not get too
large, see Fig.~\ref{fig7}. Then, using (\ref{distrib}) one obtains
\begin{equation} G(a) = V \sigma(1) [ \sigma(a)+a-2]! / [ a! \, \sigma(a)! ]
\nonumber \end{equation} 
For $a, \sigma \gg 1$
\begin{eqnarray} G(a) & \propto & a^{-2.5} \exp( \alpha_c a) \nonumber \\
\textrm{with} \quad \alpha_c & = & 
\ln [ (\lambda + 1)^{\lambda + 1}/\lambda^\lambda ]
\nonumber \end{eqnarray} 
A similar expression for $G(a)$ is obtained if $P_a(s)$ is a Poisson 
distribution. As a consequence
\begin{eqnarray} m(a) & \propto & a^{-2.5} \exp\{(\alpha_c - \alpha) a\} 
\nonumber \\
\textrm{with} \quad \alpha & = & \ln [ 1 / \rho (1-\lambda)^\lambda]
\nonumber \end{eqnarray} 
$\alpha$ is minimum for $\alpha = \alpha_c$ at
\begin{equation} \rho_c = 1/(1+ \lambda) 
\nonumber \end{equation} 
Then $m(a) \propto a^{-2.5}$ for $\rho = \rho_c$.
One again finds the manifestation of a power law behaviour, which is 
realized in the case of finite systems.

  If $V$ is large $\sigma(a)$ lies close to $2(d-1)a+2$, hence
$\lambda=2d-2$ for cubic lattices. Then
\begin{equation} \rho_c=1/(2d-1)=A_0/V 
\nonumber \end{equation} 
In fact $\rho_c$ is determined by the crossing of the two topological 
limits fixed by the small and large cluster sector. Hence in finite systems 
the onset of a ``critical'' regime is correlated with a topological 
constraint which, in a finite but large volume, corresponds to cluster 
surfaces with a maximum in the vicinity of $\rho_c V = V/(2d-1)$.

  In the case of large $\rho$ it is sensible to parametrize the surface 
following (\ref{sigma}) above. Then
\begin{equation} m(a) \propto q^{A-a}/(A-a)!  
\nonumber \end{equation} 
with
\begin{equation} q=\{(V-A)/\big(A+K(V-A)\big)\}^{1-K}A(1-K)K^{-K}
\nonumber \end{equation} 
This distribution is monotonically rising with $a$ if $K < K_{cr} \sim
A/(A+1)$ and reaches a maximum at $a=A$. In all other cases it peaks at 
$q \simeq A-a$. For $s_{min}(a) > V-a, \ m(a)$ is of course zero.

  In summary, simple considerations lead to a different behaviour of 
cluster size distributions in different density sectors of finite systems.
This behaviour is essentially governed by the mass conservation constraint
(\ref{mass}) imposed by the finite volume in which the system is located.
For small densities the size distribution decreases essentially as an 
exponential. There exists a density $\rho_c$ for which the distribution is a 
power law or close to it, with a fixed index irrespective of the 
dimensionality of space. For $\rho > \rho_c$ one observes both a decreasing 
light cluster sector and a large size sector building up.

  The present results are general. As we shall show below they are in 
qualitative agreement with explicit simulation results and experimentally 
observed cluster size distributions.

\subsection{Models related to percolation and first confrontations with the 
experiment}
\indent

  The first applications of percolation concepts to the description of nuclear 
fragmentation were proposed by Bauer and collaborators \cite{Bauer1, Bauer2},
Campi and Desbois \cite{Campi}.

  The starting point of Bauer et al. was the observation of the power law 
distribution of intermediate mass fragments \cite{Minich, Panagiotou,
Panagiotou1} obtained by means of fragmentation processes
\begin{equation} A_p+A_t \to A_f+X \nonumber \end{equation} 
where $A_p, A_t, A_f$ are the mass of the projectile, target and a 
specific fragment and $X$ the rest of the reaction outcome.
 
  In the absence of any precise information about the reaction mechanism 
the authors \cite{Bauer1, Bauer2} suggested a minimal-information 
description, following the same spirit as the one which prevailed in 
ref. \cite{Campi1}. They introduced a nucleus lattice model (NLM) in which 
a set of particles occupy lattice sites in a finite space. The presence of 
bond clusters is tested by means of a bond percolation algorithm as 
described above applied to a $3d$ cubic lattice model in a finite volume. 
The so called standard percolation theory (SPT) deals with infinite systems. 
The physical framework here is an $A+1$ particle system obtained by means 
of energetic collisions of a proton on a nucleus. The penetration of the 
proton generates a cylindrical tube of radius $r$ with very excited 
matter (fireball), in an initial spherical 
nucleus at an impact parameter $b$. The nucleons of the fireball are kicked 
out, the remaining nucleons are sitting on lattice sites and experience a 
strong disturbance which reflects in the definition of a bond breaking 
probability $p$ characterizing the link between the remaining spectator 
nucleons. In this description the exact number of particles is not fixed and 
there is no information about the energy imparted to the system. The bond 
breaking probability $p$ is supposed to be the larger, 
the larger the kinetic energy of the incoming projectile proton. 
A Monte Carlo percolation algorithm applied to 
the system for different values of $p \in [0,1]$ leads to different shapes of 
the fragment (cluster) size distribution, ranging
from an $U$-shape for small $p$ to a typical exponential fall-off with no 
heavy fragments available for large $p$, as predicted in ref. \cite{Biro}.
For some $p \simeq p_c \simeq 0.74$ the width of the mass distribution gets 
maximum and the mean yield shows a power law behaviour for light and 
intermediate mass fragments, with an exponent $\tau \simeq 2.21$. This 
value is somewhat lower than the one corresponding to the infinite system.

\begin{figure}[b!]
\centerline{\epsfig{figure=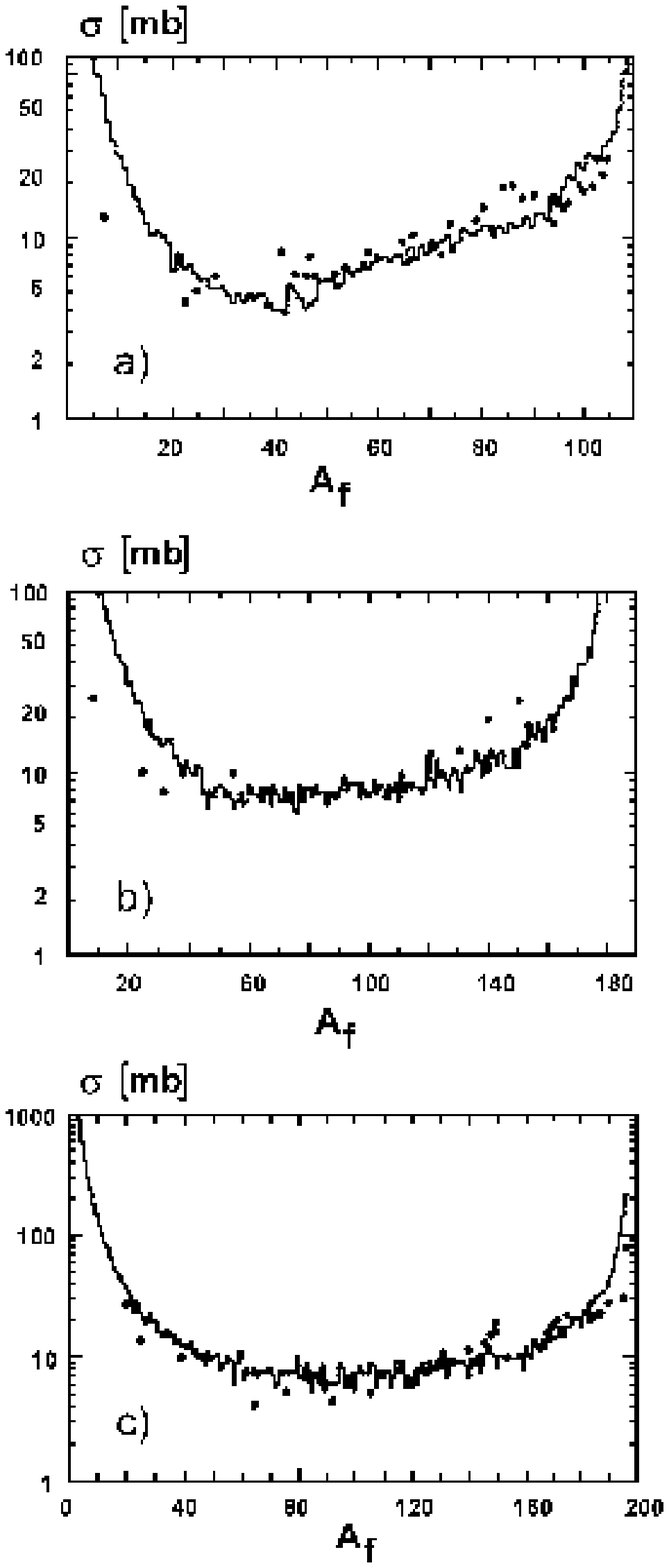,height=160mm}}
\caption{Mass yields obtained with the parametrization of the bond 
probability $p(b)$ given in the text for 
(a) $p+Ag$ at $11.5 \,GeV$, $p_0 = 0.65$;
(b) $p+Ta$ at $5.7  \,GeV$, $p_0 = 0.69$;
(c) $p+^{197}Au$ at $11.5  \,GeV$, $p_0 = 0.70$ \cite{Bauer2}.}
\label{fig8}
\end{figure}

  This bond-breaking model was confronted with experimental results 
concerning the reactions $p+Ag$, $p+Ta$ and $p+Au$, for proton energies 
of several $GeVs$. If the bond probability $p$ is parametrized in terms of 
the impact parameter as
\begin{equation} p(b) = p_0 / [1 + \exp \big( (b-R)/d \big) ] 
\nonumber \end{equation} 
where $R$ is the radius of the system, $b$ the impact parameter an $d$ a 
diffusion coefficient, one finds a very good agreement between calculated 
and experimental fragment size distributions, see Fig.~\ref{fig8}, 
by sampling over 
the whole range of impact parameters, $b$ = 0 to $R$. In order to get a 
closer link between percolation calculations it is sensible to replace $p$ 
which is not an observable by the average over realizations of the fragment 
multiplicities $m$ obtained for a fixed value of $p$. This allows to relate 
$\tau$ extracted from a power law fit of the fragment size distribution to 
an average multiplicity at the ``critical'' point, $<m>_c$.

  The strikingly good agreement between the application of simple 
percolation concepts and the outcome of inclusive observables extracted 
from complex collision processes raises several questions. The model does 
not care about mass and energy conservation, possible geometrical 
deformation effects when the fragmenting system is generated through heavy 
ion collisions. The bond probability is necessarily related to the 
microscopic dynamics. It may be asked why this unique parameter 
parametrizes so perfectly the underlying interaction between the
nucleons in the excited nuclear medium. We shall come back to this point 
in section 5.

  In a similar spirit, Campi and Desbois \cite{Campi} developed a continuum 
site percolation approach.$\ A_T$ particles occupy an ensemble of points 
$ \{ \vec r_i, \vec p_i, i = 1,\ldots,A_T \}$ in phase space chosen in a 
classical one-body Wigner distribution $ f_W(\vec r,\vec p)$ determined by a 
one-body Hamiltonian with a Saxon-Woods potential. The collision process 
leaves $pA_T$ nucleons of the original nucleus, $(1-p)A_T$ nuleons are 
kicked off the original nucleus, $p \in [0,1]$ depends on the 
characteristics of the reaction like impact parameter, energy, target
size~\ldots Clusters are generated by means of phase space prescriptions 
which decide whether particles belong to fragments or not. A given number 
of particles forms a bound cluster if
\begin{equation} d_{ij} = |\vec r_i - \vec r_j| \cdot |\vec p_i - \vec p_j|
\leq 2.5 \hbar \label{dij} \nonumber \end{equation} 
for every pair of nearest neighbours belonging to the supposed cluster, 
where the maximum action on the r.h.s. of (\ref{dij}) corresponds to the 
value obtained for the deuteron. This condition is supplemented by a 
compactness constraint which imposes that the mean square radius $R_A$ and 
mean square momentum $p_A$ verifies
\begin{eqnarray} R_A & \leq & (1+ \epsilon) r_0 A^{1/3} \label{R-A} \\
p_A& \leq & (1+ \epsilon) \Big( \frac{3}{5} \Big)^{1/2} p_F^{1/2} 
\nonumber \end{eqnarray} 
where $p_F$ is the Fermi momentum and $\epsilon \simeq 0.12$. If the 
constraints (\ref{R-A}) are not fulfilled the longest link $d_{ij}$ is cut 
and the determination of clusters starts again. The first condition takes 
care of the balance between volume and surface energy of the cluster, the 
second eliminates nucleons with too large momenta. The model is constructed 
with percolation concepts which do however not correspond to the standard 
ones, essentially because the generation algorithm is not fixed by a single 
and simple bond probability parameter. The cluster multiplicity obtained in 
this way can be fitted by means of parametric analytical 
expressions \cite{Campi}. The generated clusters are excited objects whose 
excitation energy $E^*$ can be estimated. Particle emission from fragments 
and their fission are taken into account. 

  In order to enable the confrontation of the model with experiment, the 
quantity $p$ which fixes the number of remaining nucleons after the 
collisions is related to the number of nucleon-nucleon collisions 
which are experienced in the system for a fixed impact parameter $b$ and 
nucleon mean free path $\lambda$ which varies with the energy. This leads 
to a distribution for $p,\ \rho(p)$, for $p$ larger than a value $p_{min}$ 
corresponding to $b=0$. Depending on the physical conditions, $p$ can reach 
or not the percolation threshold. One can then calculate explicitly 
inclusive mass yields for fragments of size $A_F$
\begin{equation} \sigma(A_F)=\int_{p_{min}}^1 \ud p \,\rho(p)\, m(p,A_F)
\nonumber \end{equation} 
where $m(p,A_F)$ is the multiplicity of fragments of size $A_F$, for a 
fixed value of $p$, and the total reaction cross section
\begin{equation} \sigma_R=\int_{p_{min}}^1 \ud p \,\rho(p) 
\nonumber \end{equation}
  The model reproduces very well the fragment size yields $\sigma(A_F)$ for 
low and high energy data corresponding to $^{20}Ne+^{181}Ta$ at
$400\,MeV\!\cdot\!A$, $p+^{181}Ta$ at $5.7\,GeV$ and $p+^{181}Ta$ at
$340\,MeV$. In the lower energy data one observes the opening of 
the fission channel which is predominant along with particle evaporation at 
this energy. For high energy reactions, $p+Xe$ at $80-350\,GeV$, both the 
slope and absolute value of $\sigma(A_F)$ are well reproduced. 
The trend of the apparent ``critical'' exponent 
$\sigma(A_F) \propto A_F^{-\tau(E)}$ where $E$ is the bombarding energy 
shows a minimum $\tau \sim 2.4$ at $E=50\,GeV$. This is understood in the 
following way. When $E$ is small $p_{min}$ is large, $p_c$ cannot be 
reached, hence the system is essentially made up of one large cluster and a 
set of small particles. When $E$ increases $p_{min}$ decreases, $p_c$ can 
be reached and one gets a power law distribution over the whole range of 
fragments sizes, except for finite size effects. For large value of $E$ 
$p_{min}$ continues to decrease, more and more smaller fragments are 
generated and the apparent slope of the distribution increases again. At 
very high energy the nucleon mean free path $\lambda(E)$ becomes constant, 
hence $p_{min}$ and $\tau$ saturate.

\subsection{Relevant observables and percolation analysis of experimental 
data}
\indent

  The first extensive confrontation between standard percolation theory and 
nuclear fragmentation was initiated by Campi \cite{Campi2}. The phenomenon 
was interpreted as a bond-breaking process between bound nucleons due to 
the energy supplied to the system under violent collision conditions. As 
discussed in section~4.3, the moments $m_k$ of the fragment size 
distribution get singular at $p=p_c$ for $k \geq 2$, so do the normalized 
moments $S_k=m_k/m_1$. Hence one expects that they show a maximum in the 
vicinity of $p \sim p_c$. Since $m_k \propto \exp(-\mu_k)$ with
$\mu_k=(\tau-1-k)/\sigma$ (see (\ref{momk}) and (\ref{muk})) 
at $p_c$, there should exist a 
linear relation between $\ln S_k$ and $\ln S_l$ at $p_c$. For $k=2$ and 
$l=3$ the slope can be written as $\ln S_3 =  \lambda \ln S_2$ with
$\lambda=1+1/\sigma \gamma$ and one can in principle read off the value of 
the power law exponent $\tau$. 

  The introduction of $m_k$ and $S_k$ allows for a direct comparison between 
the moments obtained from bond percolation models (here $3d$ cubic) and 
those obtained from the experiment, $m_k^j$, which can be determined from 
the fragment size distributions of a set of events labelled by the index 
$j$ by taking an average over all available events. Another interesting
observable is given by \cite{Campi3}
\begin{equation} \gamma_2 = \frac{m_2 \cdot m_0}{m_1^2} 
\label{gamma2} \end{equation} 
which is related to the variance $\sigma$ of the fragment size distribution by
\begin{equation} \gamma_2 = 1+\sigma^2\cdot m_0^2/m_1^2
\nonumber \end{equation} 
and should also show a sizable enhancement at the ``critical'' threshold.

  Finally the last observable proposed in ref. \cite{Campi2} concerns the 
behaviour of the largest cluster $A_{max}$ in the system. Indeed, in the 
infinite system, $A_{max}$ should get infinite at criticality, hence it is 
expected to get large when $S_2$ gets large. Hence the correlation between 
$A_{max}$ and $S_2$ is expected to contain information of interest. 

\begin{figure}[b!]
\centerline{\epsfig{figure=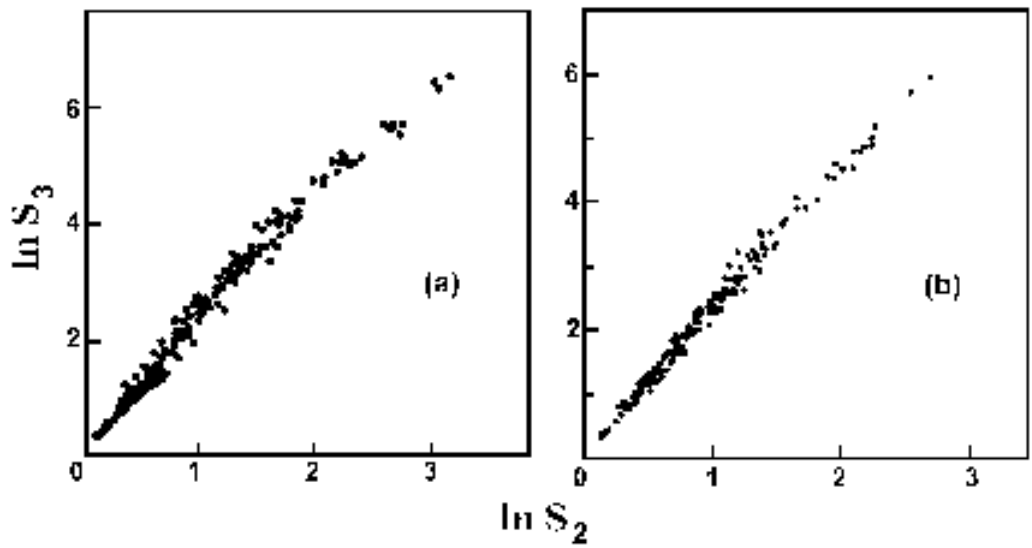,width=140mm}}
\caption{$\ln S_3$ as a function of $\ln S_2$ for single events. Part (a) 
corresponds to the break-up of gold nuclei, part (b) to a cubic bond 
simulation with 216 sites and bond probabilities $0 < p < 1$,
\cite{Campi2}.}
\label{fig9}
\end{figure}

\begin{figure}[b!]
\centerline{\epsfig{figure=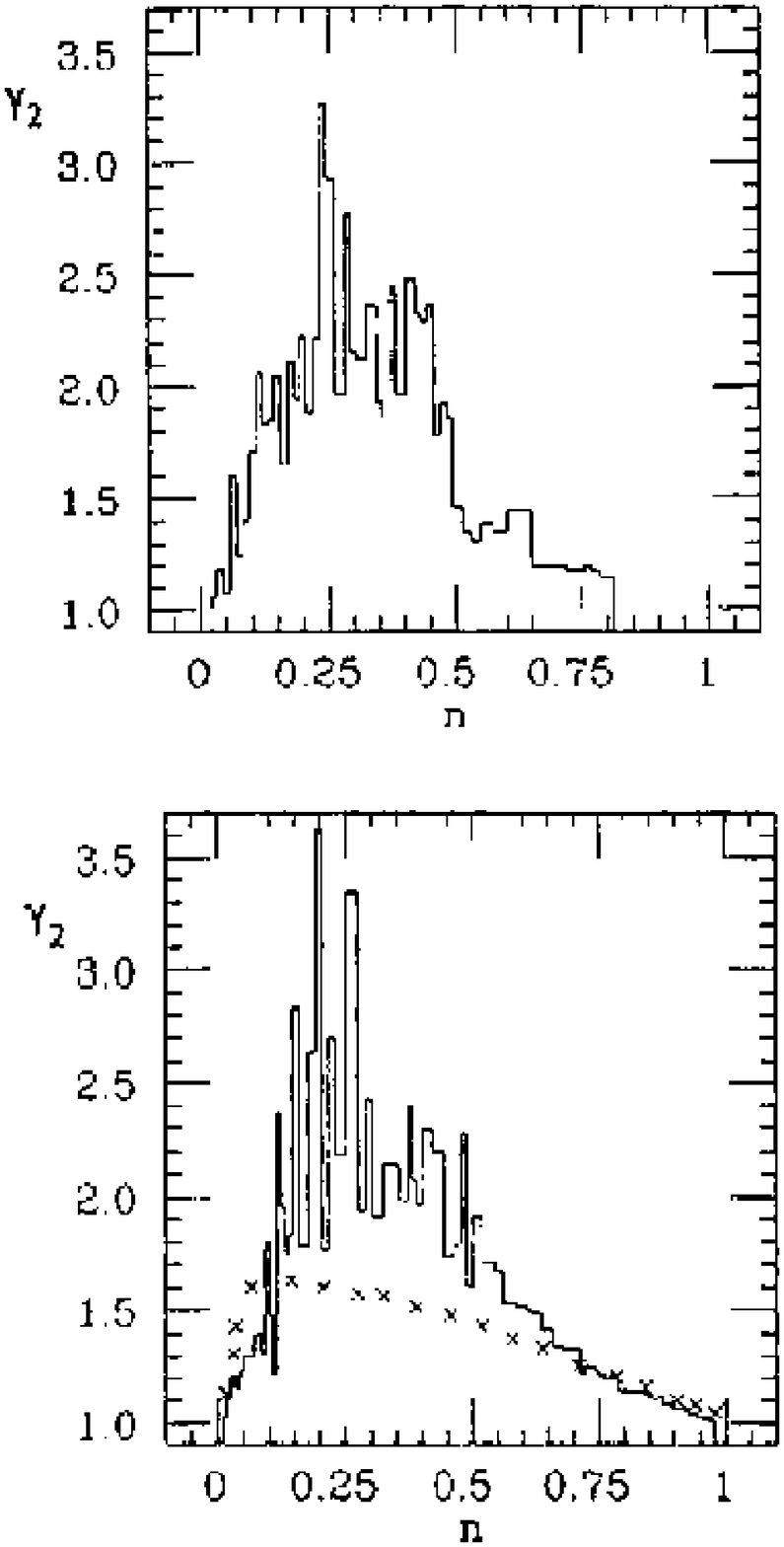,height=160mm}}
\caption{The quantity $\gamma_2$ defined in (\ref{gamma2}). The upper 
part corresponds to the fragmentation of gold and the lower part to 
percolation simulations, \cite{Campi3}.}
\label{fig10}
\end{figure}

\begin{figure}[b!]
\centerline{\epsfig{figure=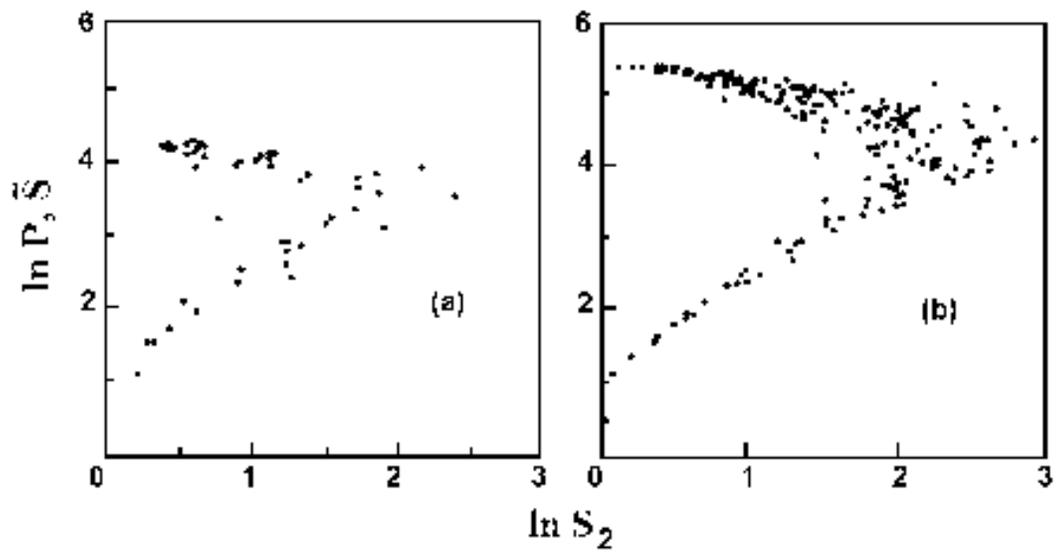,width=140mm}}
\caption{Size of the largest fragment for different events as a function of 
$S_2$. (a) Experimental results obtained from 376 events. (b) Bond 
percolation results in a cubic volume with 216 sites, 4000 realizations
\cite{Campi2}.}
\label{fig11}
\end{figure}

  Bond percolation calculations in a cubic lattice in a finite system were 
performed for different sizes and $S_2, S_3, \gamma_2$ and $A_{max}$ were 
calculated. The average quantities were confronted with the outcome of a 
few hundred events generated by the fragmentation of $^{197}Au$ nuclei in 
emulsions \cite{Waddington}. Typical fission events involving two large 
fragments were excluded. Results are shown in 
Fig.~\ref{fig9} which relates $S_3$ and $S_2$ for both experimental 
events and percolation simulations. ``Critical'' events correspond to 
large values of $S_2$ and $S_3$. Violent $(p=0)$ and smooth $(p=1)$ 
collision events correspond to small values of $S_2$ and $S_3$. One 
observes that there is a striking agreement between the two results. It 
should be noticed that in both cases one mixes events which correspond to 
different values of $p$ (as far as calculations are concerned) and 
multiplicities which are related to the degree of violence of the process 
(as far as the experiment is concerned). They both show the expected linear 
behaviour which can be related to the presence of ``critical'' events. It 
is also remarkable to notice that the slopes $\lambda \simeq 2.22$ are in 
both cases very close to the value $\lambda = 2.25$ corresponding to bond 
percolation in the infinite system. The same type of agreement can be 
observed when dealing with $S_5$ vs. $S_2$ \cite{Campi2}. In Fig.~\ref{fig10}
$\,\gamma_2$ is drawn as a function of the average multiplicity $m_0$. The
similarity of simulation results and experiment is again remarkable. One
observes an enhancement in the neighbourhood of $m_0 \simeq 0.25$ which, when 
related to the bond probability $p$, corresponds to $p \sim p_c$ in $3d$ 
cubic systems. The fact that $\gamma_2 > 2$ for $m_0 \simeq 0.25$ indicates 
a power law distribution. Finally Fig.~\ref{fig11} compares the behaviour of 
$A_{max}$ as a function of $S_2$ in both cases. ``Critical'' events are 
concentrated in the right corner of the figure. One observes again a strong 
agreement between calculations and experiment.

  The possibility to extract values of $\lambda$ and $\tau$ and the fact 
that these are close to standard percolation results may induce the 
temptation to conclude about the universality class of the apparent 
critical phenomenon unravelled by the present studies. One must however be 
careful, because different types of transitions may lead to exponents which 
lie close to each other but correspond to different classes. The 
determination of exponents needs some care. We shall come back to the 
problem concerning the determination of critical exponents in section 5.

\subsection{Extensive tests on peripheral collision data : analysis of 
ALADIN experiments}
\indent

  Many experimental efforts have been made over the last decade by the 
ALADIN collaboration of the GSI in order to collect a large amount of 
significant information about particle and fragment multiplicities
generated in the fragmentation of the projectile by means of energetic
peripheral heavy ion collisions. Such precise and complete information
allowed for a careful
comparison with theoretical approaches, in particular percolation and 
percolation-inspired models. The first measurements which were performed 
established correlations between the multiplicity of light particles 
$M_{lp}$, the charge $Z_{max}$ of the largest fragment in an event and the 
sum of the charges of the fragments with $Z \geq 2, \ Z_{bound}$ \cite{
Hubele}. It was found that the correlation between the multiplicity of 
intermediate mass fragments (IMFs), i.e. fragments with $3 \leq Z \leq 30$ 
and $Z_{bound}$ was independent of the target nucleus. This was interpreted 
as a sign for the equilibration of the projectile remnant (spectator 
particles) before decay. The first detailed investigation of the average 
multiplicity of IMFs, $<M_{IMF}>$, showed that with increasing energy 
transferred to the projectile in reactions of $Au$ on $C, Al$ and $Cu$ 
targets $<M_{IMF}>$ first increases to a maximum and then
decays \cite{Ogilvie}. This could be interpreted in the following way. For 
small energies peripheral collisions lead essentially to one large 
projectile remnant and a few light particles. With increasing energy, the 
projectile breaks up into more and more IMFs. If the energy gets very large 
most fragments are very small, $(Z \leq 2)$ and hence $<M_{IMF}>$ decreases 
again. These results were then compared to theoretical investigations of $<
M_{IMF}>$ vs. $Z_{bound}$ \cite{Hubele1} by means of the statistical models 
of Copenhagen \cite{Bondorf1, Bondorf2, Barz} and Berlin \cite{Gross3}, and 
a sequential decay scenario \cite{Charity}. It is seen that sequential 
evaporation cannot reproduce the data.

  An extensive analysis concerning the reactions with $Au$ on $C, Al, Cu$ 
and $Pb$ at $600\,MeV\!\cdot\!A$ was performed in ref. \cite{Kreutz}. The 
charge distribution was fitted by a power law and the exponent $\tau$ showed 
a minimum somewhat below 2 when represented as a function of $Z_{bound}$. 
The average largest fragment charge $<Z_{max}>$ increases with 
increasing $Z_{bound}$. Other correlation functions like 
$\gamma_2$ (see above) and
\begin{equation} A_3 = [(Z_{max}-<Z>)^2 + (Z_2-<Z>)^2
+ (Z_3-<Z>)^2]^{1/2} / \sqrt{6} \nonumber \end{equation}
\begin{equation} \textrm{where} \quad <Z> = \frac{1}{3}(Z_{max}+Z_2+Z_3)
\nonumber \end{equation} 
and $Z_2$ and $Z_3$ are the second and third largest fragment charges were 
determined. 

\begin{figure}[b!]
\centerline{\epsfig{figure=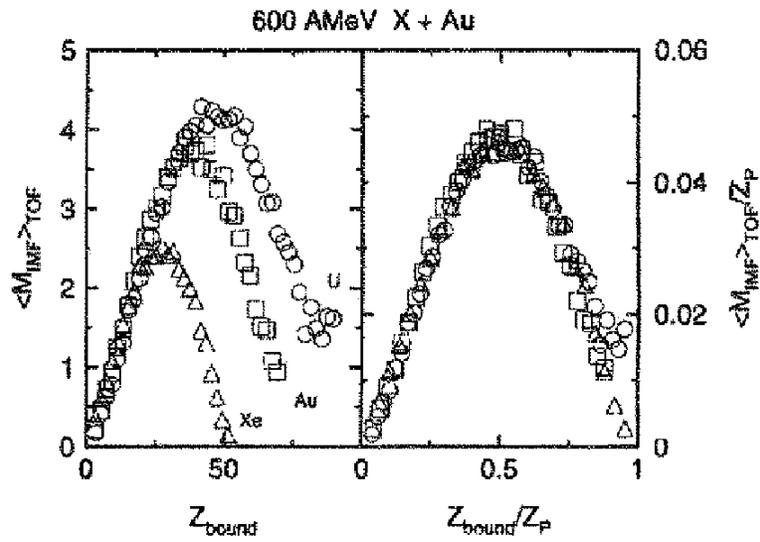,width=100mm}}
\caption{Left: measured mean multiplicities $<M_{IMF}>$ as a function of 
$Z_{bound}$ for $^{238}U$ on $^{197}Au$ (circles), $^{197}Au$ on $^{197}Au$ 
(squares) and $^{129}Xe$ on $^{197}Au$ (triangles) for $E = 600\,MeV\!\cdot
\!A$. Right: same quantities but rescaled by normalization with respect to 
the number of charges $Z_p$ of the projectile \cite{Schuttauf}.}
\label{fig12}
\end{figure}

  All results showed the striking feature that it is a simple-minded $3d$ 
site-bond percolation model with an occupation probability $p_s$ for the 
site and a fixed bond probability $p_b=0.45$ obtained through a fit to the 
$Z_{max}$ distribution which reproduces the best the available data.
Masses in the percolation calculations were converted into charges by means 
of a semi-empirical mass formula \cite{Kreutz}. The investigation of 
projectile fragmentation was finally extended to other projectiles like
$^{129}Xe$ and $^{238}U$ with $Be, C, Al, Cu, In, Au$ and $U$ targets and 
energies ranging from 400 to $1000\,MeV\!\cdot\!A$ \cite{Schuttauf}. The
striking feature about the results concerns the energy independence of 
$<M_{IMF}>$ as a function of $Z_{bound}$ and its scaling 
properties, i.e. all curves get superposed on each other if $Z_{bound}$ is 
rescaled to $Z_{bound}/Z_p$, where $Z_p$ is the charge of the projectile as 
it can be seen in Fig.~\ref{fig12}. Statistical multifragmentation models 
reproduce qualitatively if not quantitatively the experimental results.

\begin{figure}[b!]
\centerline{\epsfig{figure=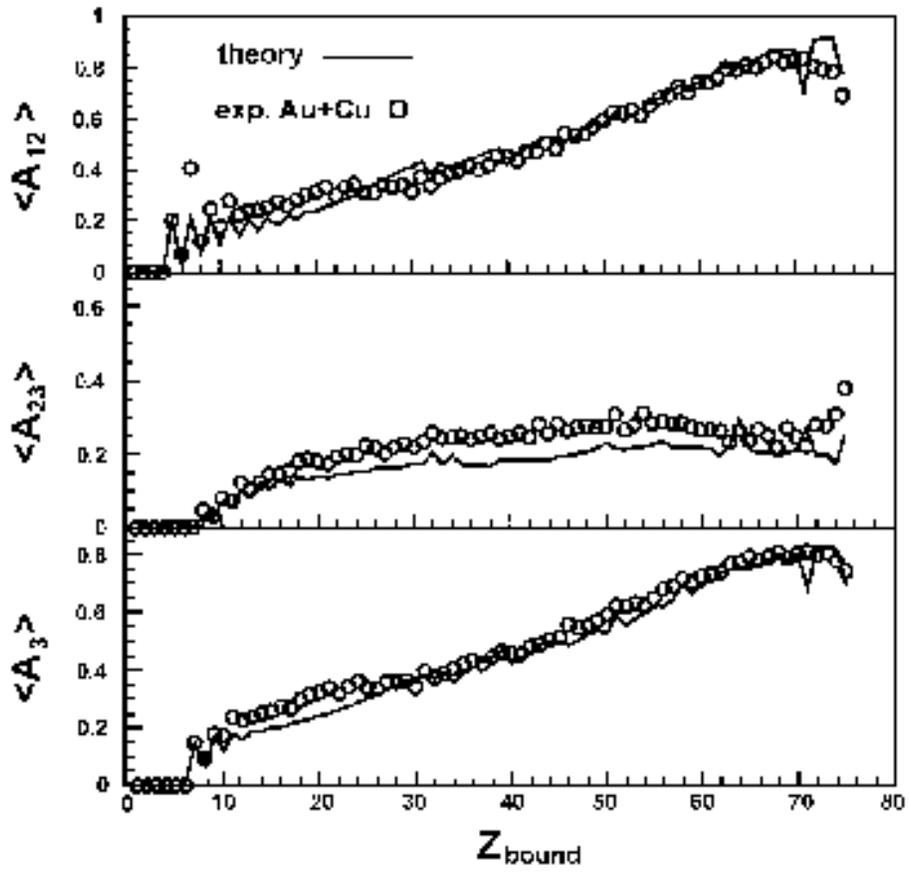,height=120mm}}
\caption{Average values of different observables defined in the text as a 
function of $Z_{bound}$ \cite{Zheng}.}
\label{fig13}
\end{figure}

  Recent ALADIN data concerning fragment size multiplicities for 
$^{197}Au$ on different target reactions at $600\,MeV\!\cdot\!A$ were 
analysed by means of a standard $3d$ cubic bond percolation 
model \cite{Zheng}. Depending on the value of the impact parameter, the 
number of spectator nucleons which remains in the projectile remnant is of 
course different. In order to decide about this number, a simulation by 
means of a BUU transport equation of the reaction process was performed. 
The spectator nucleons were arranged on a cubic lattice in a compact 
spherical arrangement and a bond percolation algorithm was introduced. 
Masses were converted into charges by means of the relation 
\begin{equation} Z=A/(1.98+0.0155A^{2/3}) \nonumber \end{equation} 
for $A \geq 4$. A filter was applied to the results of the calculations in 
order to allow for a direct confrontation between the model calculations 
and the experimental data. A whole set of observables was analysed in this 
way. Fig.~\ref{fig13} shows the average value of 
$A_{12}=(Z_{max}-Z_2)/(Z_{max}+Z_2),\ A_{23}=(Z_2-Z_3)/(Z_2+Z_3)$ 
and $A_3$ defined above as a function of 
$Z_{bound}$. As one can see the agreement with the experiment is nearly 
perfect and this is also the case for all other observables.

  A new analysis of the data \cite{Campi4} confirms this statement. The 
authors also use a standard $3d$ bond percolation algorithm. The number of 
nucleons $Z_{PS}$ which is taken into account \cite{Campi5} is parametrized 
in terms of $Z_{bound}$. Simulations are done for fixed $Z_{PS}$ and 
observables are determined for fixed $Z_{bound}$. The analysis is again 
parameter free. The calculated quantities are the fragment charge 
distributions, the average size and fluctuations of the largest fragment, 
and the largest fragment distribution. The confrontation also 
shows that there appears a ``critical'' regime, characterized by a power 
law charge distribution with $\tau \simeq 2.2$. In the neighbourhood of the 
``critical'' point the distribution of fragments with charge $z$ can be 
parametrized as 
\begin{equation} n(z)=n_c(z)f(z/\overline{z}) 
\nonumber \end{equation} 
where $\overline{z}=m_3/m_2, \ m_k$ being the moment of order $k$ defined 
in section 4.3.

  The authors consider two interpretations of these results. The first 
relies on the fact that a fragmenting nucleus is made of particles which 
interact essentially as nearest neighbours through a short range two-body 
potential. If the relative kinetic energy between pairs of neighbouring 
particles is larger than the potential energy they experience a bond is 
broken and hence independent fragments are generated if the excitation 
energy is large enough. The second interpretation requires the assumption 
of thermodynamic equilibrium. At different fixed densities it is the 
temperature $T$ of the system which fixes its fragment content. For a given 
density $\rho$ of the system, there exists a temperature $T(\rho)$ where 
the distribution looks like a power law. We shall come back to explicit 
classical microscopic model illustrations of both interpretations in 
section 5 below. It is clear that thermodynamic equilibrium is not a 
necessary prerequisite for an accurate description of the fragment content
of energetic nuclear collisions.

\subsection{Comparison with other fragmentation models}
\indent

  The intriguing and impressive success of classical percolation models in 
the description of finite nuclear fragmentation events and the existence of 
a continuous phase transition in infinite percolation systems triggered 
many attempts to compare and possibly relate different models with the 
percolation approach. Such attempts started with the Statistical 
Multifragmentation Model SMM \cite{Barz}. The authors tried to establish a 
link between the percolation bond probability $p$ and the excitation energy 
available in the fragmented system described by SMM \cite{Bondorf2} at 
freeze-out. They fixed $p=V_0/(V_0+V_f)$ where $V_0$ is the volume of the 
initial compact system and $V_f$ the volume at freeze-out. The 
investigations revealed that such a link is not clearly established and led 
to the conclusion that inclusive fragment size distributions may not contain 
enough information in order to distinguish the purely statistical features 
of the fragmentation process from those which concern physical information 
carried in phase space models.

  A detailed comparison of the fragment content obtained in the framework 
of the MMMC model \cite{Gross} with percolation was carried out in 
ref. \cite{Jaqaman3} by means of an analysis of 
the relation between the charge of the 
heaviest fragment and the second moment of the charge 
distribution \cite{Campi2}. Close similarities between MMMC and percolation 
models were found, in particular the ``critical'' zone which was discussed 
in section 4.6. The main qualitative difference between the two approaches 
is the manifestation of the presence of fission events (two heavy fragments) 
which are absent in the percolation simulations. This was interpreted as 
being due to the presence of the long range Coulomb interaction which is not 
taken into account by the percolation model. Three-body correlations 
between the three heaviest fragments were also in agreement with the 
percolation predictions. So called cracking events with more than two large 
fragments emphasized the role and importance of the Coulomb interaction.

  Moment correlations \cite{Campi1, Campi2} have been used in order to test 
different models and possibly discriminate between them \cite{Campi6}. The 
models which are introduced are  an equal probability model for the 
generation of different partitions \cite{Aichelin2, Sobotka}, 
time-dependent sequential decay processes \cite{Richert, Ziff} and a 
standard bond percolation model on a finite cubic lattice. 
All models agree qualitatively with percolation 
calculations but do not reach quantitative agreement. The peak 
corresponding to ``criticality'' in percolation is always present, but it 
is shifted in the multiplicity and its height is not reproduced. As 
expected, the agreement between percolation calculations and experiment is 
excellent.

  A model analysis of data has also been performed by Kreutz et 
al. \cite{Kreutz} with a statistical decay model \cite{Charity}, SMM and, 
as already discussed above, a site-bond percolation model. It appears again
that all models lead to quantitatively different results. Sequential decay 
fails to reproduce charge correlation measurements, $<\gamma_2>$, charge 
distributions and IMF multiplicities come out more or less closely to 
experiment in SMM, although the average largest fragment charge is too 
small.

\subsection{Final remarks}
\indent

  If one restricts the information concerning the fragmentation of highly 
excited nuclei to the description of the asymptotic fragment size 
distribution it appears clearly that parameter free standard percolation 
models are the only ones which are able to reproduce the experimental data 
quantitatively. It raises the up to now unanswered question why this is so. 
It shows that a minimum information model is sufficient in order to explain 
the a priori complex nuclear fragmentation process. The agreement is by 
essence independent of any detail concerning explicit two-body potentials. 
As it will be seen later, simple models governed by short range 
interactions are able to reproduce percolation results. The long range 
Coulomb interaction which induces fission-like fragmentation is out of 
reach. The fact that its absence does not affect the physics of the process
may be understood if the excitation energy is large enough. The 
influence of the Coulomb interaction will be more extensively discussed in 
the framework of lattice models in section 5. Taking a pessimistic 
attitude, it is tempting to conclude that the success of percolation
concepts is the sign that the 
information content of the process is poor. But there remains the fact that 
the result is not trivial since the model is not.

  The success of a percolation description raises the problem of the 
presence of a phase transition in the infinite nuclear system, its origin 
and its nature. The formal similarity with Fisher's model \cite{Fisher} is 
striking, since even the order of magnitude of the critical power law 
exponent $\tau$ is reproduced. Fisher's model leads to a thermodynamic 
phase transition. Does there exist a connection of the second order 
percolation phase transition and the thermodynamic properties of the system ?

  The outcome of the analysis of finite systems \cite{Biro} should be kept 
in mind. Mass distributions of finite systems are able to show a behaviour 
which corresponds or is close to a power law behaviour, which may 
simply be induced by particle number 
conservation in a system confined in a finite volume, suggesting that a
power law behaviour may not necessarily be the sign for the existence of
a critical phenomenon.

  Experimentally, charge distributions are only a small part of the 
available information on nuclear fragmentation.
This information has to be reproduced. It has to be 
related to the properties of observables related to the energy, in 
particular to the thermodynamic properties of the system if it is in 
equilibrium. This point has already been considered in section 4.8. It will
be raised again in section 5. There we introduce generic microscopic 
cellular and lattice models which are close in spirit to percolation models 
but also hopefully realistic enough to be able to reproduce the 
thermodynamic properties of excited fragmenting nuclei.
  
\newpage

\section{Lattice and cellular model approaches to nuclear fragmentation}
\indent

  Standard percolation provides a surprisingly good description of the 
asymptotic fragment multiplicity distributions generated in energetic 
nuclear collisions. By essence, percolation models cannot describe the 
thermodynamic properties of the decaying systems since they do not 
explicitly integrate the properties related to thermodynamic quantities 
like energy, volume, pressure. In the preceding section we 
presented models which try to link the purely geometrical aspects of 
percolation models to considerations related to phase space \cite{Barz, 
Bauer1, Bauer2, Campi, Campi1, Jaqaman3}. This link introduces 
phenomenologically reasonable but arbitrary parametrizations of the bond 
probability, generally in terms of excitation energy. The procedure may 
not be totally satisfactory. A hint to a more consistent approach can be 
found if one remembers the intimate relation between standard percolation 
models and spin models like the Ising model and its generalizations \cite{
Fortuin, Kunz}. These last models introduce explicit Hamiltonians and are 
generic for many physical systems. The aforementioned formal relation 
induced the introduction of classical microscopic models, starting with 
Ising itself. These models allow for a thermodynamic description and the 
study of the fragment content of the system in a unified framework. They 
can be considered as a simplified though hopefully realistic description of 
a nucleonic system.

  In the sequel we aim to introduce and discuss different types of cellular 
and lattice models which have been proposed and studied by different groups 
in the recent past. In order to put this study in a comprehensive 
perspective we first recall some fundamental and useful concepts concerning 
statistical ensembles and thermodynamic phase transitions in finite and 
infinite systems at equilibrium.

\subsection{Ensembles, thermodynamic stability, phase transitions in finite 
and infinite systems}
\indent

  We summarize here some general definitions and concepts which will be 
used for later developments and discussions presented below. 
They can be found in textbooks, see f.i. refs. \cite{Stanley,Yeomans,Uzunov}.

  If one considers a large classical microscopic system which is in 
thermodynamic equilibrium, each particle is characterized by its phase 
space coordinates $\{\vec r_i, \vec p_i, i = 1,\ldots,N\}$. If $\{\vec r_i, 
\vec p_i\}$ is considered as a point in this space, there can exist a large 
number of points which correspond to the same macroscopic 
state. Each set $\{\vec r_i, \vec p_i\}$ defines a realization of the 
system. A collection of realizations which correspond to the same 
macroscopic thermodynamic state is called an ensemble with an arbitrary 
fixed number of physical properties such as a fixed number of particles, 
energy, temperature, angular momentum and possibly other conserved 
quantities.

  In statistical mechanics there are essentially three ensembles which are 
used in practical applications. The most common one is the canonical 
ensemble which corresponds to a closed isothermal system with a fixed 
number of $N$ particles and which exchanges energy with an external ensemble 
generally called reservoir. If the system is made of $N_t$ subsystems of 
$n_i$ subsystems with energy $E_i \ (i=1,\ldots,N_t)$
\begin{eqnarray} \sum_i n_i & = & N_t \label{cstrN} \\
\sum_i n_iE_i & = & E \label{cstrE} \end{eqnarray} 
  The thermodynamic probability which characterizes a macroscopic state is
\begin{equation} W(\{n_i\}) = N_t! \, / \prod_i n_i! \nonumber \end{equation} 
and the entropy reads 
\begin{equation} \Sigma = k \ln W_{max} \nonumber \end{equation} 
where $k$ is the Boltzmann constant and $W_{max}$ is the maximum of $W$ 
consistent with the constraints (\ref{cstrN}) and (\ref{cstrE}). The 
average entropy $S=\Sigma/N_t$ can be written as 
\begin{eqnarray} S & = & -k \sum_i P_i \ln P_i \nonumber \\
P_i & = & n_i/N_t = e^{-\beta E_i}/ \mathcal{Z} \nonumber \\
\mathcal{Z} & = & \sum_i e^{-\beta E_i} \nonumber \end{eqnarray} 
$P_i$ is the weight of the subsystem $i$ and $\beta$ can be identified with 
the inverse
thermodynamic temperature $T^{-1}$. The knowledge of the entropy allows for 
the determination of all thermodynamic functions and quantities such as the 
free energy of a system in a volume $V$
\begin{equation} F(N,V,T) = U - TS = -kT \ln \mathcal{Z} \nonumber 
\end{equation} 
where $U$ is the energy.

  The concept of temperature has been extensively used in the canonical 
framework in low energy nuclear physics, in particular in Hartree Fock 
models \cite{Quentin} and compound nucleus theory \cite{Mahaux}. Its use 
implicitly implies that the energy is not a rigorously fixed quantity. It 
raises also the question of the concept of temperature in very small 
systems like nuclei \cite{Feshbach}.

  In the grand canonical ensemble both energy and number of particles are 
only fixed in the average by supplementing eqs. (\ref{cstrN}) and 
(\ref{cstrE}) with 
\begin{equation} \sum_i n_i N_i = N \nonumber \end{equation} 
where $N_i$ is the number of particles in subsystem $i$. Then
\begin{eqnarray} S & = & -k \sum_i \mathcal{P}_i \ln \mathcal{P}_i 
\nonumber \\
\textrm{where} \quad \mathcal{P}_i & = & e^{-\beta E_i + \beta \mu n_i}
/ \Xi, \nonumber \\
\Xi & = & \sum_i e^{-\beta E_i + \beta \mu n_i} \nonumber \end{eqnarray}
and $\mu$ is the chemical potential which fixes the average number of 
particles of the system. One can again work out all thermodynamic functions,
in particular the Gibbs free energy $G=F+PV$ where $P$ is the pressure in 
the system.

If the system is closed, with fixed energy and fixed number of particles, 
it is described in the framework of the microcanonical ensemble. The 
temperature $T$ is then no longer a natural concept. Mathematically 
speaking the partition function with fixed temperature is the Laplace 
transform of the partition function with fixed energy, hence energy and
temperature are conjugate through a Laplace transformation. The temperature 
can be introduced through the thermodynamic relation
\begin{equation} T^{-1} \equiv \bigg( \frac{\partial S}{\partial E} \bigg)_V
\nonumber \end{equation} 
where the entropy
\begin{equation} S=-k \sum_i P_i \ln P_i \nonumber \end{equation} 
and $P_i$ is the weight of subsystem $i$. The principle of equiprobability 
of occupation stipulates that 
\begin{eqnarray} P_i & = & 1/\Omega \quad \textrm{for} \quad 
E-\Delta E < E_i < E \nonumber \\
& = & 0 \quad \textrm{otherwise} \nonumber \end{eqnarray} 
where $\Omega$ is the partition sum of the system, the number of systems in 
the ensemble.

Then
\begin{equation}  S = k \ln \Omega \nonumber \end{equation} 
This shows that the central quantity to be known is the partition function 
which counts the number of states at energy $E$ or the directly related 
entropy $S$. 

\subsubsection{Stability of thermodynamic systems}
\indent

  Microscopic many-body systems can be microscopically unstable. The \\
Le Chatelier principle stipulates that any spontaneous change in the 
parameters of a system which is in stable equilibrium will give rise to a 
process which tends to restore the system to equilibrium \cite{Stanley}. 
This leads to the inequalities 
\begin{eqnarray} C & = & \frac{\ud Q}{\ud T} > 0 \label{ineQ} \\
K & = & -\frac{\ud V}{\ud P} > 0 \label{ineV} \end{eqnarray} 
for the heat capacity $C$ and the compressibility $K$ which characterize 
the change in the quantity of heat $Q$ when the temperature changes and the 
change of the volume $V$ of the system if the pressure $P$ varies. Eq. 
(\ref{ineQ}) concerns thermal equilibrium and (\ref{ineV}) mechanical 
equilibrium. These inequalities show that the free energy $F$ is a concave 
function of the temperature $T$ and a convex function of the volume whereas 
the Gibbs function $G$ is a concave function of both $T$ and $V$. Indeed, 
in particular
\begin{eqnarray} \bigg( \frac{\partial^2 F}{\partial T^2} \bigg)_V & = &
-\bigg( \frac{\partial S}{\partial T} \bigg)_V = - \frac{1}{T} C_V < 0 
\nonumber \\
\bigg( \frac{\partial^2 G}{\partial T^2} \bigg)_P & = &
-\bigg( \frac{\partial S}{\partial T} \bigg)_P = - \frac{1}{T} C_P < 0
\nonumber \end{eqnarray} 
for fixed volume $V$ and pressure $P$ respectively.

  If (\ref{ineQ}) and (\ref{ineV}) are not realized the considered system 
is unstable and cannot be described in the framework of equilibrium 
thermodynamics.

\subsubsection{Some essential reminders concerning phase transitions}
\indent

  Phase transitions can occur in many types of systems. In the traditional 
acceptance of the word they are observed in infinite systems. Only there 
can the correlation length between different points in the system or 
thermodynamic quantities like the heat capacity get infinitely large. In 
practice the phenomenon can be observed and mathematically extrapolated to 
infinite systems in macroscopic large systems like those which are at hand 
in condensed matter physics where the phenomenon has been studied for all 
kinds of materials.

  There exist different types of phase transitions in nature. They have 
been classified in universality classes. Thermodynamic transitions are said 
to be of first order if there exists a discontinuity in one or more first 
derivatives of the appropriate thermodynamic potential. If, for instance, 
the entropy is not continuous as a function of temperature then the 
transition is associated with latent heat generation.

  If the first derivatives are continuous but second or higher ones are 
discontinuous the transition is said to be of second order or continuous. 
There appears then divergences in quantities like the susceptibility, 
correlations get infinite and decay as power laws. The order of a 
transition is a universal characteristic of the phenomenon. Different 
though closely related systems can experience different types of 
transitions. Such is for instance the case of the Ising model which shows a 
second order transition at some fixed value $T_c$ of the temperature and a 
first order transition for $T < T_c$ for $B=0$ if the system experiences a 
magnetic field $B$. Then the magnetization which is the derivative of the 
free energy with respect to $B$ shows a jump. It is 
continuous at $T = T_c$ but shows an infinite slope there.

  Phase transitions can be characterized by order parameters. These are 
useful physical quantities which are a help in order to distinguish between 
different phases. 
They can be chosen arbitrarly among the extensive variables. In 
spin systems the magnetization works as an order parameter. For each phase 
there exists a characteristic value of the order parameter, the magnetization 
is zero above $T_c$ and finite for $T < T_c$. Order parameters behave 
either continuously or jump at a transition crossing point. This is the 
reason for their usefulness when trying to find the order of a transition.

\subsubsection{Critical points, singularities, universality and critical 
exponents}
\indent
  
  Phase transitions occur at specific values or intervals of values of the 
thermodynamic variables such as the temperature. As already seen in section 
4 these values which correspond to so called critical points, lines, 
surfaces, are marked by divergences in physical observables. In the 
case of a thermodynamic transition one expects that in the vicinity of a 
critical point $T_c$ observables may behave like
\begin{equation} \lim_{t \rightarrow 0} F(t) = A|t|^ \lambda (1 +
pt^{ \lambda_1} + \cdots ) \nonumber \end{equation} 
with $\lambda_1>0$, as a function of the dimensionless parameter 
$t=(T-T_c)/T_c$. 

\begin{figure}[b!]
\centerline{\epsfig{figure=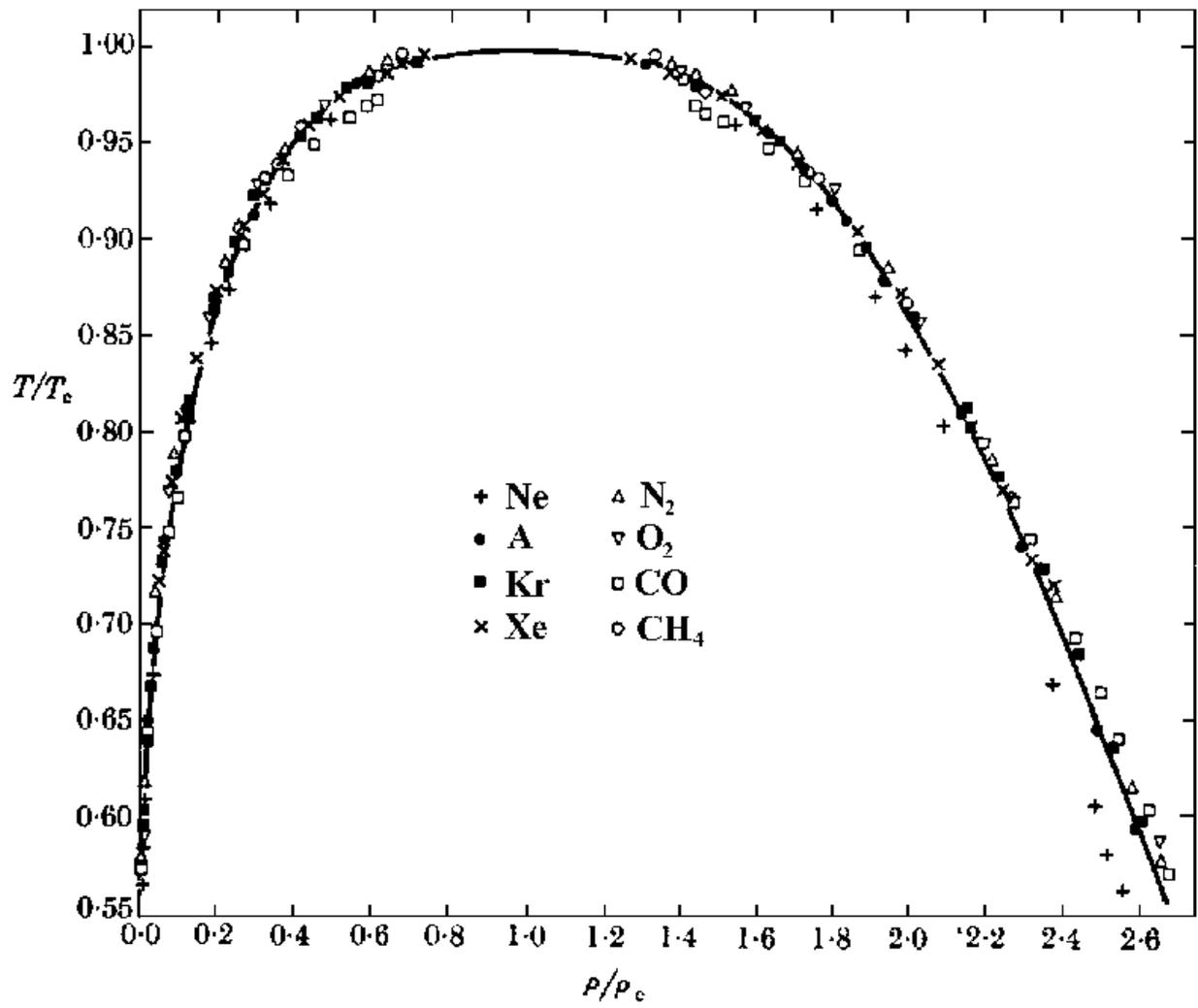,height=140mm}}
\caption{Phase diagram showing a first order transition for eight fluids 
indicated on the figure \cite{Guggenheim}. The collapse of the coexistence 
curve shows the universality character of the phenomenon.}
\label{fig14}
\end{figure}

  For instance for a fluid system one observes that the specific heat at 
constant volume $C_V$ behaves like $C_V \sim |t|^{ -\alpha}$, the 
liquid-vapor density difference $(\rho_l - \rho_g) \sim (-t)^ \beta$, the 
isothermal compressibility $K_T \sim |t|^{ -\gamma}$ and the correlation 
length $\xi \sim |t|^{ -\nu}$, where all the indices are positive. The 
interest in the knowledge of critical exponents lies in the fact that they 
possess a universal character. While the location of the critical exponent 
itself may depend on the details of the interparticle interaction, the 
exponents are universal in the sense that they depend only on quantities 
like the dimensionality $d$ of space and the symmetry properties of the 
order parameter. Evidence for this has been verified number of times, in 
particular for the first time by Guggenheim \cite{Guggenheim} in the phase 
separation line of the liquid-gas transition for different species, 
see Fig.~\ref{fig14}, 
for which it can be seen that the exponent $\beta = 1/3$. The 
very same value for $\beta$ comes in the 3-dimensional Ising model, the 
magnetization $M$ at $T=T_c$ behaves like $\simeq (-t)^{1/3}$. These 
properties lead to define universality classes for systems showing the 
same behaviour at phase transitions. They allow to classify a priori very 
different physical systems by means of their critical exponents which 
characterize the behaviour of their order parameters at a critical point. 
Scaling relations which govern the behaviour of physical quantities when 
one looks at systems at different scales allow to establish inequality or 
equality relations between the exponents. This has already been shown in 
section 4 in the framework of percolation, see also below. 

  The importance of critical exponents raises the question of their 
explicit determination. Generally this can only be done numerically. Since 
critical exponents only make sense in infinite systems it is necessary to 
find methods which lead to an extrapolation of the behaviour of the critical 
observables to the infinite limit in order to determine them. There 
exists different, more or less sophisticated methods and numerical 
techniques which allow for very precise determinations of exponents, see 
f.i. refs. \cite{Adler,Debierre}. 

  A whole class of methods relies on the so called finite 
size scaling assumption (FSS). The scaling hypothesis implies that in the 
neighbourhood of a critical point a system shows self-similarity properties 
when it is examined at different scales. These properties are reflected in 
the mathematical behaviour of thermodynamic functions and physical 
observables which show so called homogeneity properties, i.e.
\begin{equation} f(ab) = f_1(a)f_2(b) \nonumber \end{equation} 
where $f_1(a) = a^p$, $p$ is a real number, and for an arbitrary 
number of variables
\begin{equation} f( \lambda x_1, \lambda x_2, \ldots \lambda x_n) = 
\lambda^p f(x_1,x_2, \ldots x_n) \nonumber \end{equation} 
If, for the case of two variables, one chooses $\lambda = x_1^{-1}$ then
\begin{equation} f(x_1,x_2) = x_1^p f(1,x_2/x_1) = x_1^p g(x_2/x_1)
\nonumber \end{equation} 
which shows that in practice $F$ does not depend on the two independent 
variables $x_1$ and $x_2$, but on $x_1$ and the ratio $x_2/x_1$ whatever 
$x_2$. For fixed $x_1$, $g$ is a scale independent function of $x_2/x_1$,
$g( \lambda x_2/ \lambda x_1) = g(x_2/x_1)$. The FSS assumption allows to 
relate the behaviour of an observable at some scale to its behaviour at 
another scale.

  As an example of application consider an observable $O_L( \Delta t)$ 
which depends on a variable $\Delta t = (t_c-t)/t_c$ where $t=t_c$ is a 
critical point of the infinite system and describes a finite system of 
linear dimension $L$. If one defines $x(L,t) = \xi_L(t)/L$ where $\xi_L(t)$
is the correlation length in the finite system for a fixed value of t in 
the neighbourhood of $t_c$, FSS stipulates that \cite{Kim}
\begin{equation} O_L( \Delta t) = O( \Delta t) Q_0 \big( x(L,t) \big)
\label{FSS} \end{equation} 
where $O( \Delta t)$ corresponds to the value of the observable in the 
corresponding infinite system and $Q_0 \big( x(L,t) \big)$ is a scaling 
function which is universal in the sense that it does not depend on $t$ but 
only on the ratio $x$ in the neighbourhood of $t_c$, i.e. all points 
corresponding to different $t$'s lie on the same curve when $Q_0$ is 
plotted as a function of $x$. If $O_L( \Delta t)$ converges to a finite 
$O( \Delta t)$ with increasing $L$, $Q_0(x)$ can be determined from 
(\ref{FSS}) and close to $t_c$ one expects that
\begin{equation} O( \Delta t) = O_L( \Delta t) / Q_0(x)
\sim (t_c-t)^{- \alpha} \label{FSS1} \end{equation} 
 if $O$ shows a critical behaviour, and $\alpha$ is a critical exponent.
If $Q_0(x)$ can be obtained numerically it is generally fitted to a 
polynomial in $x$
\begin{equation} Q_0(x) = 1 + b_1 x + b_2 x^2 + \cdots \nonumber \end{equation} 
with $Q_0(0) = 1$ if $\xi_L(t)/L$ goes to zero when $L$ goes to infinity. 
This type of methods has been applied in different places, in particular to 
percolation $(t \equiv p)$ and spin systems \cite{Kim, Elattari, Carmona}. 
Their use leads generally to reliable and accurate values of $t_c$ and 
critical exponents. Their practical implementation shows however that the 
numerical determination of these quantities can be a cumbersome task. Other 
methods have been proposed by different authors \cite{Elliott, Bonasera3}, 
in particular in order to fix the exponents which describe experimental 
fragment size distributions in the vicinity of the expected critical point 
where the distribution should follow a power law in the corresponding 
infinite system. However these methods rely on expressions which are valid 
for large systems and do not extrapolate to the thermodynamic limit. Even 
though the information on criticality extracted this way looks reasonable 
it remains suspicious to use it in order to conclude about the determination 
of the universality class of the phase transition which is the ultimate 
goal of such an analysis. On the other hand, it is 
difficult to believe that the use of scaling assumptions implemented by 
means of FSS can be used in practice in the framework of the analysis of 
experimental events since, among other severe problems, the number of 
points necessary to construct scaling functions would be rather restricted 
because of the small interval of sizes which can be reached with nuclei.

  Critical exponents are rather refined quantities. There remains hope to 
be able to detect the existence and eventually the order of phase 
transitions in nuclear matter by other means. Considering thermodynamic 
aspects of fragmentation, this can possibly be reached restricting ones 
considerations to finite systems. The question concerning the order of a 
phase transition has been raised recently on theoretical 
grounds \cite{Borrmann,Lee} in the framework of the Lee and Yang 
theory \cite{Yang} and the approach of Grossmann and 
collaborators \cite{Grossmann}. The criterion which allows to distinguish 
between different orders concerns the location of the zeroes of the grand 
canonical and the canonical partition function in the complex temperature 
plane. But these considerations are theoretical, possibly difficult to 
implement in practice for interacting systems  and the connection with 
experiment is lacking. In section 6 we shall show that there may exist 
simple operational means to overcome the problem.

  Finite size effects can hide or even delude the truth with respect to the 
order of a phase transition. Different ensembles may lead to a different 
behaviour of observables when applied to finite systems. We shall indeed 
see below that characteristic features of observables when calculated in 
the framework of a specific ensemble may be of some help in order to decide 
about the nature of the transition. The assertion has however again to be 
taken with a grain of salt since there does not exist any guarantee that 
these characteristic features will maintain their validity in the 
thermodynamic limit.

\subsection{Lattice models}
\indent

  We present and discuss here different lattice models which have been 
introduced and studied by different groups. The essential point of interest 
concerns the existence of signs for the existence
of one or several phase transition(s) in nuclear matter.

\subsubsection{The Ising model as a random cluster model}
\indent

  The success of percolation raised the question for the reasons of its 
achievements. As we saw in section 4, classical lattice systems described 
by the Ising, Ashkin-Teller and Potts models are generally close to 
percolation models \cite{Kasteleyn, Fortuin}. They describe Hamiltonian 
systems, hence they introduce an explicit, physical description of an 
$N$-body system. This was the main motivation for the introduction of the 
Ising model in order to interpret the cluster size distributions of an excited, 
fragmented system \cite{Samaddar, Samaddar1}.

  In order to be able to get a clearer understanding of the behaviour of 
observables the model was first studied in $d=2$ which allows the use of 
analytical techniques \cite{Samaddar}. The Hamiltonian reads
\begin{equation} 
H = - J_1 \sum_{n=1}^N \sum_{m=1}^M \sigma_{nm} \sigma_{n+1m}
    - J_2 \sum_{n=1}^N \sum_{m=1}^M \sigma_{nm} \sigma_{nm+1}
\nonumber \end{equation} 
where $J_1$ and $J_2$ are positive constants and $\sigma_{nm}$ is a 
classical spin $(= \pm 1)$ located at the position $(n,m)$ on a $2d$ 
lattice with $N$ lines and $M$ columns.

  An ensemble of sites are said to form a cluster if the corresponding 
spins show the same value and each site of the cluster has at least one 
neighbour site with a spin of the same value (connexity). The perimeter of 
a cluster is the ensemble of nearest neighbour sites whose spin takes the 
value which is opposite to the spins in the cluster. With this definition 
and after a mapping of the classical system into a quantum system of 
independent fermions \cite{Schultz} it is convenient to introduce 
projection operators along lines $(l)$ and columns$(c)$
\begin{eqnarray}
P_+^{(l)} & = & \frac {1}{2}(\textrm{\bf{\large 1}}
+ \sigma_{nm}^x \sigma_{nm+1}^x) \nonumber \\
P_+^{(c)} & = & \frac {1}{2}(\textrm{\bf{\large 1}}
+ \sigma_{nm}^x \sigma_{n+1m}^x)
\nonumber \end{eqnarray} 
and
\begin{eqnarray}
P_-^{(l)} & = & \frac {1}{2}(\textrm{\bf{\large 1}}
- \sigma_{nm}^x \sigma_{nm+1}^x) \nonumber \\
P_-^{(c)} & = & \frac {1}{2}(\textrm{\bf{\large 1}}
- \sigma_{nm}^x \sigma_{n+1m}^x)
\nonumber \end{eqnarray} 
where $\sigma^x$ is the standard Pauli matrix. The correlation operators
$P_+$ and anticorrelation operators $P_-$ have a classical interpretation, 
$P_+^{(l,c)} = 1$ and $P_-^{(l,c)} = 0$ for aligned spins, and
$P_+^{(l,c)} = 0$ and $P_-^{(l,c)} = 1$ for anti-aligned spins.

  These operators allow to define the probability of presence of a cluster 
of a given size and geometry.
\begin{equation} \Pi^{(i)} = < P_+^{L_+^{(i)}} \cdot P_-^{L_-^{(i)}} >
\label{proba} \end{equation} 
where the index $i$ characterizes the linked cluster which is considered,
$L_+^{(i)}$ is the number of bonds between cluster sites and $L_-^{(i)}$ the 
number of bonds between the surface of the cluster and the perimeter sites.
For a given cluster size $N_S$, one finds
\begin{equation} 2 L_+^{(i)} + L_-^{(i)} = 4 N_S 
\nonumber \end{equation} 
The brackets in (\ref{proba}) indicate a trace over the many-body quantum 
states \cite{Samaddar}.

  The multiplicity corresponding to clusters of a fixed number of sites 
$N_S$ and a fixed topology $(i)$ is given by
\begin{equation} M^{(i)}(N_S) = D^{(i)}(N_S,N_T) \Pi^{(i)}
\nonumber \end{equation} 
where $D^{(i)}(N_S,N_T)$ is the degeneracy of cluster of size $N_S$, and
$N_T$ is the total number of sites. The total multiplicity of clusters of 
size $N_S$ reads
\begin{equation} \overline{\mathcal{M}}(N_S) = \sum_{(i)} M^{(i)}(N_S)
\nonumber \end{equation}
Since the number of spins with a given sign is not fixed in the model (see 
discussion below) one introduces a normalization factor $\mathcal{N}$ and
\begin{equation} \overline{M}(N_S) = \frac{1}{\mathcal{N}} 
\overline{\mathcal{M}}(N_S) \nonumber \end{equation} 
such that 
\begin{equation} N_T = \mathcal{N} \sum_{N_S = 1} ^{N_T} N_S 
\overline{\mathcal{M}}(N_S) \nonumber \end{equation}
The interest of this approach lies in the fact that it is easy to work out 
the evolution of quantities like the average size of clusters as a function 
of the temperature
\begin{equation} <N_S(T)> = \sum_{N_S = 1} ^{N_T} N_S \overline{M}(N_S,N_T) 
/ \sum_{N_S = 1} ^{N_T}\overline{M}(N_S,N_T) \nonumber \end{equation}
the corresponding variance
\begin{equation} \sigma^2(T) = <N_S^2(T)> - <N_S(T)>^2 
\nonumber \end{equation} 
and the fragment size distributions. The expected qualitative 
features obtained in a finite system when the temperature of the system 
increases is observed.
The results can be interpreted in terms of site percolation 
concepts. ``Occupied'' sites in the percolation framework correspond to 
cluster sites in the present case and ``empty'' sites to perimeter sites. 
The relative amount of ``occupied'' sites is governed by the temperature. 
The number of perimeter sites increases with increasing temperature and the 
relative amount of large clusters diminishes. In the percolation picture 
the number of empty sites increases with decreasing occupation probability 
$p$, leading to the same trend.

\begin{figure}[b!]
\centerline{\epsfig{figure=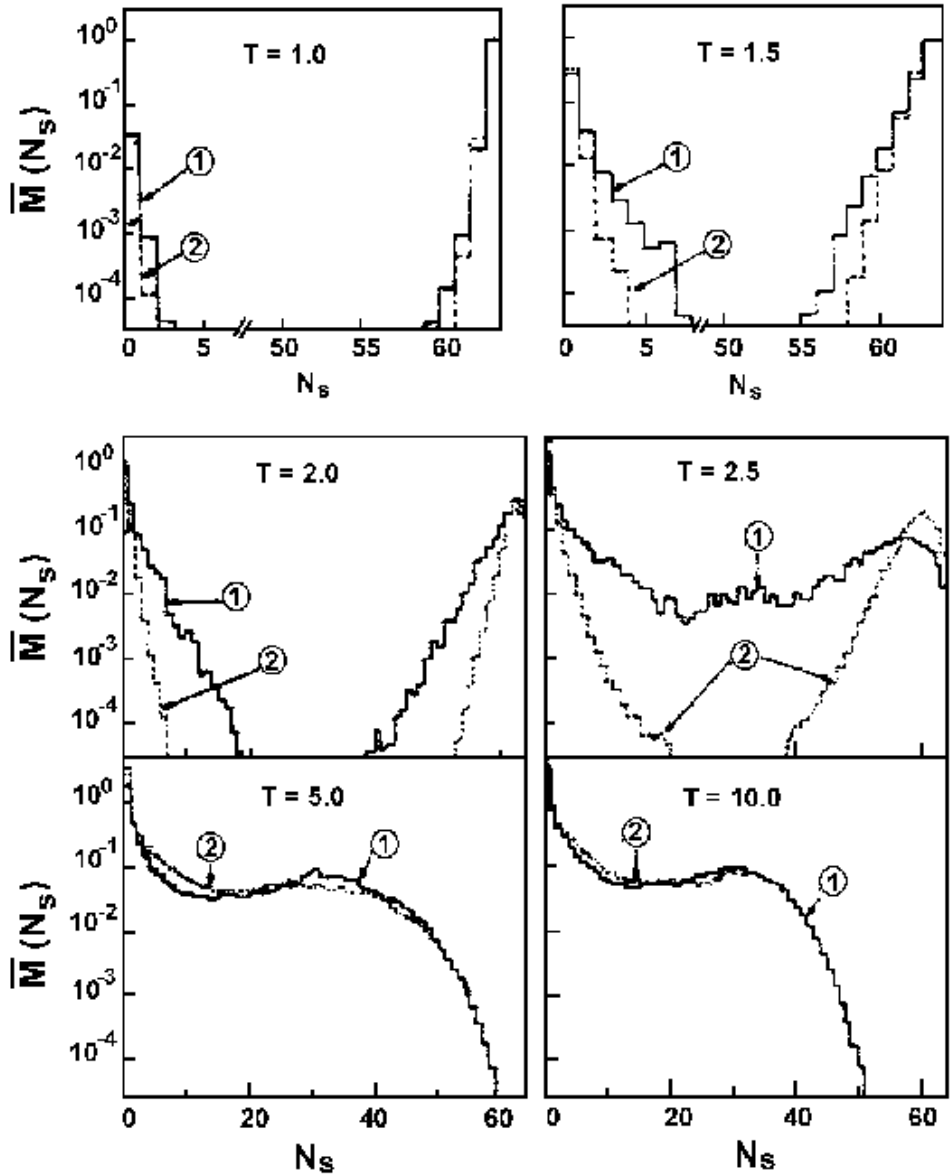,height=160mm}}
\caption{Multiplicities of clusters made of parallel spins in the Ising 
model for (1) short range and (2) long range interactions and different 
temperatures \cite{Samaddar1}.}
\label{fig15}
\end{figure}

  The model which deals with short range nearest neighbour interactions has 
been extended to include long range interactions \cite{Samaddar1}. 
Numerical studies show that the range of the interaction in finite systems 
does not change the qualitative behaviour of multiplicity distributions in 
finite systems, except for the fact that a long range attractive 
interaction favours the survival of large clusters over a large temperature 
interval. At high temperature both distributions collapse onto each other, 
as it can be seen in Fig.~\ref{fig15}. The presence of ``critical'' events is 
seen in the behaviour of the second moment of the cluster size distribution. 
Similar results are obtained with $3d$ systems. The a priori general 
difficulty to relate lattice models with nuclear fragmentation will be 
discussed later. The use of the Ising model presents a 
further weakness which stems from the fact that the identification 
of clusters in terms of spins which are aligned in a given direction breaks 
the $\mathcal{Z}_2 \ (\pm 1)$ symmetry which is present by construction and 
leads to non conservation of the number of ``particles''. The remedy to 
this disease consists of the introduction of a term governed by a Lagrange 
multiplier. We shall come to this point below.

  The present models were thought to build up a bridge between the purely 
geometrical concepts introduced by percolation models and sophisticated 
liquid-gas descriptions. They triggered the construction of more refined 
models aimed to describe and understand the physics of fragmentation and 
critical behaviour in finite nuclei.

\subsubsection{The grand canonical lattice gas model (LGM)}
\indent

  The preceeding study was aimed to look for the connection between 
the geometrical and probabilistic concepts which define percolation models 
and the thermodynamic properties of spin models which are defined in terms 
of classical Hamiltonians. It did however not explicitly define the link 
between bond probabilities, energy and temperature.

  This link was established by Jicai Pan and Das Gupta \cite{Pan, 
Pan1, DasGupta1} and Campi, Krivine and collaborators \cite{Campi7, Campi8, 
Campi9}. In ref. \cite{Campi7} the authors consider a classical lattice 
gas model defined by the Hamiltonian
\begin{equation} H = \sum_{i=1}^N n_i \frac{p_i^2}{2m} - 
\epsilon \sum_{<i,j>} n_i n_j \nonumber \end{equation} 
where $n_i=0,1$ is the occupation number of the lattice site $i$, $N$ the 
total number of lattice sites and $\epsilon$ the energy between nearest 
neighbours $<i,j>$. The number of particles $A = \sum_i n_i$ is conserved 
in the average by means of a constraint induced by a Lagrange multiplier 
procedure.

\begin{figure}[b!]
\centerline{\epsfig{figure=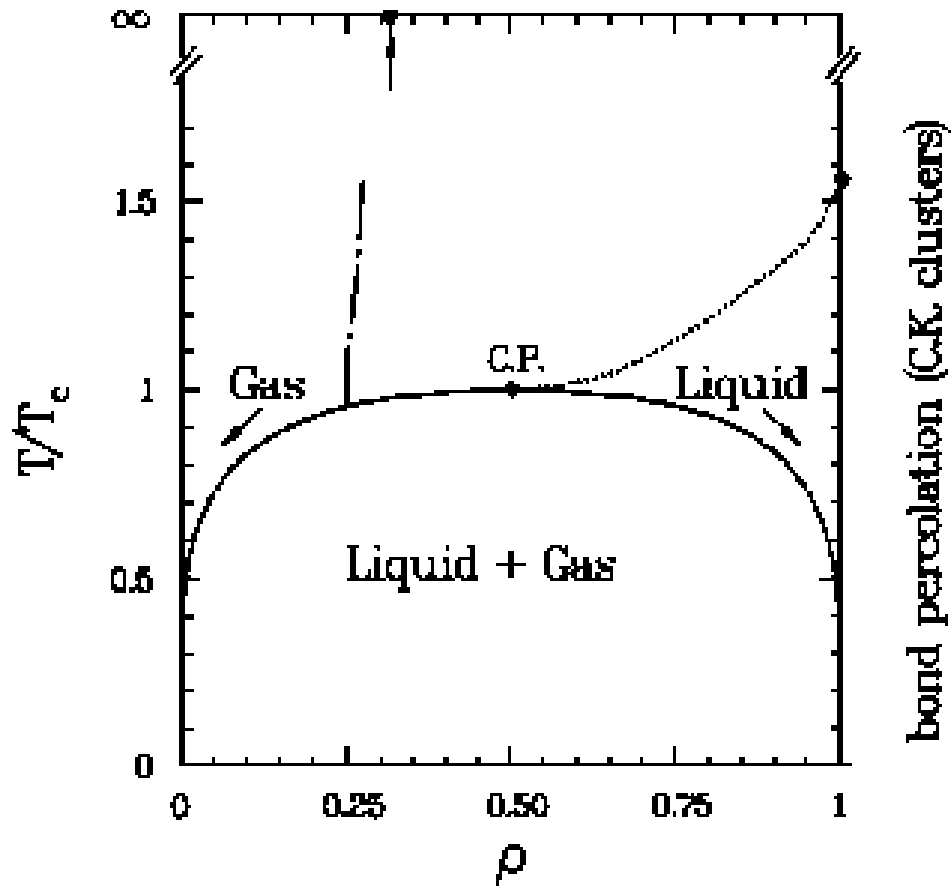,height=120mm}}
\caption{Typical phase diagram corresponding to a lattice gas model 
\cite{Campi7}. See discussion in the text.}
\label{fig16}
\end{figure}

  The model is treated by means of Metropolis Monte Carlo simulations, 
in the framework 
of the grand canonical ensemble which can be formally related to the 
canonical description of the Ising model with magnetic field \cite{Lee1}.
The exploration of the properties of the system in the density-temperature 
plane $(\rho,T)$ leads to the typical phase diagram shown in Fig.~\ref{fig16}.
One observes the presence of a full line which separates two phases above the 
line, a gas at low $\rho$ and a liquid at large $\rho$. Below the 
separation line gas and liquid coexist in an inhomogeneous phase forming 
domains (clusters) of different sizes. In the thermodynamic limit the 
transition line corresponds to a first order transition. At $\rho=1/2$ in 
units of the normal density $\rho_0$ there appears a critical point at $T = 
T_c$ which corresponds to a continuous transition characterized by critical 
exponents which are universal, i.e. do not depend on the detail of the 
interaction between particles.

  The main point of interest in the study of this model lies in the 
analysis of the cluster content of the system for different values of 
$\rho$ and $T$. The identification of clusters raises the question of their 
definition. One gets confronted with two problems \cite{Campi7}. The most 
disturbing one concerns the fact that the thermodynamic critical point does 
not coincide with the percolation critical point at which the cluster size 
distribution shows a power law behaviour if clusters are defined as an 
ensemble of particles occupying connex (nearest neighbour) sites \cite{Sykes}.
If clusters are defined in a purely geometric way as made of connected 
occupied sites percolation events generate a whole line of points in the 
phase diagram, an infinite cluster is obtained in the infinite limit on 
the right hand side of the dot-dashed line shown in Fig.~\ref{fig16}. 
This is of course unphysical, since at high temperature one expects 
essentially small clusters and isolated particles. A remedy to this problem 
was proposed by Coniglio and Klein \cite{Coniglio} who introduced the 
energy concept in order to define clusters as made of connected ensembles
of nearest neighbour particles with a bond probability smaller or equal than
\begin{equation} p_{CK}(\epsilon/T) = 1 - exp(-\epsilon/2T)
\label{pck} \end{equation} 
If one uses this definition the thermodynamic and percolation critical
point coincide and there exists a whole continuous line of points in the
$(\rho,T)$ plane called the Kert\'esz line \cite{Kertesz}, along which the 
cluster size distribution shows a power law behaviour. The important point 
concerns the fact that the definition (\ref{pck}) is shown to be very close 
to the bond probability prescription between neighbour particles given by
\begin{equation} p_b \bigg( \frac{\epsilon}{T} \bigg) =  
Prob \bigg[ \bigg( \frac{p^2}{2m} - \epsilon \bigg) < 0 \bigg] = 
1 - \frac{4}{\sqrt{\pi}} \int_{\sqrt{\epsilon / T}} ^ 
\infty \ud u \, u^2 e^{-u^2}
\label{pbound} \end{equation} 
which stipulates that a pair of particles is bound if 
$(p^2/2m - \epsilon) < 0$ where the momentum $p$ is drawn from a Maxwell 
distribution \cite{Pan1, Campi7}.
It is shown \cite{Campi7} that if (\ref{pbound}) is chosen as 
a bond criterion then the separation energy $S(A)$ of one particle from a 
bound cluster of $A$ particles satisfies approximately the inequality
\begin{equation} S(A) = B(A-1) - B(A) < 0 \nonumber \end{equation} 
where $B(i)$ is the binding energy of a system of $i$ bound particles, and 
the ``critical'' percolation line coincides approximately with the 
Kers\'etz line defined above. 

  Following this prescription one obtains cluster size distributions in 
different part of the $(<\rho>,T)$ plane, where $<\rho>$ is the average 
density in the grand canonical ensemble. These distributions show the same 
properties as those obtained in the framework of percolation models, i.e. a 
power law behaviour corrected for finite size effects along the Kers\'etz 
line, with a power law exponent $\tau \simeq 2.2 - 2.3$, an exponential 
decay at low density and high temperature, the characteristic large cluster 
peak at high density. One may notice also that contrary to the case of 
multifragmentation models \cite{Gross, Bondorf} in which clusters are by 
construction spherical the cluster along the Kers\'etz line have surfaces 
with fractal properties which allows for a freeze-out density larger than 
the ``critical'' density. Furthermore one observes that the excitation 
energy per particle above the ground state keeps approximately constant 
along the critical line \cite{Campi9}. Further work relying on lattice
models in the framework of nuclear fragmentation was developed in the 
recent past. It concerns the study of the two types of phase transitions 
(thermodynamic and percolation) mentioned above 
which may be of different orders (first and second), the importance of 
the Coulomb interaction, problems related
to the finite nature of the systems. We examine different approaches and 
their contribution to the understanding and explanation of these questions 
in the next subsections.

\subsubsection{Lattice models and their application in the framework of 
nuclear fragmentation}
\indent

  The generic character of lattice gas models (LGM), their kinship with 
percolation models and classical systems of particles with short range 
interactions lead to a large amount of theoretical investigations.

  Das Gupta and Jicai Pan proposed different variants of this type of 
models \cite{Pan,Pan1} and applied them for direct comparison with 
experimental data \cite{DasGupta1}. They worked out the canonical partition 
function  $\mathcal{Z}$ on the lattice in an analytic form by using the 
Bragg-Williams approximation for the calculation of the potential 
contribution to the free energy. In this approximation the number of sites 
which are occupied in the neighbourhood of a fixed ocupied site is fixed in 
the average as $z n/N$, where $z$ is the number of neighbours (coordinance) 
and $n/N$ the average number of occupied sites ($n$) in a lattice with $N$ 
sites. The approximate analytical expression of $\mathcal{Z}$ allows to 
determine different equations of state, in particular the relation between 
the pressure and the volume which comes out as a van der Waals-type 
equation \cite{Pan}. In a further development \cite{Pan1} the authors 
introduced a grand canonical description, the mean field Bragg-Williams 
approximation was replaced by the Bethe-Peierls approximation which 
corresponds to the introduction of blocks of lattice sites inside which 
occupied nearest neighbour sites interact. An interaction 
$\overline{\epsilon}$ between blocks is taken into account and fixed along 
with the chemical potential of the particles by means of a self-consistent 
procedure. This further improvement brings in fact rather small changes 
when compared with the Bragg-Williams approximation and is rather close to
the outcome of a mean field approach using a realistic Skyrme 
interaction. In all cases the descriptions show signs for the existence of 
a first order phase transition for all values of $\rho$, a second order 
transition at $\rho_c = 0.5 \rho_0$ where $\rho_0$ is the normal nuclear 
matter density.

  Emphasis was put on the study of the cluster content of finite lattice
gas systems. It was shown by means of explicit calculations that the second 
moment of the fragment size distribution shows a maximum at different 
values of the temperature for different values of the density \cite{Pan}. 
In an interpretation  in terms of percolation concepts this is the sign 
for the existence of a Kert\'esz line. If one introduces
\begin{eqnarray} p^2_r/2\mu - \epsilon & \le & 0 \nonumber \\ 
\textrm{or} \quad & > & 0 \nonumber \end{eqnarray} 
(where $p_r$ is the relative momentum between neighbouring particles, $\mu$ 
the reduced mass and $\epsilon$ the strength of the interaction) as a 
criterion to decide whether the considered particles are bound or not the 
bond probability is the same as the one given by (\ref{pbound}) in section 
5.2.2 \cite{Pan1}. Effective exponents $\tau$ extracted from the experiment 
in the spirit of refs. \cite{Panagiotou,Panagiotou1} (see section 2.1) were 
compared with lattice calculations in an attempt to obtain informations 
about characteristic freeze-out densities of systems formed at different 
beam energies. The freeze-out density determined in this way came out to be
$\rho \simeq 0.4 \rho_0$. 

  One very important point in nuclear fragmentation concerns the role and 
importance of the Coulomb interaction. 
This point has been emphasized in the framework of 
multifragmentation models \cite{Gross}. A calculation in the framework of 
an LGM for a large system obtained through the collision of $Au$ on $Au$ 
shows that the Coulomb interaction may have a sizable effect and may lead 
to large discrepancies between LGM and the experiment \cite{DasGupta1}. One 
may introduce two types of binding energies between nearest neighbours 
which correspond to the interaction between identical and different types 
of particles (protons and neutrons) \cite{Pan2}. This has been done by 
means of Monte Carlo simulations and the results were compared to the 
outcome of molecular dynamics (MD) calculations which coincide with the 
lattice gas results in the absence of the Coulomb interaction. It is no 
longer the case when this interaction is taken into account. The LGM 
results deviate from the MD results in large systems with a sizable amount 
of protons. We shall come back to the Coulomb problem below.

  Further comparisons with different models have been 
performed \cite{DasGupta2} such as percolation and multifragmentation 
models. The analysis of different observables shows that these models 
share common features, in particular the statistical multifragmentation and 
lattice gas approaches, except for quantities like the specific heat $C_V$ 
at low temperature. The multifragmentation approach possesses quantum 
mechanical properties and $C_V$ increases linearly with $T$ while in the 
LGM $C_V$ stays at the constant value $3/2$. It should be mentioned that 
the algorithm used in the simulations \cite{Pan, Pan1, DasGupta1} does not 
sample the events in a rigorous fashion \cite{Muller}. This should however 
not qualitatively affect the results and change the conclusions.

\subsubsection{Extensions of the LGM in the framework of the canonical 
ensemble}
\indent

  Lattice gas models were further developed and analysed in the framework 
of the canonical ensemble by Gulminelli and Chomaz \cite{Gulminelli1,
Chomaz1, Gulminelli2, Duflot}, Borg, Mishustin and Bondorf \cite{Borg} and 
Carmona, Richert and  Taranc\'on \cite{Carmona}.

\begin{figure}[b!]
\centerline{\epsfig{figure=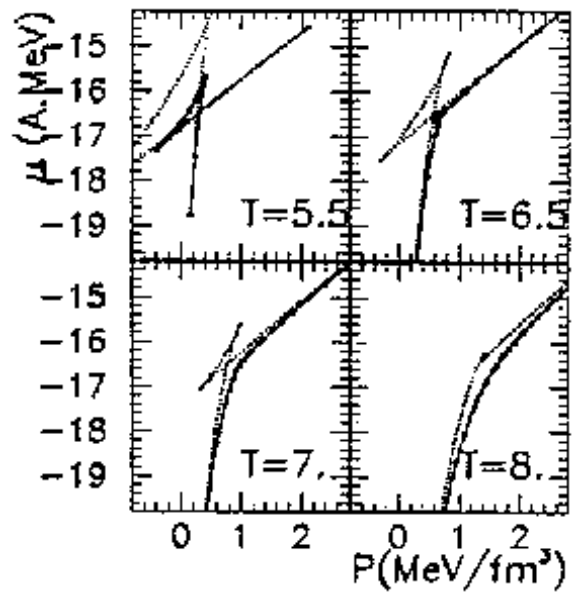,height=80mm}}
\caption{Exact (full lines) and mean field (dotted lines) chemical 
potentials as a function of pressure for different temperatures 
\cite{Gulminelli2b}.}
\label{fig17}
\end{figure}

  Gulminelli and Chomaz \cite{Gulminelli1, Gulminelli2} constructed the 
phase diagram of a finite system by working out the compressibility 
coefficient. This quantity shows an anomalous behaviour, it becomes 
negative at some temperature $T$ over some density interval, the sign for 
the existence of a first order phase transition in the infinite system.
Fig.~\ref{fig17} shows the behaviour of the chemical potential represented as 
a function
of pressure exemplifies this phenomenon. It has also been observed that the 
fragment size distribution possesses a separation line (Kert\'esz line) which 
crosses the phase separation line and enters the coexistence zone at some 
low value of the density. This is a priori somewhat surprising 
since it means that in the thermodynamic limit one would observe the 
existence of a second order transition in the coexistence zone. However, it 
is found out that for large systems the separation line stops effectively 
at the thermodynamic critical point located at $\rho / \rho_0 = 0.5$, hence 
the authors interpret the effect as a typical finite size effect. They 
comfort this point of view by showing that in this region of the phase 
diagram the fragment size distribution does not go through a power law 
distribution when one crosses the separation line in the coexistence zone. 
If this is so, it means that the system never shows the signs for a 
critical behaviour for subcritical densities.

  Lattice gas models were also recently extended to the case where two 
types of particles, protons ($p$) and neutrons ($n$) coexist. The 
refinement is implemented by the choice of different nearest neighbour 
interaction strngths, $\epsilon_{np} \ne \epsilon_{nn} = \epsilon_{pp}$. 
These constant quantities were chosen such as to reproduce the nuclear 
binding energies of the corresponding quantum systems. One may ask for the 
consequences of the introduction of the isospin degree of freedom 
\cite{Chomaz1}. In fact, the extension of the model has only a small 
effect on the phase diagram, so that in practice there exists only one 
relevant order parameter, $\rho_L - \rho_G$, where $\rho_L$ and $\rho_G$ 
are the liquid and gas density respectively. Critical exponents do not 
depend on the isospin degree of freedom either, and the critical 
temperature shows a quadratic dependence on asymmetry. The sign for isospin 
effects may perhaps be found in the fact that in the coexistence region the 
vapour mode of small fragments is more isospin asymmetric than the liquid 
mode of larger fragments. This can in fact be observed in the experiment
\cite{Veselsky, Xu}.

  A similar model with protons and neutrons has been studied in 
ref. \cite{Borg}. The thermodynamic properties of the system are determined 
for open boundary conditions and the caloric curve is constructed at 
constant density and constant pressure. For fixed pressure one observes 
clearly the appearance of a plateau which is the sign for a first order 
transition. For fixed density, the caloric curve is monotonously increasing 
with energy. Fragment size distributions are analysed along isobars and 
isochores. They show the usual trend, with increasing temperature the 
large fragment yield shrinks and the small fragment and light particle 
contribution increases continuously.

  Recently the effect of the shape of the surface of the fragmenting system 
on the fragment size distribution has also been investigated \cite{Duflot}. 
Cubic shapes with and without periodic boundary conditions as well as 
spherical shapes with sharp and diffuse surfaces have been considered. 
These shapes have no real influence on the thermodynamic properties and 
fragment size distributions of small systems with typically $100$ particles 
at the critical density. Similar tests have been performed on other
models, see below \cite{Elattari1}. It raises the question wether the shape 
dependence of fragmentation events may be detected in experimental data.

\subsection{Cellular models of nuclear fragmentation}
\indent

  As already mentioned the success of percolation concepts for the 
reproduction of experimental fragment size distributions raised the 
question of the correlations between theses quantities and the actual 
dynamics of highly excited nuclei \cite{Samaddar, Samaddar1}. Peripheral 
collisions \cite{Trautmann1} indicate that the spectator remnant behaves 
essentially as an excited system of particles which is thermalized. It 
leads to the idea that fragmenting nuclei can essentially be described as 
completely disordered systems of interacting particles, at least under the 
experimental conditions described above. Furthermore, it is a non trivial 
but established fact that disorder may be connected with complexity and 
universality \cite{Wigner}. This property reflects the fact that 
disordered systems are not sensitive to the details of their dynamical 
characteristics but show features which are common to many of them such as 
the scaling properties discussed above.

  In the present section we introduce and discuss so called cellular models 
which are aimed to relate the phase space properties of a classical system 
of particles to bond probabilities and hence to the generation of bound 
clusters. They can be considered as extensions of lattice models.

\subsubsection{Disordered systems and cluster identification}
\indent

  A fragmenting nucleus is considered as an assembly of interacting 
particles contained in a finite three dimensional volume $V$ divided into 
$A$ cubic cells, $V = Ad^3$ where $d$ is the linear size of each cell. Each 
cell contains one particle \cite{Elattari1, Elattari2}. Each particle 
inside its cell is characterized by its position $\vec r_i \,
(i = 1,\ldots,A)$ and linear momentum $\vec p_i$, both phase space 
coordinates being taken at random and obeying $|\vec r_i - \vec r_j|
|\vec p_i - \vec p_j| \geq \hbar$ which has to be verified for particles 
located in neighbouring cells. The interaction is chosen as a short range 
two-body potential adapted to a classical description of the 
nucleus \cite{Wilets, Lenk}. It acts between nearest neighbours only. The 
total energy of two such particles can be split into a centre of mass and a 
relative contribution
\begin{equation} E = P^2/2M + p^2/2\mu+V(r) = P^2/2M + \epsilon
\nonumber \end{equation} 
with $\vec P = \vec p_i + \vec p_j,  \, \vec p = \vec p_i - \vec p_j,  \, 
r = |\vec r_i - \vec r_j|, \, M = 2m$ and $\mu = m/2$, where $m$ is the 
mass of the nucleon. Each coordinate $\vec r_i$ is defined as $x_i = x_{i0}
\pm \eta d/2$ (same for $y_i$ and $z_i$), where $x_{i0}$ locates the centre 
of the cell and $\eta$ is drawn uniformly in the interval $[0,1]$. The 
momenta $\{\vec p_i\}$ are taken uniformly in a sphere of radius $p_{max}$. Two 
neighbouring particles are said to be bound to each other if $\epsilon < 0$ 
and unbound if $\epsilon > 0$. By means of numerical simulations the 
criterion allows to construct a bond probability $p(E) = n(E) / 
\mathcal{N}(E)$ where $\mathcal{N}(E)$ is the total number of draws in an
energy interval $[E, E+\delta E]$ and $n(E)$ the total number of events out 
of $\mathcal{N}(E)$ for which the two-particle system is bound at energy $E$.
The violation of the inequality relating relative distances and momenta 
defined above leads to the rejection of the corresponding event. 

  Three different procedures have been introduced in order to identify 
bound clusters. In the first called A2, the energy $E$ of each pair is 
drawn from the distribution $\mathcal{N}(E)$ and the corresponding $p(E)$ is 
compared to a random number $\eta$ belonging to $[0,1]$. If $p(E) > \eta$ 
the pair is considered as bound and unbound in the reverse case. This 
procedure is close to a percolation algorithm but $p(E)$ is not chosen in
a uniform distribution. It neglects also the correlations which exist 
between pairs sharing common particles. 
  
  The second procedure (A1) fixes the energy of a pair $E = E(\vec r_i, 
\vec p_i; \vec r_j, \vec p_j)$ of neighbouring particles $i$ and $j$ by 
means of their phase space coordinates which are fixed as described above. 
This fixes also $p(E)$ and the bond criterion is the same as for A2. 
Finally, correlations are fully taken into account by means of algorithm A0 
for which the generated phase space $\{\vec r_i, \vec p_i, i = 1,\ldots,A\}$
is used to calculate explicitly $\epsilon$ for each pair. It corresponds to 
a deterministic determination of the clusters for each generated event.

  It is important to notice that none of these descriptions implies that 
the considered systems are necessarily in any kind of equilibrium.

\subsubsection{Physical observables. Confrontation with bond percolation}
\indent

\begin{figure}[b!]
\centerline{\epsfig{figure=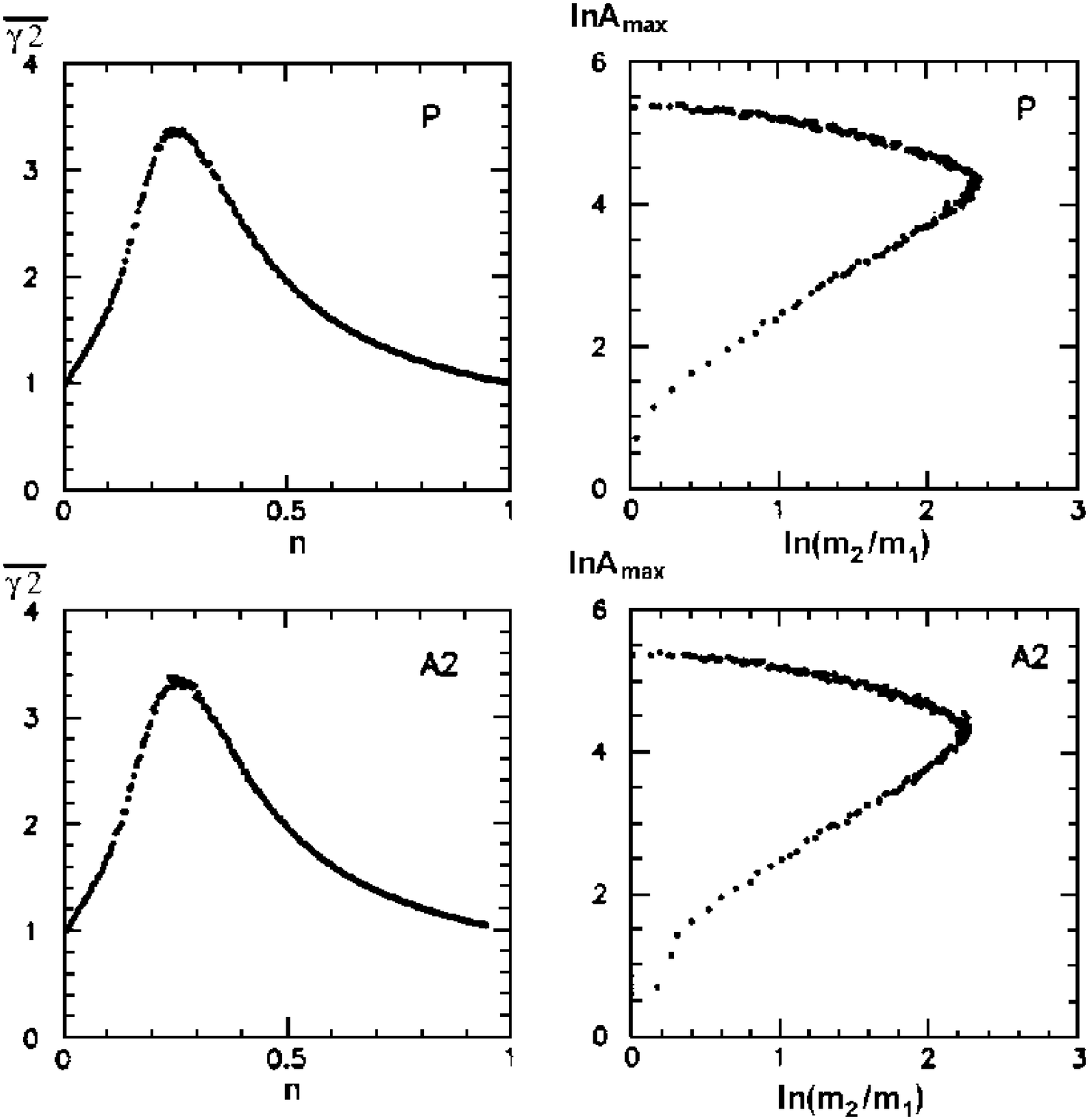,height=160mm}}
\caption{Comparison of percolation results (upper part) with A2 algorithm
(lower part). \cite{Elattari1}.}
\label{fig18}
\end{figure}

  The fragment content can be characterized by the mass distribution, its 
moments and associated observables like $\gamma_2$ which has already been 
defined above. In order to compare directly with the experiment these 
observables are determined as a function of the multiplicity $m$ which can 
be measured and is univoquely related to the bond probability $p$.
For each event $k$ 
\begin{eqnarray} m_2^{(k)} & = & {\sum_i} ' \, i^2 P_i^{(k)}(m) 
\nonumber \\
\textrm{or} \quad S_2^{(k)}(m)  & = & m_2^{(k)}(m) /m_1^{(k)}(m) \nonumber
\end{eqnarray} 
where $P_i^{(k)}(m)$ is the number of clusters of size $i$ and the prime 
index indicates that the heaviest fragment is omitted. Similarly
\begin{eqnarray} \gamma_2^{(k)}  & = & m_0^{(k)}(m) \cdot m_2^{(k)}(m)
/m_1^{(k)^2}(m) \nonumber \\
  & = & 1 + \sigma^{(k)^2}(m) \cdot m_0^{(k)^2}(m)/m_1^{(k)^2}(m) 
\nonumber \\
\textrm{where} \quad \sigma^{(k)^2}  & \equiv & \bigg( m_0^{(k)}(m) \cdot
m_2^{(k)}(m) - m_1^{(k)^2}(m) \bigg)/ m_0^{(k)^2}(m)\nonumber
\end{eqnarray} 
Average quantities are defined as 
\begin{eqnarray} <m_2(m)> & = & N_{ev}^{-1} \sum_{k=1}^{N_{ev}}
 m_2^{(k)}(m) \nonumber \\
<\gamma_2(m)> & = & N_{ev}^{-1} \sum_{k=1}^{N_{ev}} \gamma_2^{(k)}(m)
\nonumber \end{eqnarray} 
  Numerical investigations have first been performed with the algorithm A2. 
The comparison with bond percolation is shown in Fig.~\ref{fig18}. 
One observes an 
amazing agreement between the two descriptions although bond probabilities 
are taken from different distributions and this agreement is very robust 
with respect to the strength of the two-body potential. In fact, if 
$\rho(\eta)$ is the probability distribution of $\eta = p(E_0)$ for a 
given $E_0$ the equivalent probability in bond percolation is
\begin{eqnarray} \tilde{\eta} = 1 - \int_0^{\eta} \ud \eta ' \rho (\eta ')
\nonumber \end{eqnarray} 
Through this relation $m$ and $\tilde{\eta}$ are related to each 
other in the same way as $m$ and $p$ in bond percolation. In particular, 
one observes a critical value $\tilde{\eta}_c$ for which an infinite 
cluster appears in the infinite system. 

  The agreement survives also if protons and neutrons are differentiated by 
means of different bond probabilities $p_{pp}(E) = p_{nn}(E) \neq p_{pn}(E)$
\cite{Elattari1}. Qualitatively similar but quantitatively different 
results are obtained with algorithm A1, in particular the location of the 
maxima of $<m_2>$ and $<\gamma_2>$ are shifted to higher values of $m$. The 
deviations are indicated by the correlations which are built in A1 and 
absent from A2. One expects that the effect is due to the hybrid nature of 
the algorithm which mixes probabilistic and deterministic properties. 

  Algorithm A0 provides a priori the most realistic description since it 
takes all correlations consistently into account. It allows also for an easy 
treatment of the long range Coulomb interaction. For a fixed value of the 
length $d$ of the cubic cells the moments $<m_2>$ and $\overline{\gamma_2}
= <m_2> \cdot <m_0> / <m_1>^2$ can differ sensibly from the percolation 
results. If, however, one averages over events corresponding to values of 
$d$ which are uniformly distributed over a given interval of densities 
$\rho$, say from $5 \times 10^{-3}$ to $0.17\,fm^{-3}$, one obtains 
mass distributions, $<m_2>$ and $\overline{\gamma_2}$ which are remarkably 
close to those obtained with A2 and robust with respect to 
modifications of the strength of the short range potential as it is the 
case with A2. The agreement between the results obtained with A0 and A2 is 
not easy to understand. It seems to indicate that averaging over densities 
acts as a mean to wash out correlation effects.

\subsubsection{Confrontation with experimental data}
\indent

\begin{figure}[b!]
\centerline{\epsfig{figure=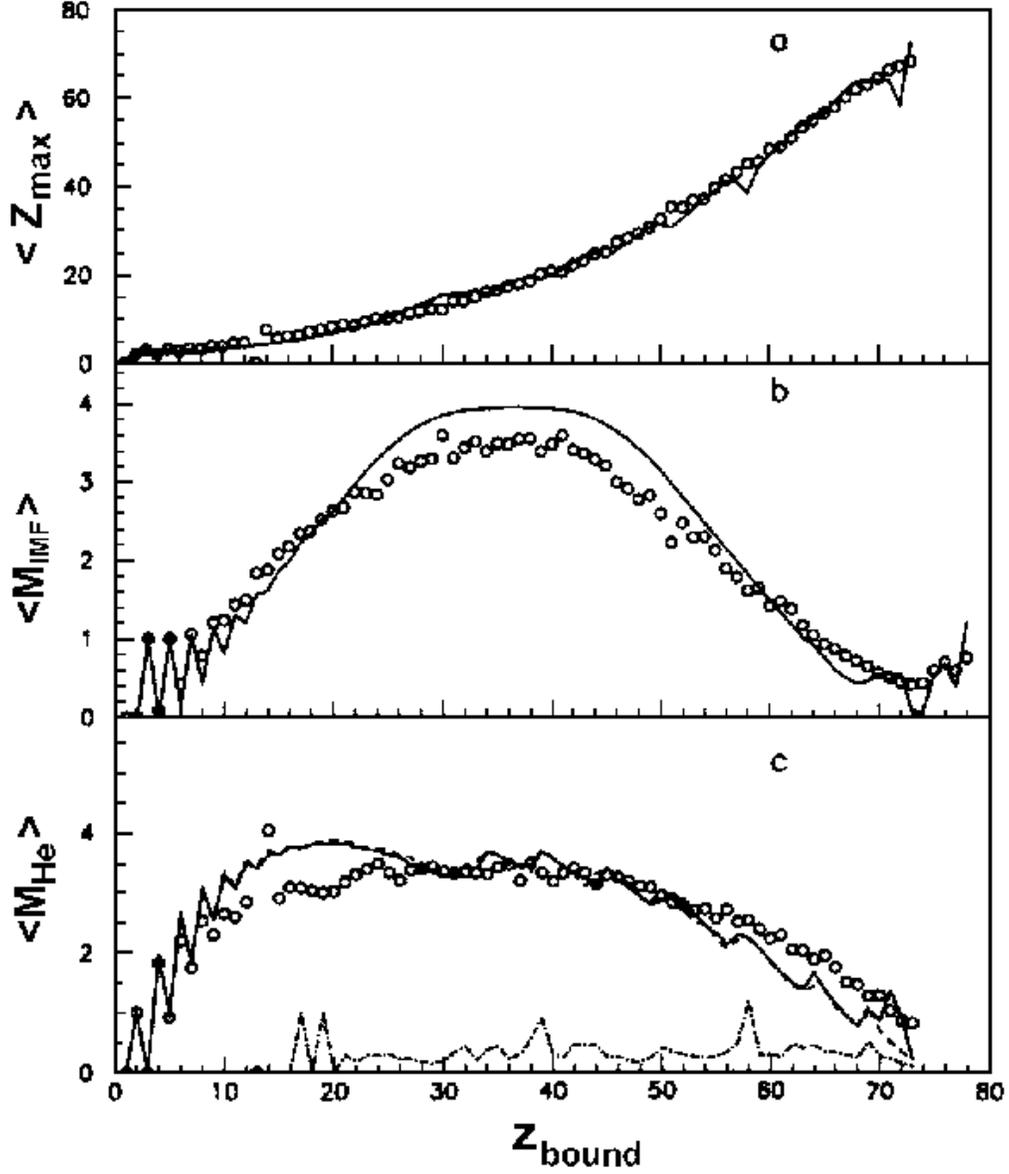,height=160mm}}
\caption{Different observables represented as a function of $Z_{bound}$. 
The experimental average largest cluster $<Z_{max}>$, IMF multiplicities
$<M_{IMF}>$ and $He$ multiplicities $<M_{He}>$ represented by dots are 
compared to calculations performed with the A0 algorithm, see text 
\cite{Elattari1}.}
\label{fig19}
\end{figure}

  Multifragmentation data available from experiments made by the ALADIN
collaboration \cite{Lynen, Kreutz, Trautmann2} have been confronted with 
the outcome of the A2 algorithm in the same way as it was done for the 
percolation approach in section 4.7 \cite{Elattari1, Elattari2}. The charge 
$Z$ to mass $A$ relation used is the same as in this section. For fragments 
of mass $A=3$ the isobaric ratio $^3H / ^3He$ is fixed to $2.2$. The mass 
of the spectator part of the fragmented system depends on the impact 
parameter, the number of particles involved is divided into $10$ different 
sets corresponding to $10$ intervals for the impact parameter. Each system 
with $A_i \, (i=1,\ldots,10)$ particles is put in a cube of $7^3$ cells in 
as compact and symmetric a geometry as possible. Each calculated observable 
$O$ is averaged over the whole impact parameter range
\begin{equation} <O>=\sum_{i=1}^{10} b_i \Delta b_i O(b_i) \ \bigg/
\sum_{i=1}^{10} b_i \Delta b_i \nonumber \end{equation} 
Results of this analysis are shown in Fig.~\ref{fig19} 
for different observables, 
the average value of the charge of the largest fragment $Z_{max}$, the IMF
$(3 \leq Z \leq 30)$ multiplicity, the $He$ isotope multiplicity which are 
represented as a function of $Z_{bound}$ and the average number of fragments 
with charge $6 \leq Z \leq 30$. As it can be seen, the agreement between 
experiment and simulations is excellent. This is not really surprising 
since bond percolation and A2 lead to very close results and, as it has 
been seen in section 4, percolation reproduces experimental results.

\subsubsection{Geometric and collective effects}
\indent

  The use of simple cubic shapes for the description of the fragmenting 
system may appear as too restrictive. This point has been investigated by 
introducing other shapes, spheres, parallelepipeds, streaked, crenellated 
and crumpled surfaces which were generated by means of a random growing 
process leading to fractal surface structures which may introduce empty 
cells in the bulk of the system \cite{Elattari1}. The moments of the mass 
distributions show deviations from the cubic case, but these deviations 
are always surprisingly small, they never lead to qualitative changes in 
any observable related to fragment size distributions.

\begin{figure}[b!]
\centerline{\epsfig{figure=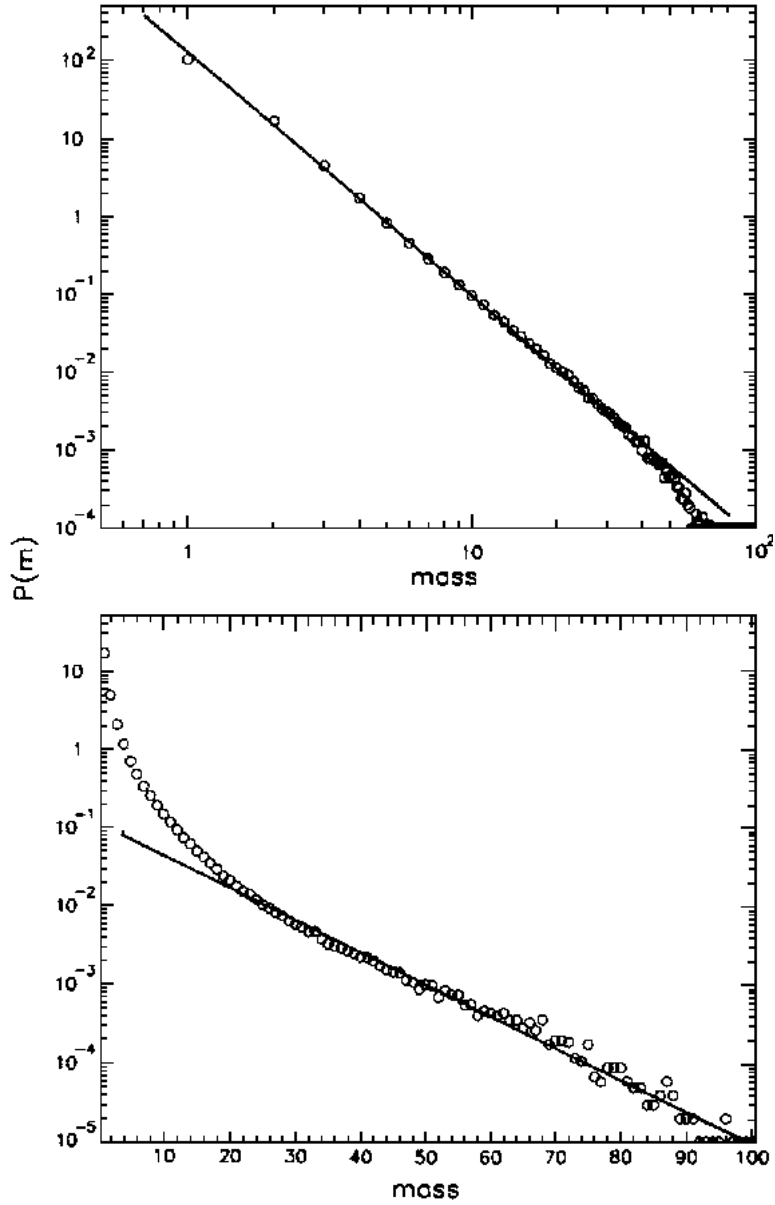,height=160mm}}
\caption{Mass distributions for a $6^3$ system. The upper curve corresponds 
to a calculation using the A0 algorithm, the lower corresponds to the 
simulation of collective effects, see text \cite{Elattari1}.}
\label{fig20}
\end{figure}

  Cellular models describe disordered systems which show a power law 
behaviour with deviations due to finite size effects at a value of the 
fragment multiplicity corresponding to criticality in the infinite 
system. Experimentally this situation prevails for peripheral 
collisions  \cite{Lynen, Kreutz, Trautmann2}. In the case of more central 
collisions fragment size distributions show rather an exponential fall off 
as a function of the mass number \cite{Kunde}. These collisions are also 
characterized by a sizable collective radial flow \cite{Hsi, 
deSchauenburg}. It is out of scope to reproduce this dynamic effect with 
the present simple-minded approach. It is however possible to mimic it. 
This was done with algorithm A0, mixing events with linear cell size 
$1.8 \leq d \leq 6\,fm$ for $A=216$ particles. The linear momenta strengths 
of the particles are chosen randomly but their directions are fixed along a 
line which joins the centre of the system to their actual position, 
pointing out of the system. The typical behaviour of the mass 
distribution in this case is shown in Fig.~\ref{fig20}, along with the case 
where collective effects are absent. The shape which is power law like in 
the case where the directions of the momenta are random gets very close to 
an exponential for sizes between $20$ and $90$ units of mass. Deviations 
are observed for light and heavy fragments. Such a behaviour has been 
confirmed in ref. \cite{deSchauenburg} were a more detailed analysis on a 
more sophisticated approach has been performed. This result comforts the 
idea that peripheral collisions induce the generation of full disorder 
and are very weakly affected by collective effects whereas central 
collisions show sizable collective effects.

\subsubsection{Thermodynamic properties of systems described by cellular 
models : an analytical model}
\indent

  The encouraging results concerning fragment multiplicities obtained in 
this type of simple models raises the question whether the present 
description supports the existence of a thermodynamic phase transition as 
it is the case for the lattice gas models described above. The problem which 
comes up is the fact that the generated events do not need to possess the 
property of being in thermodynamic equilibrium since the algorithm which is 
used does not guarantee it. Indeed, a necessary condition for thermodynamic 
equilibrium in the canonical ensemble is the determination of an absolute 
minimum in the energy for a fixed temperature through the application of 
detailed balance. This has not been implemented. However there are reasons 
to believe that the generated systems are close to if not at equilibrium. 
This led to the analysis of their thermodynamic properties in the framework 
of the microcanonical ensemble \cite{Richert2}. The entropy $S(E)$ has been 
determined as a function of the energy $E$ and the caloric curve 
constructed by defining the temperature as $T^{-1}=(\partial S/\partial E)_V$ 
where $V$ is the volume of the system. The outcome of the study is the 
fact that $T$ increases monotonously with $E$ and there is no sign for the 
presence of a transition. An analytical investigation which describes the 
system in a way which is close to the one which is obtained through 
numerical simulations confirms this result \cite{Richert2}. 

  The absence of a transition is in fact qualitatively understandable if 
one goes back to the lattice gas models of section 5.2. Indeed, the 
discussed cellular models correspond to the extreme case where the system 
is homogeneously filled with particles, hence $\rho / \rho_0 = 1$ in
Fig.~\ref{fig16} where the system is in a pure liquid phase.

  In order to study the behaviour of a system for every density one may 
introduce a simple 3 dimensional cubic lattice whose volume $V$ is 
divided into $N^3$ cells of linear size $d$, $V = (Nd)^3$. This geometry 
can be straightforwardly extended to a parallelepiped $V = N_1 N_2 N_3 d^3$
\cite{Richert3}. The restriction to attractive nearest-neighbour 
interactions leads to a Hamiltonian
\begin{equation} H_0 = \sum_{i=1}^A \frac{p_i^2}{2m} - 
V_0 \sum_{<n.n>_1}s_is_j \label{Hamilton} \end{equation} 
where $A$ is the total number of particles, $<n.n>_k$ stands for the ``$k$th
nearest-neighbour'' sites$(i,j)$, $V_0$ is the strength of the interaction 
and $s_k = 0$ (resp. 1) corresponds to a cell which is empty (resp. 
occupied by a particle). The model which describes an inhomogeneous system 
can be trivially mapped into a spin-model by means of the change of 
variables 
\begin{equation} \sigma_k = 2s_k-1 \nonumber \end{equation} 
Particle number conservation requires
\begin{equation} 1/2 \sum_{k=1}^{N^3} (1 + \sigma_k) = A
\nonumber \end{equation} 
In terms of the new variables the Hamiltonian reads
\begin{equation} H_0 = \sum_{k=1}^A \frac{p_k^2}{2m} - \frac{V_0}{4} \Big[
\sum_{<n.n>_1} \sigma_k \sigma_l + 6\sum_{k=1}^{N^3}\sigma_k + 3N^3 \Big]
\nonumber \end{equation} 
which corresponds to an Ising model with fixed magnetic field $h=3V_0/2$. In 
order to work out its thermodynamic properties one goes over to the continuum 
limit by replacing $\sigma_k = \pm 1$ by the continuous variables $u_k$ in
$[-\infty,+\infty]$. This leads to a description in the framework of the 
spherical model which is exactly solvable in any number of dimensions for 
different types of interactions and even in the presence of disorder 
\cite{Berlin, Stanley1, Baxter, Singh, Frachebourg}. Constraining the 
continuum variables $\{u_k\}$ by means of
\begin{equation} \sum_{k=1}^{N^3} u_k^2 = N^3
\label{constr} \end{equation} 
and introducing the Lagrange multipliers $\lambda$ and $\mu$ leads to the 
Hamiltonian 
\begin{equation} H = H_0 + \lambda \sum_{k=1}^{N^3} u_k^2 + \frac{\mu}{2}
\Big( \sum_{k=1}^{N^3} u_k + N^3 \Big) \nonumber \end{equation} 
and a canonical partition function
\begin{equation} \mathcal{Z}(\beta,A,N^3) = \int_{-\infty}^{+\infty}
\prod_{k,i} \ud p_k^i \int_{-\infty}^{+\infty}\ud u_k \exp[-\beta H(\{u_k\})]
\nonumber \end{equation} 
where $p_k^i$ corresponds to the $i$th component of the moments $p_k$ of 
the particle $k$. The parameters $\lambda$ and $\mu$ are fixed by 
\begin{eqnarray} -\beta^{-1} \partial \ln \mathcal{Z} /
\partial \lambda & = & N^3 \label{partD1} \\
-\beta^{-1} \partial \ln \mathcal{Z} / \partial \mu & = & A \label{partD2}
\end{eqnarray} 
The physical justification for the introduction of the continuum limit 
variable $\{u_k\}$ is the fact that the interaction in $H_0$ can now 
be interpreted as taking
a continuum of values, not only $\pm V_0/4$, which is sensible in the 
case of cellular models, since the location of the particles in cells is 
not restricted to the centre of the cells. Because of the constraints 
imposed by the volume constraint (\ref{constr}) the important contributions 
to $V_{<n.n>_1}$ remain finite. The procedure does not take care of the 
fact that positions of a set of interacting particles are correlated which 
induces correlations on the two-body contributions. These correlations are 
not taken into account here, like in the use of algorithm A2 in numerical 
simulations. 

  The partition function can be worked out explicitly. It can be cast in 
the form
\begin{equation} \mathcal{Z}(\beta,A,N^3) \propto
\exp[-\beta N^3 f(\beta,A,N^3)] \nonumber \end{equation} 
with the free energy per cell
\begin{eqnarray} f(\beta,A,N^3) & = & [(\mu-3V_0)/2]^2/[4(3V_0/2- \lambda]
+ (1/2\beta N^3) \cdot \sum_{i=0}^{N^3-1}
\ln[\beta(\lambda - V_0 \alpha_i/2)] \nonumber \\
\textrm{where} \quad \alpha_i & \equiv &
\alpha_{i_1i_2i_3} = \sum_{k=1}^3 \cos \xi_{i_k} \nonumber \\
\xi_{i_k} & = & 2\pi i_k/N \quad (i_k=0,1,\ldots,N-1)
\nonumber \end{eqnarray} 
The constraints (\ref{partD1}) and (\ref{partD2}) lead to the relation 
\begin{equation} \sum_{i=0}^{N^3-1} [\lambda - V_0 \alpha_i/2]^{-1}
= 2 N^3 \beta (1 - M^2) \label{disp} \end{equation} 
which fixes $\lambda$. Then
\begin{equation} \mu = 4(3V_0/2 - \lambda)M + 3V_0
\label{defmu} \end{equation} 
where
\begin{eqnarray} M & = & -[1+2(\beta N^3)^{-1} \partial \ln \mathcal{Z}
/ \partial \mu] \nonumber \\
  & = & -(\mu-3V_0) / [4( \lambda - 3V_0/2)] \nonumber \\
  & = & -1 + 2A/N^3 \nonumber \\
  & = & 2\rho/\rho_0 - 1\nonumber \end{eqnarray} 
which shows that $M \in [-1,+1]$.

  The dispersion relation (\ref{disp}) has $N^3$ poles. There exists one 
solution which corresponds to $\lambda > V_0 \alpha_{max}/2$ with 
$\alpha_{max}=3$ and guarantees that the corresponding value of $\lambda$ 
leads to an expression of $\mathcal{Z}$ which makes physical sense.

\begin{figure}[b!]
\centerline{\epsfig{figure=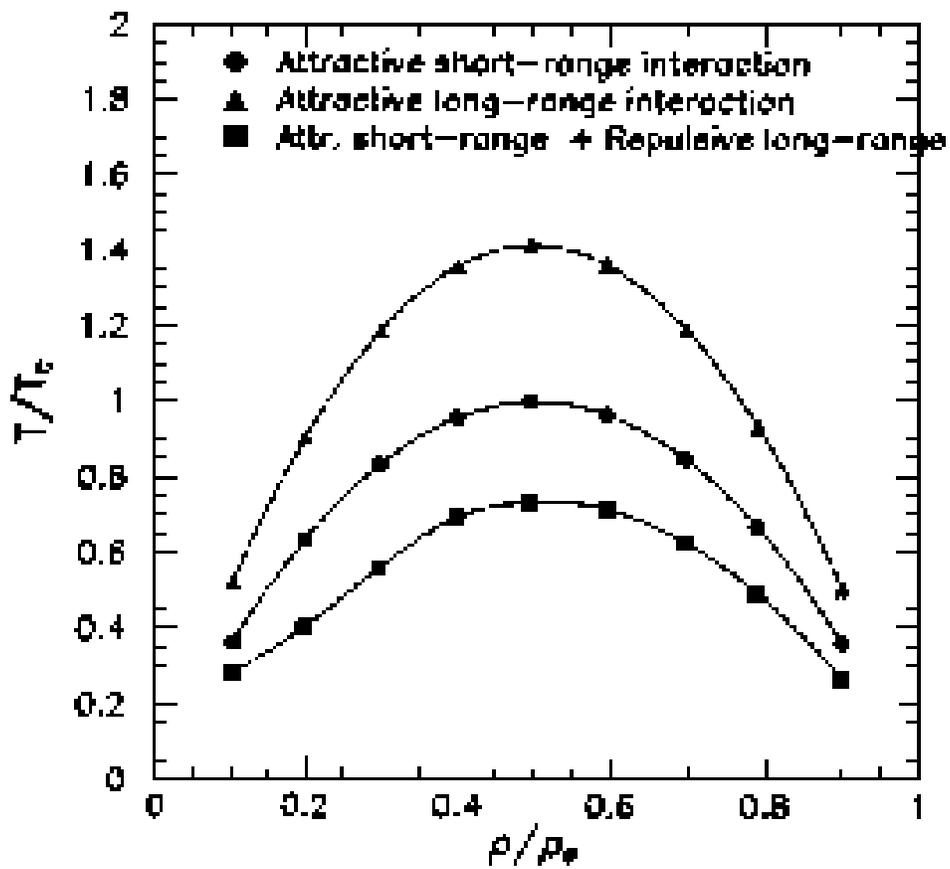,height=120mm}}
\caption{Phase diagrams for different types of interactions, see 
explanations in the text \cite{Richert3}.}
\label{fig21}
\end{figure}

  It can be shown \cite{Singh} that for $M=0 \ (\rho / \rho_0) = 0.5)$ the 
free energy per unit volume gets singular for a given value of the 
temperature $T=T_c(0)=V_0/2W_3(0)$ where $W_3(0)=0.2527$\cite{Singh, 
Richert3}. The same behaviour of the free energy is observed for all other 
values of $M$ leading to critical temperatures $T_c(M)$. The specific heat 
is given by
\begin{equation} C_V=-\beta^2 \ud \epsilon / \ud \beta 
\nonumber \end{equation} 
where $\epsilon=E/A$ is the energy per particle. $C_V$ shows a maximum for 
every value of $M$ at $T_c(M)$ which signals a continuous phase transition 
between two phases. The corresponding phase diagram is shown in 
Fig.~\ref{fig21}. 
The analytic expression of $\mathcal{Z}$ allows for the determination of 
all thermodynamic quantities and the construction of the caloric curve 
$T=f(E)$ which is a monotonously increasing function. This is due to the
fact that $C_V$ remains finite at the transition point for any value of 
$\rho / \rho_0$ \cite{Richert3}. Only the derivative of $C_V$ shows a 
discontinuity at the transition points.

  The model can be extended to the case where particles interact through 
short or/and long range interactions
\begin{eqnarray} H = K & - & [V_{01} \!\! \sum_{<n.n>_1} \!\! s_i s_j
+ V_{02} \!\! \sum_{<n.n>_2} \!\! s_i s_j + \cdots
+ V_{0m} \!\! \sum_{<n.n>_m} \!\! s_i s_j] \nonumber \\
 & + & \mu \sum_i s_i + \lambda \sum_i (2s_i - 1)^2
\nonumber \end{eqnarray} 
where $K$ is the kinetic energy and $<n.n>_k$ has been defined below 
(\ref{Hamilton}). A sensible parametrization in the case of nucleus would 
correspond to
\begin{eqnarray} V_{01} & = & V_s + V_l^{(1)} \nonumber \\
V_{02} & = & V_l^{(2)} \nonumber \\
& \vdots & \nonumber \\
V_{0m} & = & V_l^{(m)}
\nonumber \end{eqnarray} 
where $V_s > 0$ stands for the short range nuclear potential and 
$V_l^{(k)} < 0$ for the long range repulsive Coulomb interaction 
parametrized in the following form
\begin{equation} V_l^{(k)} = (Z_p/A)e^2/kd \nonumber \end{equation} 
where $Z_p$ is the total number of protons.

  The partition function can be worked out in the framework of the 
spherical model in the same way as in the former case \cite{Richert3}. 
Fig.~\ref{fig21} shows typical phase diagrams for different cases involving 
long range interactions. As one can see, they lead to qualitatively similar 
results.  
  
  In the usual spherical model in which the constraint (\ref{partD2}) is 
not implemented $\mu$ plays the role of a fixed magnetic field. Then, for 
$\mu \neq 0$, the magnetization can jump from $-M$ to $+M$ which 
corresponds to a first order transition. For $\mu = 0$ the transition is 
continuous. This is not the case here since the magnetization $M$ 
(density $\rho$) is fixed, and not $\mu$ which is determined through 
(\ref{partD2}). It explains why the transition is continuous in the present 
case for any value of $M$ or $(\rho / \rho_0)$.

\subsubsection{Thermodynamic properties of systems described by cellular 
models : numerical simulations}
\indent

  The former approach is a simplified even though quite realistic 
thermodynamic description of a disordered system. However, it does not 
allow to work out the fragment content of the system, which are accessible 
by means of numerical simulations \cite{Carmona1}. This has been performed 
in the framework of the canonical ensemble for particles which experience 
short range nearest-neighbour interactions.

  The Hamiltonian is written as 
\begin{equation} H = \sum_{i=1}^A \frac{p_i^2}{2m} s_i +
\sum_{<i.j>_1} v_{12}(r_{ij}) s_i s_j \nonumber \end{equation} 
where $v_{12}$ is the potential which acts between particles located in 
neighbour cells \cite{Wilets} at a distance $r_{ij}$ from each other, $A$ 
is the number of particles, $\rho = A/L^3$ the density of particles in a 
volume $V = L^3$ and the $s_i$'s are occupation numbers defined above. 
Periodic boundary conditions are imposed in the numerical simulations.

  A Metropolis Monte Carlo procedure is used in order to generate events 
corresponding to different densities $\rho$ and temperatures $T$. For each 
$(\rho,T)$ the total number of events is $6 \cdot 10^4$ and for each event 
about 5000 Metropolis iterations have been realized. Each iteration 
consists of either the move of a particle from its original cell to an 
empty cell, or its displacement inside a cell, or the determination of a 
new momentum. Each operation is realized with a probability $1/3$.

  The latent heat is defined as 
\begin{equation} C_V = \frac{1}{A} \big( <U^2> - <U>^2 \big)
\nonumber \end{equation} 
where the average is an ensemble average and $U$ the potential energy of 
the system.

  The calculations have also been extended to the microcanonical ensemble 
with fixed $A$ and energy $E$, by means of an algorithm described in ref.
\cite{Ray} which is well adapted to the present type of models. The 
probability for the existence of an event which has a configuration $s$
characterizing the positions and momenta of the particle is given by
\begin{equation} W_E(s) \propto \big( E - U(s) \big)^{\frac{dV}{2}-1}
\Theta\big( E - U(s) \big) \nonumber \end{equation} 
where $d$ is the space dimension, $V$ the volume and $\Theta$ the Heaviside 
function. Events are selected by means of a Metropolis procedure which 
accepts a configuration $s'$ starting from a configuration $s$ with a 
probability
\begin{equation} W_{s \to s'} = \min \bigg( 1, \frac{W_E(s')}{W_E(s)} \bigg)
\nonumber \end{equation} 
The temperature is obtained from
\begin{equation} T=2<K>/Vd \nonumber \end{equation} 
where $<K>$ is the average kinetic energy of the particles.

\begin{figure}[b!]
\centerline{\epsfig{figure=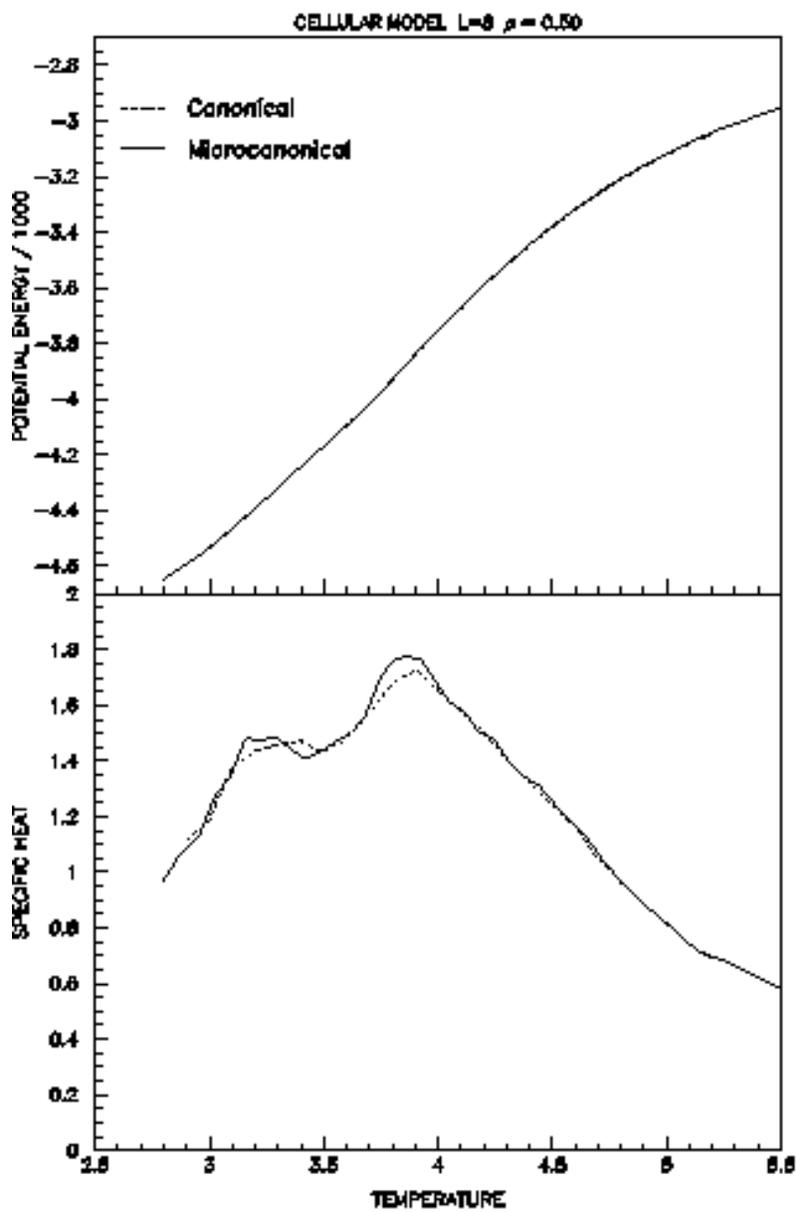,height=160mm}}
\caption{Caloric curve and specific heat for a finite size system, 
$L = 8^3$, and density $\rho / \rho_0 = 0.5$. Canonical and microcanonical 
simulations coincide \cite{Carmona1}.}
\label{fig22}
\end{figure}

  Typical examples of the behaviour of $C_V$ and the caloric curve are 
shown in Fig.~\ref{fig22}. The specific heat presents a more or less pronounced 
enhancement for some value of $T$. This enhancement is taken as the finite
size restriction of a point which determines the phase separation line 
between a liquid and a gas in the thermodynamic limit. The numerical 
simulations show a smooth monotonous increase of the energy with 
temperature for fixed volume which is very similar to the results obtained 
in the framework of liquid gas models. For these small values of $A$ and 
$V$ it is not possible to identify the order of the transition. It is also 
remarkable to see that canonical and microcanonical calculations are very 
close to each other, even though the size of the system and the volume 
which contains it are small. We shall come back to this point and develop it 
in section 6.

  The identification of bound clusters leads to the separation of the 
fragment content into two areas of the $(\rho, T)$ phase diagram
separated by the equivalent of a Kert\'esz 
line, above which the system contains essentially individual particles and 
small fragments, whereas large clusters are present below it. The 
separation line is expected to correspond to a power law fragment mass 
distribution in the limit of an infinite system. The presence of the power 
law behaviour is detected by analysing the calculated mass distributions.

\begin{figure}[b!]
\centerline{\epsfig{figure=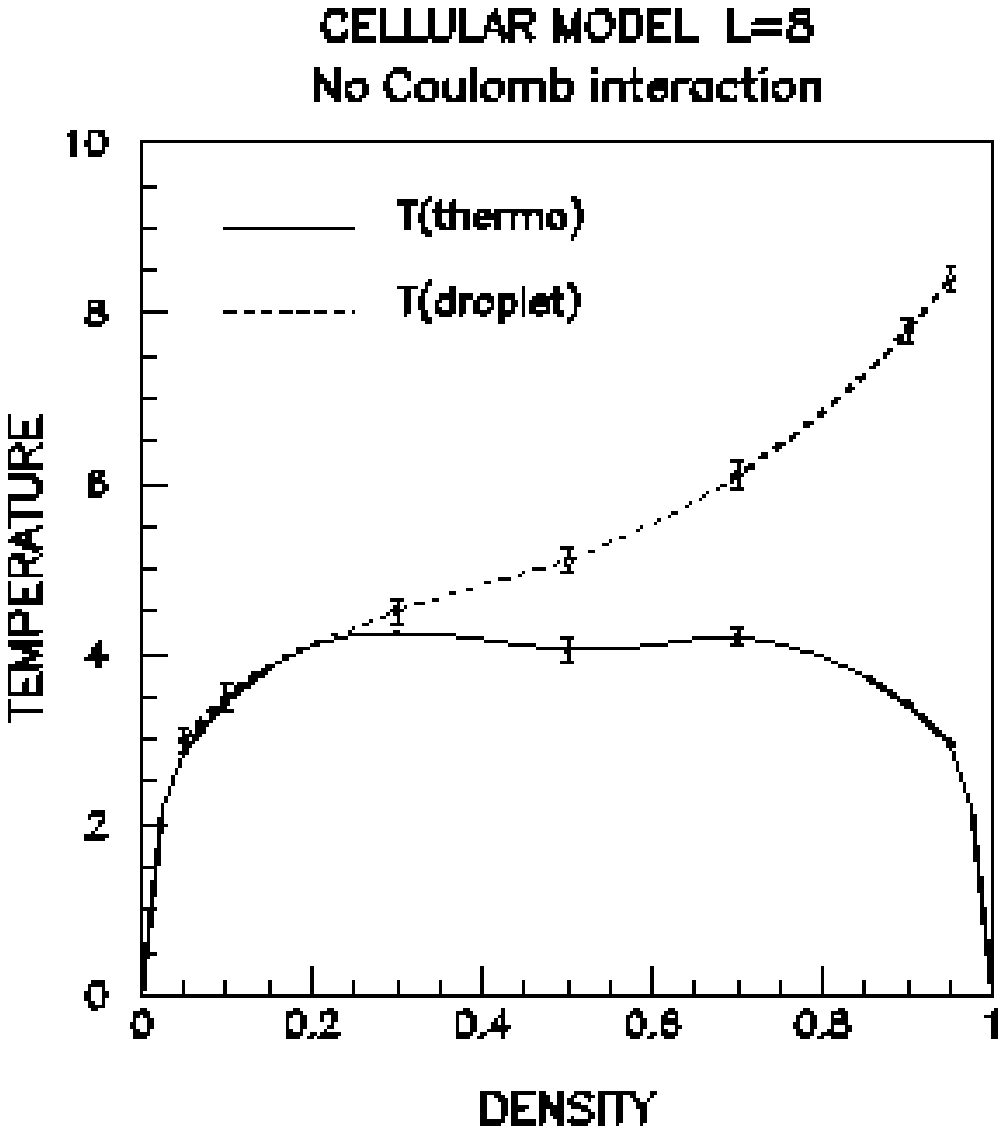,height=160mm}}
\caption{The phase diagram obtained with the cellular model described in 
the text \cite{Carmona1}. See discusion in the text.}
\label{fig23}
\end{figure}

  The results are shown in the phase diagram of Fig.~\ref{fig23}. 
Several remarks 
are in order. The thermodynamic separation line shows a dip at $\rho = 0.5$.
This is a finite size effect. A study of this point on a spin model which 
presents the same property confirms this point, see below in section 6. The 
separation line itself is not rigorously determined, there exists other 
means to fix it and furthermore it does not correspond to the limit of the 
infinite system. Contrary to the result of ref. \cite{Gulminelli1}, 
the Kert\'esz line stays above the separation line for all values of $\rho$ 
but does not end at $\rho = 0.5$. Furthermore, at $\rho = 0.5$ the 
thermodynamic critical point does not coincide with the corresponding point 
on the Kert\'esz line. The reasons for this are not easy to identify clearly.
It makes sense to believe that all these points are related to finite size 
effects and possibly also the numerical precision with which the relevant 
quantities can be obtained. The determination of critical exponents by 
means of finite size scaling arguments has not been performed. The analysis 
and outcome would be similar to the ones obtained in the framework of another 
model which will be discussed below.

  The remarkable point which may be retained from this study is the fact 
that the a priori more sophisticated cellular models discussed here are 
very close to the more schematic lattice gas models. If there are 
quantitative differences it is, at present, out of scope to judge which is 
quantitatively closer to the experiment. 

\newpage

\section{Finite nuclei and phase transitions : present experimental and 
theoretical status}
\indent

  The experimental quest for a thermodynamic phase transition in nuclear 
matter received a new impetus in 1995 with the work of Pochodzalla et al. 
who tried to construct the so called caloric curve \cite{Pochodzalla}. If 
nuclear matter would behave like a Fermi gas one would expect an increase 
of the temperature $T$ with excitation energy $E^*$ such that $T \propto 
E^{*1/2}$. Deviations could indicate the existence of a phase transition 
which, following classical arguments, is expected to be of the liquid-gas 
type.

  However, nuclei are finite and the asymptotic limit of infinite matter is 
experimentally out of reach. The situation which is common to other fields 
like physics of aggregates and mesoscopic systems in condensed matter 
physics needs the development of new tools which may allow to detect the 
existence of the transition through the study of the behaviour of physical 
observables which characterize finite systems.

  In the present section we try to give an overview of the recent 
experimental efforts which have been made and the theoretical 
investigations which have been initiated, specifically with the help of 
lattice-type models in order to detect a possible phase transition and 
unravel its nature in excited nuclear matter under the assumption that 
experimental measures are made on thermodynamically equilibrated systems.

\subsection{Caloric curves and phase transitions : experimental status}
\indent

  The caloric curve which relates the internal energy of an excited system 
at thermodynamic equilibrium to its temperature is a priori the simplest 
experimental tool to look for the existence of a phase transition. The 
first determination attempted by the ALADIN collaboration \cite{Pochodzalla} 
collected the outcome of $^{197}Au$ on $^{197}Au$ collisions at $600\,MeV\!
\cdot\!A$ with data obtained by means of less energetic collisions. The 
excitation energy was determined by following the procedure prescribed in 
ref. \cite{Campi5} and the corresponding temperature fixed by means of 
arguments relating this quantity to the so called double ratio procedure 
\cite{Albergo}, in the present case the ratios of $^3He/^4He$ and 
$^6Li/^7Li$ isotopes. The corresponding curve showed the features of a rather 
strong first order transition, with a characteristic close to constant 
temperature $T$ over a large energy interval lying between 3 and $10\,MeV$ 
excitation energy per nucleon, which may be interpreted as a sign for the 
generation of latent heat and followed by a strong increase of $T$ with 
excitation energy above $10\,MeV$. A critical discussion followed this 
observation, in which the hypotheses and simplifications underlying the 
definition of the temperature, the freeze-out density and the increase of 
$T$ at high excitation energy were examined \cite{Campi10}.

  A similar analysis by means of $1\,GeV\!\cdot\!A$ collisions of $C$ on 
$Au$ was initiated soon after this first attempt \cite{Hauger}. The 
reaction was interpreted as a two-step process, a prompt stage with the 
emission of light particles and a second step corresponding to the decay of 
an equilibrated system undergoing fragmentation. The outcome of the 
analysis lead to a smooth caloric curve from which any clear sign 
corresponding to a first order transition was seemingly 
absent, showing at most the 
possible existence of a continuous transition. In this experiment the 
temperature was determined in terms of two different isotope ratios,
$(^3He/^4He)/(^6Li/^7Li)$ and $(^3He/^4He)/(d/t)$. The results were 
essentially confirmed later in a further experiment \cite{Hauger1}.

  The INDRA collaboration considered $Ar$ on $Ni$ collisions at 52 and 
$92\,MeV\!\cdot\!A$ in order to investigate this point \cite{Ma}. 
Several double 
isotopic yield ratios were used in order to define the temperature. They 
led to different slopes in the temperature increase as a function of the 
excitation energy. The issue of the experiment and its 
interpretation have been performed by means of statistical models 
\cite{Borderie, Borderie1, Gulminelli}. A criticism has however 
been raised \cite{Lefevre1} which claims that the observation of the 
transition may have been missed because of the choice of too large steps 
in energy binnings.

\begin{figure}[b!]
\centerline{\epsfig{figure=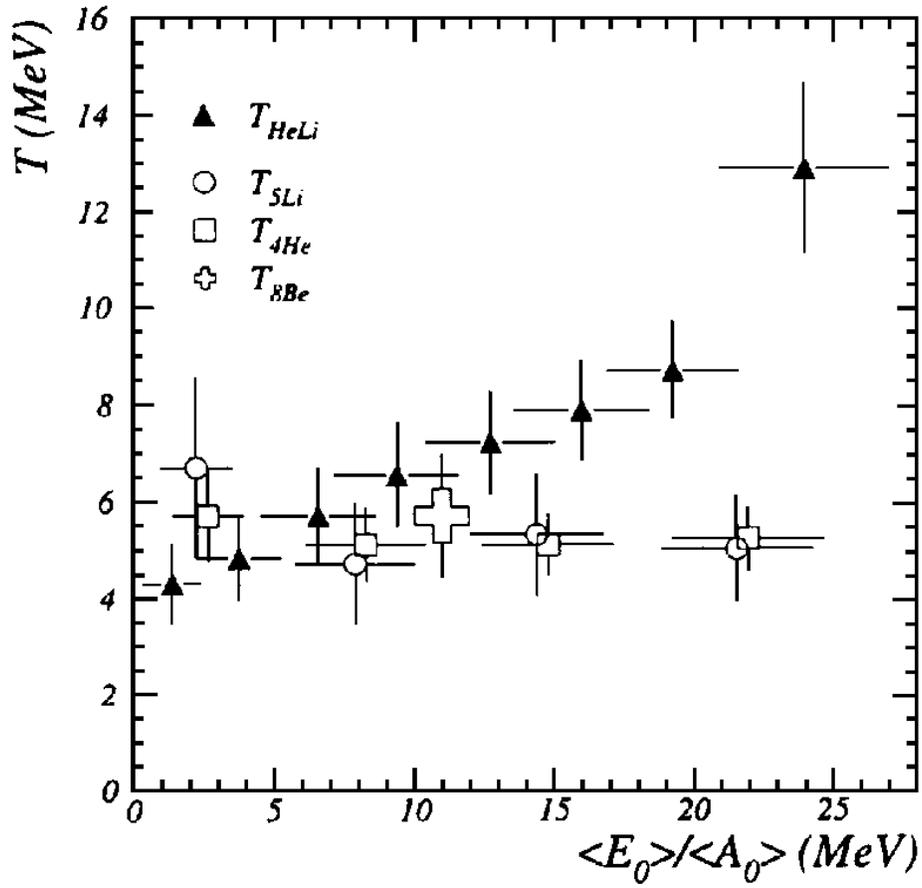,height=120mm}}
\caption{Isotope temperature $T_{He Li}$ (full triangles) and temperatures 
obtained by means of excited state population rate measurements (open 
symbols) as a function of the excitation energy per nucleon for $^{197}Au$ 
on $^{197}Au$ at $E = 1000\,MeV\!\cdot\!A$ \cite{Trautmann1}.}
\label{fig24}
\end{figure}

  At this stage the ALADIN collaboration considered the problem of the 
determination of the temperature by means of an analysis of $Au$ on $Au$ 
collisions at $1000\,MeV\!\cdot\!A$ \cite{Trautmann1} which was determined 
through different isotopic yield ratios and corrected for sequential 
evaporation from primary fragments. In this experiment, 
temperatures were also defined by means of excited state populations 
\cite{Pochodzalla1, Kunde1}. Both types of temperatures are shown in 
Fig.~\ref{fig24}. 
Those obtained from population yields are remarkably constant 
as a function of the excitation energy per particle, in contrast to the 
temperatures obtained by means of isotopic yield ratios. The interpretation 
of this apparent inconsistency leads to 
the conclusion that these temperatures 
may correspond to the final stage of fragment emission. The corrected new 
caloric curve emerging from the use of isotopic yields does no longer show 
a clear plateau-like behaviour. At the highest value of the excitation 
energy there remains however a sizable rise in the temperature which seems 
to indicate the entrance into a new phase.

  A further attempt has been performed by bombarding $Ag$ and $Au$ with 
$4.8\,GeV\ ^3He$ ions \cite{Kwiatkowski}. The temperature was fixed by 
means of the double ratio yield $(^2H/^3H)/(^3He/^4He)$. A slope 
discontinuity is observed for $E^*/A \simeq 2 - 3\,MeV$ which is followed, 
with increasing excitation energy, by a monotonic rise up to the measured 
energy $E^*/A \simeq 10\,MeV$. The curve is in qualitative agreement with 
the theoretical predictions of the Expanding-Evaporating Source model (EES)
\cite{Friedman1, Friedman2} and the Statistical Multifragmentation Model
(SMM) \cite{Bondorf, Botvina}.

  More recently, a further result has been obtained on excited systems with 
$A \simeq 110$ particles produced by means of different projectiles and 
targets with a bombarding energy of $47\,MeV\!\cdot\!A$ \cite{Cibor}. The 
temperature was obtained from the double ratios $(d/t)/(^3He/^4He)$. An 
analysis based on coalescence models \cite{Awes, Csernai, Mekjian1, Mekjian,
Sato} was used in order to follow the evolution of temperature and densities. 
Calorimetry measurements fix the excitation energies. The experimental 
points combined with former results \cite{Wada} show a steep increase of 
$T$ up to $E^*/A \simeq 3.5\,MeV$. The temperature stabilizes for $3.5 \leq 
E^*/A \leq 7\,MeV$ around $T \sim 7\,MeV$.

  The analysis and comparison of results obtained by different groups show 
hardly more than qualitative similarities and are rather non conclusive as 
far as the existence and possible order of a phase transition are concerned. 
They are plagued by different problems which are not only related to finite 
size effects or the equilibration problem, but also by the fact that in 
the experiment the system undergoes expansion, sequential decay of primary 
excited fragments, apparently temperature determinations can lead to 
inconsistent results. These problems have been investigated by several 
groups \cite{Xi1, Xi2, Odeh, Xi3, Bondorf3, Xi4, Souza}.
Further new methods for the determination of the temperature have 
been proposed \cite{Veselsky}. As the matter stands, there are signs which 
point towards the existence of a transition. The caloric curve may however 
not be the most adequate tool to confirm it. We shall discuss this point in 
the next section.
  
\subsection{Signs for phase transitions in finite systems : tests and 
applications to the nuclear case by means of lattice models}
\indent

  Nuclei are finite systems, hence experimental observations refer to small 
systems, far from the thermodynamic limit. Contrary to other aggregates or 
quantum dots whose size can be experimentally varied and controlled, there 
is no such flexibility in fixing the number of constituents over large 
scales here. It shows that 
even if the experimental determination of physical observables could be 
fully tested, there would remain the problem of the small size limit. 
This point raises the question of the detection of the existence and 
nature of a phase transition through the measurements on finite systems. 
The signs for the existence of transitions clearly exist,
they are seen in finite systems
since infinite systems do not exist, and one may speak about phase 
transitions in finite systems \cite{Gross6}. The determination of their 
order necessitates good tests and a great deal of caution. This point has 
been raised in several fields of physics in the recent past. We review some 
of the methods and tests which have been proposed in the next section 
before applying them to lattice models and the analysis of experimental 
data.

\subsubsection{Tests for the existence of phase transitions from finite 
system analysis}
\indent

  There exist several characteristic properties which may signal that a 
finite system can show a critical behaviour related to a phase transition.

  Close to a critical point observables follow scaling laws which relate 
their behaviour at different scales, i.e. for different sizes of the system. 
They allow for the determination of critical exponents which characterize 
the singular behaviour at the critical point in the limit of an infinite 
system. This property has already been presented and discussed in 
subsection 4.3 above \cite{Carmona}. We shall come back to it below.

  A second characteristic of finite systems which are close to a phase 
transition is the enhancement and universal behaviour of fluctuations in 
observables and order parameters. This type of properties is commonly 
studied \cite{Bramwell} and has been used \cite{Chomaz2} as will 
be shown below. Recently the properties of order parameters of finite 
systems close to a second order transition have been studied in the 
framework of a finite size scaling approach \cite{Botet}.
It comes out that order parameters can be characterized by 
probability densities which possess specific properties depending on their 
location with respect to the critical point. In particular, at the critical 
point, they show universal scaling and deviations from a 
gaussian behaviour in the tails. This behaviour is valid for both 
equilibrium as well as off-equilibrium processes in systems which undergo 
second order transitions. The method could work as an alternative to the 
determination of critical exponents. Applications will certainly be 
developed in the near future.

  Informations about the thermodynamic properties of small extensive and 
non-extensive systems (for which the sum of the energies and entropies 
is not the sum of energies and entropies of their parts, f.i. if long range 
interactions are present) can be very efficiently analysed in the framework 
of the microcanonical ensemble. This aspect has been extensively discussed,
see f.i. \cite{Gross6} and refs. therein. The important point of practical 
importance concerns the topology of the entropy $S(E,A)$ of a system with 
$A$ particles and total energy $E$ \cite{Gross6}. 
In the presence of a first order transition, the entropy shows 
convexity (negative curvature) in certain ranges of energy. This behaviour 
is specific to the microcanonical ensemble which is able to give a faithful 
description of inhomogeneous systems. In the caloric curve the temperature 
is a multivalued function of the energy, showing a bend (``$S$'' curve) and, 
as a consequence, the specific heat can get negative over some intervals of 
energy. The physical explanation for this phenomenon is the existence of a 
surface tension which is generated by the coexistence of two phases. Indeed, 
the creation of intra-phase surfaces generates an entropy loss $\Delta
S_{Surf}$ which can be obtained from the relation \cite{gross7}
\begin{equation} \Delta S_{Surf} = \frac{\gamma}{T_c} \cdot \frac{\Sigma}{A}
\nonumber \end{equation}
where $\gamma$ is the surface tension, $T_c$ the transition temperature, 
$\Sigma$ the surface area of all species with mass larger than $1$, and $A$ 
the number of particles. This remarkable property shown by finite systems 
will be used below as a main tool in order to detect and characterize the 
behaviour of fragmenting nuclear systems.

  The present collection of tests is certainly not exhaustive, there may 
exist other means to detect signs for phase transitions. One of them which 
grounds on the Yang and Lee theory has already been mentioned in section 
5.1.3 \cite{Borrmann}. Recently Ma \cite{Ma1, Ma2} proposed to use the 
enhancement of entropy production at the pseudo-transition point and 
claimed that Zipf's law \cite{Zipf} is verified at the transition point. 
This has been tested numerically by means of calculations involving an LGM 
model and Molecular Dynamics simulations. Ref. \cite{Umirzakov} discusses 
the correlation between an ``$S$'' curve in the caloric curve in the 
microcanonical framework, the thermodynamic temperature $T=(\partial E
/\partial S)_V$ and the temperature determined from the kinetic energies
of the constituents of the system. It is shown that the caloric curve 
obtained by means of the thermodynamic temperature shows a bend if this is 
the case for the kinetic temperature, but that the reverse is not true.

\subsubsection{Applications of the tests to models of nuclear fragmentation.
Finite size scaling and microcanonical description}
\indent

  The finite size scaling test for the determination of the order of a 
phase transition by means of the determination of 
critical exponents has been recently applied to a lattice model, the Ising 
Model with Fixed Magnetization (IMFM) \cite{Carmona}. Its canonical 
partition function reads
\begin{equation}
\mathcal{Z}_V = \sum_{[\sigma]} \exp \big( \beta V_0 \!
\sum_{ <n \cdot n>_1 } \! \sigma_i \sigma_j \big) \ \delta_{ M, \widehat{M}}
\nonumber \end{equation}
where $V_0$ is a negative potential strength, $M=2\rho/\rho_0-1$ the 
``magnetization'' related to the relative density $\rho/\rho_0$, 
$\widehat{M}$ a fixed value of $M$, $\{\sigma_i=\pm1\}$ related to the 
lattice site occupations  $\{s_i=0,1\}$ in the notation of subsection 5.3.5.

  The model is equivalent to an LGM except for the fact that the number of 
particles, $A=\sum_i s_i$, is strictly fixed and, as a consequence, 
$\rho/\rho_0$ is fixed if the volume $V$ is fixed. It is a non trivial 
model whose properties have been investigated recently \cite{Pleimling}. In 
the canonical ensemble the kinetic energy plays a trivial role, therefore 
there is no need to introduce it if one looks for critical properties of 
the system in this framework.

  The system governed by the IMFM shows a phase transition with a phase 
diagram $(\rho,T)$ which has the same topology as LGM and related models, see
f.i. Fig.~\ref{fig16}. However in the ordinary Ising model with zero magnetic 
field contribution, the magnetization jumps when the system enters the 
coexistence phase at $T<T_c$. Since the present description keeps the 
magnetization fixed, the order of the transition is a priori not clear.

  The problem has been investigated numerically, first in the canonical
framework for a system with fixed volume. The sensible quantity which has
been worked out is the specific heat
\begin{equation} C_V = 3V ( <E^2> -<E>^2 ) \nonumber \end{equation}
where $V$ is the volume and $E$ the energy of the system. At the separation
line, one looks for the asymptotic behaviour of $C_V$ in the form
\begin{equation} C_V \propto (\beta - \beta_c^{Th})^{-\alpha}
\nonumber \end{equation}
$\beta^{Th} = T^{-1}$, $\beta_c^{Th} = (T_c^{Th})^{-1}$, $T_c^{Th}$ being 
the thermodynamic transition temperature of the infinite system for fixed
density $\rho$ and $\alpha$ a real positive exponent which may help to
characterize the universality class to which the transition belongs.

\begin{figure}[b!]
\centerline{\epsfig{figure=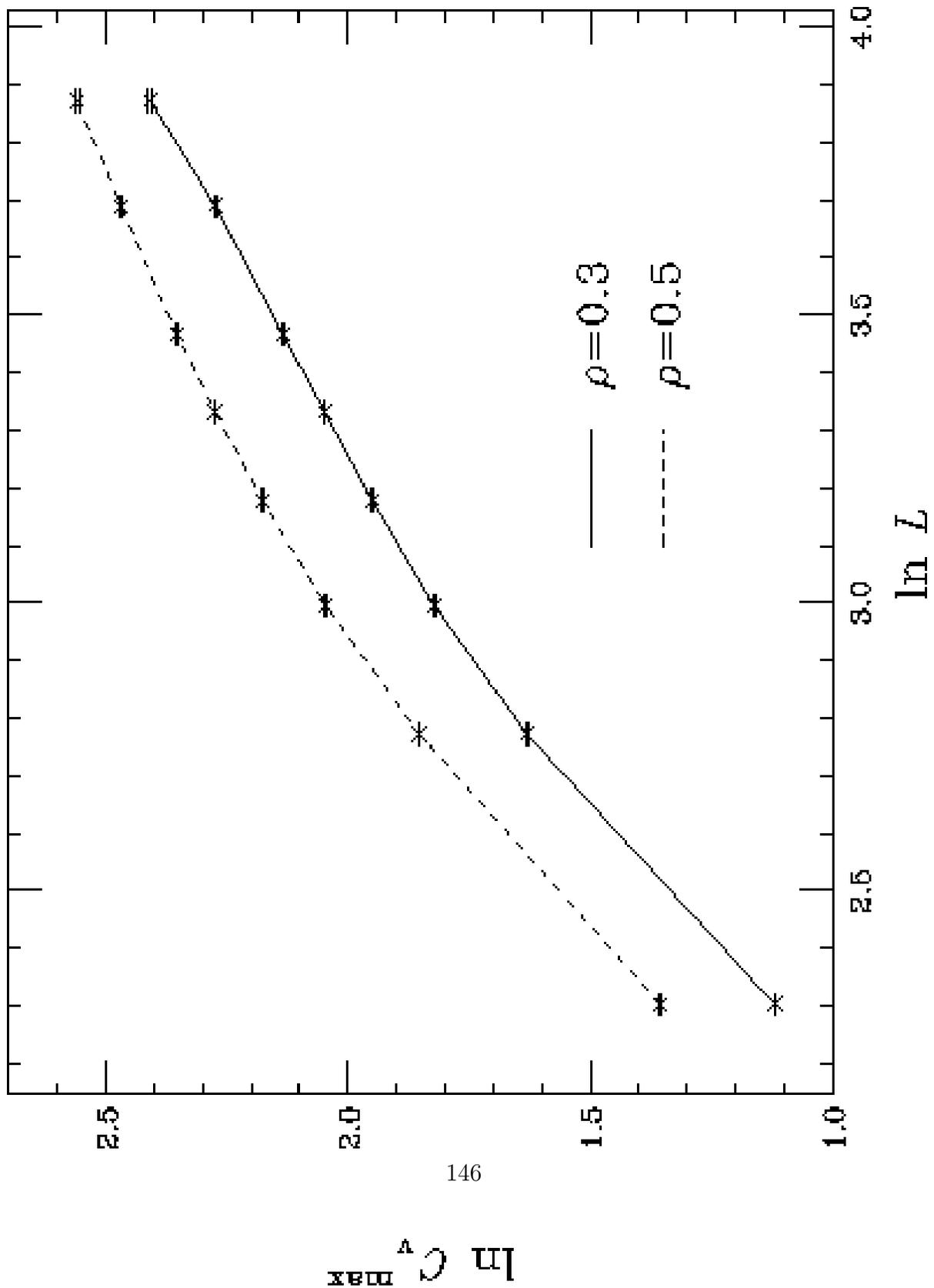,height=160mm,angle=90}}
\caption{Fit of $C_V^{max}$ as a function of the linear size $L$ of the 
system for different densities. Crosses indicate different sizes. The 
largest size is $L = 48$ \cite{Carmona}.}
\label{fig25}
\end{figure}

  Using finite size scaling arguments and writing
\begin{eqnarray} C_V(L) & = & A + B L^{\alpha/\nu} \label{CV} \\
\beta_c(L) - \beta_c(\infty) & = & \bar{A} L^{-1/\nu} \nonumber
\end{eqnarray}
where $A,B,\bar{A}$ are real constants and $L$ is the linear size of the 
system, it is possible to obtain $\nu$ and $\alpha/\nu$, hence $\alpha$, 
corresponding to the thermodynamic limit \cite{Carmona}. Typical results 
for the evolution of $C_V$ with the size of the system by identification of 
the maximum of this quantity for fixed $\rho$ are shown in Fig.~\ref{fig25}
for two 
values of $\rho/\rho_0 = (\widehat{M}+1)/2$. In this figure $\alpha/\nu$ is 
read from the slope of the curves which should behave like rigorously 
straight lines if the scaling hypothesis (\ref{CV}) is working. The 
straight line behaviour is observed for $L$ large enough $(L \geq 28)$, 
although there are deviations at close inspection. The values of $\nu$ and 
$\alpha$ for $\rho/\rho_0 = 0.3$ and 0.5 could be compatible with the 
universality class of the Ising model although this is not absolutely 
established since the behaviour for $L=48$ could still be transitory. This 
value of $L$ in $3d$ space requires already a large numerical effort.

  The non completely conclusive situation led to further investigations 
\cite{Carmona2}. First, the topology of the system on the critical line is 
very similar for different values of the density, showing a quite homogeneous 
pattern. Second, the evolution of the caloric curve shows a change of slope 
which gets more accentuated with increasing size of the system, with the 
presence of an inflexion point which is characteristic of a continuous 
transition.

\begin{figure}[b!]
\centerline{\epsfig{figure=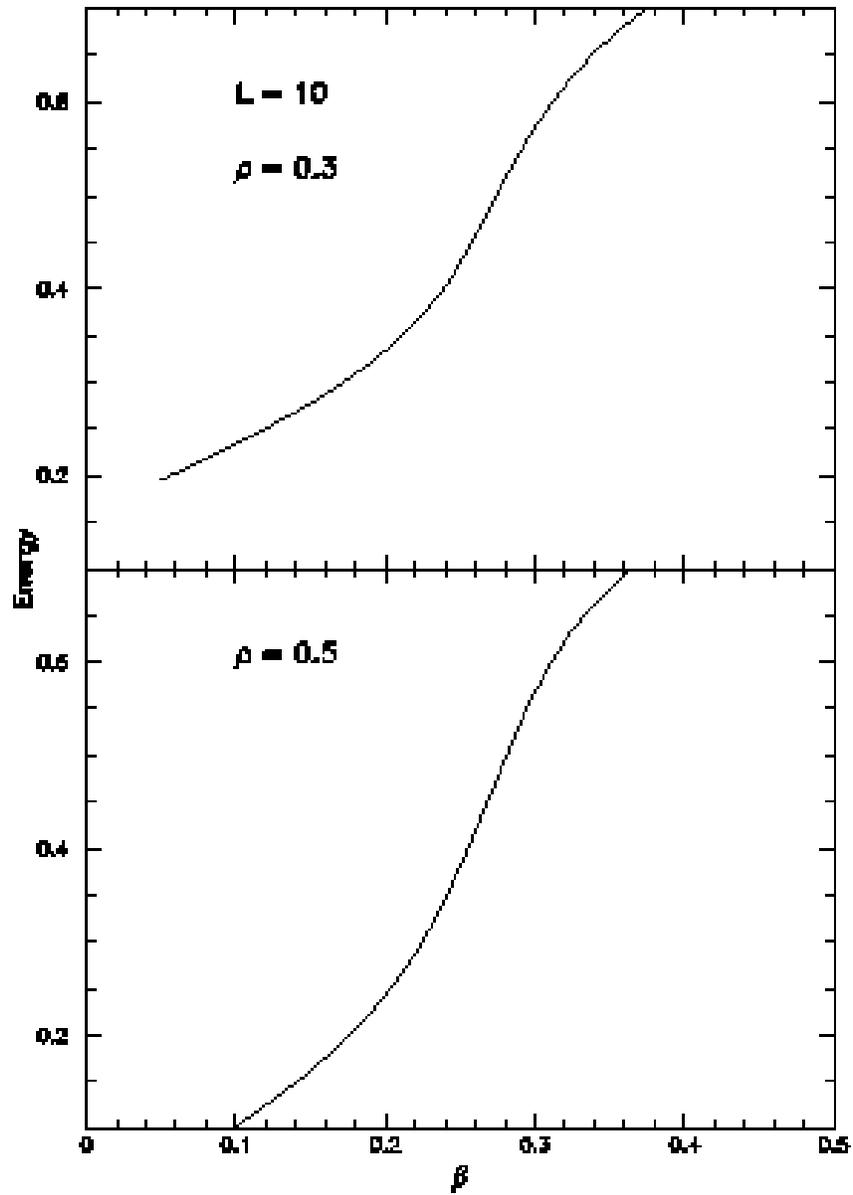,height=160mm}}
\caption{Caloric curve calculated for a 3D IMFM for a system of linear size 
$L=10$, $\rho / \rho_0 = 0.3$ (upper part), $\rho / \rho_0 = 0.5$ (lower 
part). Canonical and microcanonical calculations are indistinguishable. See 
comments in the text \cite{Carmona2}.}
\label{fig26}
\end{figure}

  A sharper test consists of the use of the microcanonical ensemble. Here 
the kinetic energy part is taken into account and the procedure is 
described in subsection 5.3.6. As discussed above, a first order transition 
should manifest itself by a temperature which is a multivaluated function 
of energy, and this is not the case if the transition is continuous. This 
phenomenon was numerically checked on the Potts model which can effectively 
be of first or second order for different numbers of spin directions on a 
given site \cite{Potts, Binder}. The application of the microcanonical 
algorithm to the IMFM does not show the expected ``$S$'' curve 
as it would be the case for a first order transition, see Fig.~\ref{fig26}.
This fact does however not lead to a definite answer because it is 
expected that for a fixed volume the caloric curve in the infinite system 
never shows a plateau in the presence of a first order liquid-gas type 
phase transition but only a more or less steep change of slope, the 
plateau arises for fixed pressure as recalled in ref. \cite{Moretto2}. 
Furthermore, as already mentioned, the finite system announces a transition 
which may not necessarily be the one which is really observed in the 
thermodynamic limit.

  Two conclusions can be drawn from this study. First, the present model 
deals with two constraints, fixed volume and fixed number of particles. 
These constraints may play a critical role in finite systems. This can 
be seen on the Potts model. For $q \geq 5$ where $q$ is the number of 
spin directions on a site, the caloric curve should show an ``$S$'' curve. 
When constraints are applied on the spin population of different spin 
directions at the different sites of the system the bend disappears
\cite{Carmona2}. 

  The importance of constraints has of course also consequences for the 
interpretation of experimental results which depend of the physical 
conditions under which data are collected. Fixed volume  is an 
illustration of the effect of constraints.

\subsubsection{Fluctuations and criticality}
\indent

  As already seen above \cite{Botet}, critical phenomena are strongly 
correlated with large fluctuations in the thermodynamic variables. This 
property has been exploited recently by Chomaz and Gulminelli \cite{Chomaz2, 
Gulminelli3, Chomaz3} in the framework of the microcanonical ensemble. 
These authors have explicitly shown on a simple model how the specific heat 
which can be defined in terms of energy fluctuations can reach negative 
values.

  A closed system with energy $E_t$ is divided into two subsystems, 1 and 2,
with energy $E_1$ and $E_2$, $E_t = E_1 + E_2$. Then the energy 
distribution in subsystem 1 can be written as
\begin{eqnarray}
P_1^{E_t}(E_1) & = & \frac{W_1(E_1)W_2(E_t-E_1)}{W(E_t)} \nonumber \\
& = & \mathcal{N} \exp[S_1(E_1) + S_2(E_t-E_1) -S(E_t)] \label{engdist}
\end{eqnarray}
where $\mathcal{N}$ is a normalization factor, $W_i(E_i)$ is the density of 
states of the system $i$ at the energy $E_i$ and $S_i(E_i)$ its entropy.
$W(E_t)$ is a convolution of partial densities
\begin{equation} W(E_t) = \int_0^{E_T} \ud E_1 \exp[S_1(E_1) + S_2(E_t-E_1)]
\nonumber \end{equation}
At equilibrium $W(E_t)$ is maximum for 
\begin{equation}
\frac {\partial S_1}{\partial E_1} \bigg |_{\overline E_1}
= \frac {\partial S_2}{\partial E_2} \bigg |_{E_t - \overline E_1}
\label{Wmax} \end{equation}
where $\overline E_1$ corresponds to the stationary point where the 
equality is realized. Relation (\ref{Wmax}) shows that the temperatures 
$T_1$ and $T_2$ are the same in both subsystems, $T_1 = T_2 = \overline T$ 
as expected for a system in thermodynamic equilibrium.

  Close to the maximum of $W(E_t)$ it is reasonable to parametrize 
$P_1^{E_t}$ by a gaussian
\begin{equation} P_1^{E_t}(E_1) = (2 \pi \sigma_1^2)^{-1/2}
\exp [-(E_1- \overline E_1)^2 / 2 \sigma_1^2] \label{Pgauss0} \end{equation}
Then
\begin{equation} \frac {\partial^2 P_1^{E_t}(E_1)}{\partial E_1^2}
\bigg |_{\overline E_1} = -(2 \pi)^{-1/2} \sigma_1^{-3}
\label{Pgauss1} \end{equation}
From (\ref{engdist}), for $\mathcal{N}=(2 \pi \sigma_1^2)^{-1/2}$
\begin{eqnarray} \frac {\partial^2 P_1^{E_t}}{\partial E_1^2}
\bigg |_{\overline E_1} & = & -\frac{1}{\overline T^2}
\Bigg(\frac {C_1+C_2}{C_1C_2}\Bigg) \label{Pgauss2} \\
\textrm{where} \quad C_i & = & -\frac{1}{T_i^2}
\Bigg(\frac {\partial^2 S_i}{\partial E_i^2}\Bigg)^{-1}
= \frac{\ud \overline E_i}{\ud E_t}
\nonumber \end{eqnarray}
is the specific heat of the system $i$. Equating (\ref{Pgauss1}) and 
(\ref{Pgauss2}) and introducing 
\begin{equation} \overline C = C_1 + C_2 \label{CVsum} \end{equation}
leads to
\begin{equation} \overline C = \frac{C_1^2}
{C_1 - \sigma_1^2 / \overline T^2} \nonumber \end{equation} \indent
  If now subsystem 1 corresponds to the kinetic energy part of the system 
with energy $E_k$ and specific heat $C_k$ and subsystem 2 to the potential 
contribution $E_p$ then
\begin{equation} \overline C = \frac{C_k^2}
{C_k - \sigma_k^2 / \overline T^2} \label{CV1} \end{equation}
where $C_k = \frac{3}{2} A$ and $A$ is the number of particles. The 
knowledge of the temperature $\overline T$ of the system and the kinetic 
energy fluctuations characterized by $\sigma_k^2$ allow then to determine 
the heat capacity of the system in the framework of the microcanonical 
ensemble. It is clear that if $\sigma_k^2$ gets larger than $C_k 
\overline T^2$, $\overline C$ gets negative which corresponds to an ``$S$'' 
bend in the caloric curve. The analysis requires however some care. There is 
a priori no justification for the distribution (\ref{Pgauss0}) to be gaussian 
nor for (\ref{CVsum}) to be strictly true in the vicinity of a critical 
point. The lack of concavity of the entropy is the sign for it. There are 
however signs that, even if (\ref{CVsum}) is not strictly correct, the 
variance of the real distribution can be assimilated to the variance of a 
gaussian distribution if the number of particles is not too small 
\cite{Gulminelli4}. The correction to the heat capacity due to higher 
moment contributions to (\ref{Pgauss1}) has been discussed in ref. 
\cite{Chomaz2}. The kinetic energy distribution is in principle 
experimentally accessible.

  The relation (\ref{CVsum}) has been applied to lattice models of the 
same type as the one whose Hamiltonian is used in subsection 5.2.2 
\cite{Chomaz3}. The partition function reads 
\begin{equation} \mathcal Z_\lambda(E) = \sum_V W_V(E) e^{-\lambda V}
\nonumber \end{equation}
where $W_V(E)$ is the density of states at energy $E$. The volume of the 
system is not fixed here but fluctuates. This is described by means of the 
introduction of a Lagrange multiplier $\lambda$ which is interpreted as a 
quantity proportional to a pressure $P = \lambda T_\lambda$ where 
$T_\lambda = \partial \ln \mathcal Z_\lambda / \partial E$.

\begin{figure}[b!]
\centerline{\epsfig{figure=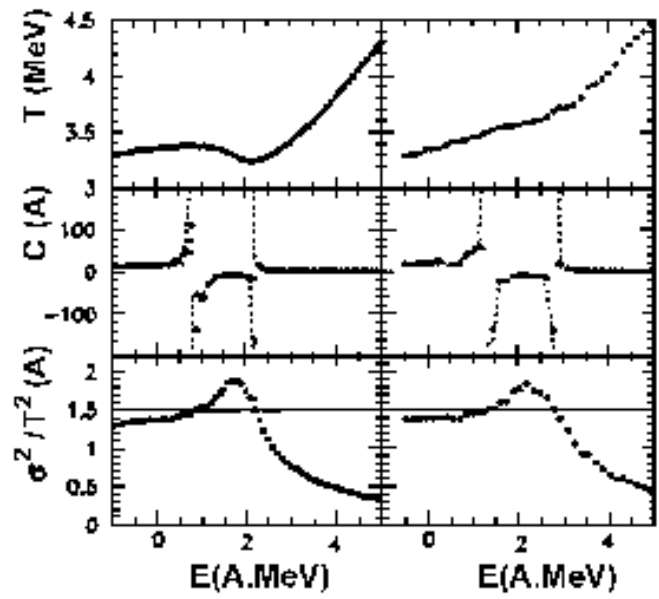,height=80mm}}
\caption{Caloric curve, specific heat and kinetic energy fluctuations at 
constant pressure (left part) and constant volume (right part) calculated 
in the framework of the microcanonical ensemble \cite{Chomaz3}. See 
comments in the text.}
\label{fig27}
\end{figure}

  The equation of state for this model has been studied numerically in the
3-dimensional space $(T,E,\lambda)$. One observes a backbending effect if 
one represents $P = \lambda T$ as a function of the average volume 
$<V>$ which is interpreted here as being the consequence of mass and energy 
conservation constraints. An ``$S$'' bend appears also in the caloric curve 
on the $(E,T)$ surface. For strictly fixed volume the caloric curve does 
not show any backbending. This is illustrated in Fig.~\ref{fig27} in which the 
specific heat is represented as a function of energy using (\ref{CV1}). 
With this interpretation the puzzle concerning the order of the transition 
which was discussed in subsection 6.2.2 \cite{Carmona2} is apparently 
explained, the shape of the caloric curve depends on the thermodynamic 
variables which are considered \cite{Moretto2}. This message is of some 
importance in the framework of experimental analyses.

\subsection{Experimental construction of the caloric curve and 
determination of the specific heat}
\indent

  Different attempts have been made since the pioneering work already 
mentioned above [ALADIN, INDRA, EOS,\ldots] in order to construct the caloric 
curve starting from experimental data.

  The reaction $^{28}Si$ on $^{100}Mo$ leading to an incomplete fused 
system has been used in order to relate the excitation energy per particle 
$e^*$ of the system to a so called apparent temperature $T_{app}$ which is 
obtained from the slopes of the energy distributions of emitted light 
particles \cite{Chbihi}.
\begin{equation} \frac{\ud \sigma}{\ud E_{kin}} \propto (E_{kin} - B) \,
\exp \Bigg[-\frac{(E_{kin} - B)}{T_{app}} \Bigg] \nonumber \end{equation}
where $E_{kin}$ is the kinetic energy and $B$ the Coulomb barrier. 
$T_{app}$ which is not the thermodynamic temperature $T_{th}$ is shown to be 
a sensible quantity, both $T_{app}$ and $T_{th}$ show the same trend as a 
function of the energy. The outcome of the experience is in reasonable 
agreement with calculations performed in the framework of the MMMC model.
Since this model indicates the existence of a phase transition, it is 
tempting to conclude that it is also experimentally true since model 
calculations and experimental data have been worked out under the same 
physical conditions. It should nevertheless be mentioned that the results do 
not constitute a full proof because there remain some differences between the 
calculations and the data analysis. Former data from the Texas A \& M-group 
\cite{Wada} are in agreement with those discussed here.

  The problem concerning the determination of the temperature has also been 
raised in the framework of a sequential decay model in which fragmentation 
evolves in time by means of successive binary break-ups of excited 
fragments \cite{Durand2} in a way which is similar to the approach of ref. 
\cite{Richert}. The analysis \cite{Gulminelli, Siwek} of different 
definitions of the temperature leads to different results, due to the 
time-dependence of the process. It comes out that the experimental caloric 
curve \cite{Pochodzalla} can be reproduced over a large range of excitation 
energies.

  Very recently an attempt to detect the signal of a phase transition 
through the determination of the specific heat has been made by the 
Multics-Miniball collaboration at the cyclotron of the Michigan State 
University \cite{DAgostino1, DAgostino2}. The excited system has been 
generated by means of $Au$ on $Au$ collisions at $35\,MeV\!\cdot\!A$.
The angular distributions of fragments show isotropy which is a sign for 
possible equilibration \cite{DAgostino1}. The conditions are such that the 
number of involved particles and the total energy are fixed, so that the 
arguments which characterize microcanonical ensembles can be applied. The 
correlation between the excitation energy $E^*$ and the temperature is 
fixed through the balance in energy at the freeze out
\begin{equation} m_0 + E^* = \sum_{i=1}^M (m_i + E_i^{int}) + E_{Clb}
+\frac{3}{2} (M-1) T \nonumber \end{equation}
where $m_0$ is the initial mass excess, $m_i$ the mass excesses of the
fragments, $E^*$ the excitation energy of the source, $E_i^{int}$ the
internal excitation energy of fragment $i$, $E_{Clb}$ the Coulomb repulsion
energy, $M$ the multiplicity and $T$ the temperature which is fixed by the
balance equation. No collective energy is taken into account. 
A backtracing procedure is used in order to correlate the 
asymptotically observed events to the freeze-out configurations. The 
comparison of $E^*(T)$ with SMM predictions are in good though not 
perfect agreement. The heat capacity shows a sizable enhancement for 
temperatures lying between 4 and $6\,MeV$.

  In a further step the experiment has been analysed following the 
microcanonical prescriptions of Chomaz and Gulminelli, see subsection 6.2.3 
\cite{Gulminelli3, Chomaz3}. The total energy $E_t$ has been decomposed 
into a kinetic and a potential contribution
\begin{eqnarray} E_t = m_0 + E^* = E_1 + E_{pot} \nonumber \\
\textrm{where}  \quad E_{pot} = \sum_{i=1}^M m_i + E_{Clb} \nonumber
\end{eqnarray}
and $E_{Clb}$ is the Coulomb energy acting between the fragments at 
freeze-out. This energy has been evaluated by means of simulations in which 
primary fragments are randomly distributed in a freeze-out volume equal to 
three times the normal volume. Two extreme hypotheses have been used. 
Either (1) all charged species including all light particles were 
considered as primary or (2) the totality of neutrons and light particles 
was distributed over the final fragments. These choices correspond to the 
procedures used in the framework of the SMM and MMMC models. In this way, 
one fixes the ensemble averaged quantities over sets of events $<E_1>, \
<E_1^2>, \ \sigma_1^2$ and $C_1 = \frac{3}{2}<(M-1)>$ where
\begin{equation} <E_1>= \Big< \sum_{i=1}^M a_i \Big> \, T^2 +
\Big< \frac{3}{2} (M-1) \Big> \, T \nonumber \end{equation}
and $a_i$ is the level density of fragments $i$.

\begin{figure}[b!]
\centerline{\epsfig{figure=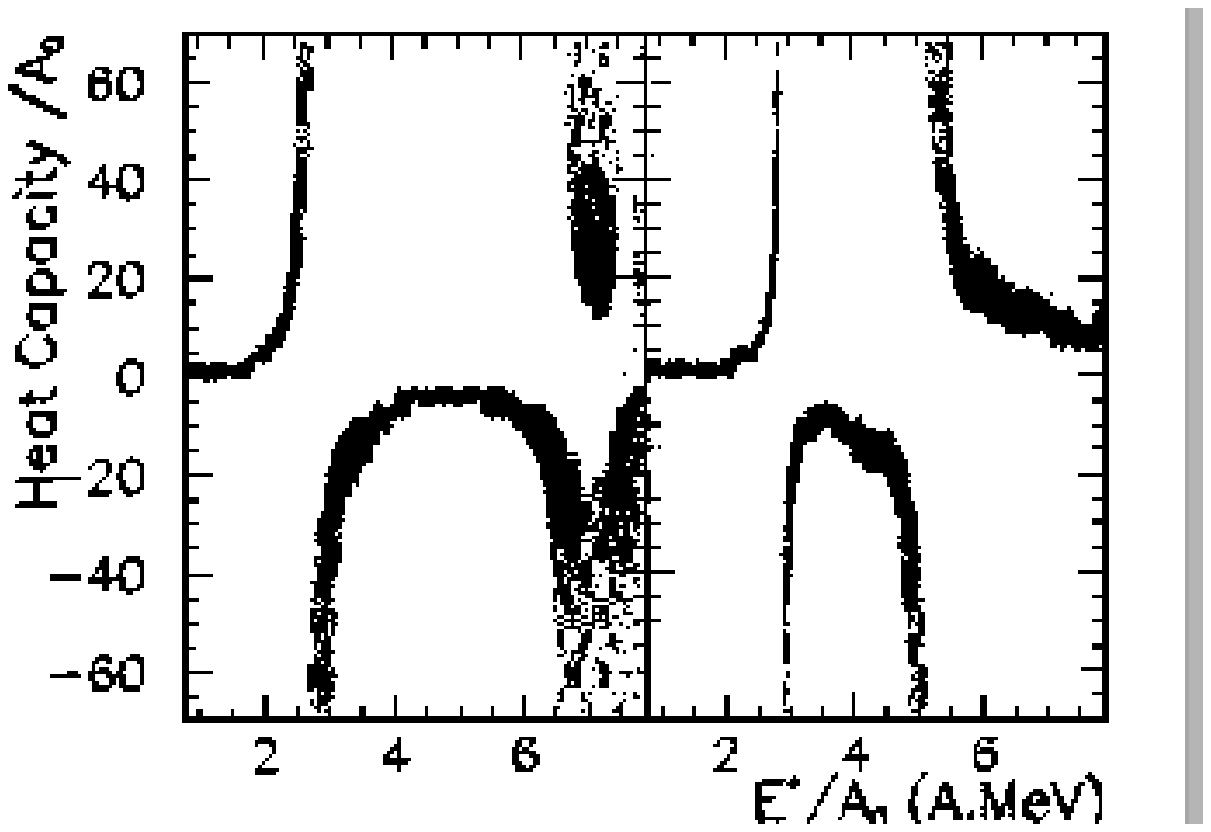,height=80mm}}
\caption{Experimental heat capacities for the two extreme hypotheses which 
where made in order to evaluate $E_{Clb}$ (see text). The dotted curve 
corresponds to the Fermi-gas calculation \cite{DAgostino2}.}
\label{fig28}
\end{figure}

  The results corresponding to cases (1) and (2) are shown in Fig.~\ref{fig28}. 
In both cases one observes the signs for a divergent behaviour of the heat 
capacity at two values of the energy and the expected negative 
contributions to the specific heat in the microcanonical framework. The 
stability of this result with respect to variations in the size of 
excitation energy bins, the assumptions on $E^*, \, E_t$ and the choice 
of the $a_i$'s have been checked.

  The same type of analysis has been performed with the INDRA detector on 
the outcome of the reaction $Xe$ on $Sn$ at 32 and $50\,MeV\!\cdot\!A$ 
central collisions \cite{LeNeindre, LeNeindre1}. In order to extract the 
kinetic energy fluctuations which enter the determination of the specific 
heat, it has been necessary to substract the collective effects which are 
present and to take care of the particle emission which takes place during 
the expansion phase of the process. One of the two divergences 
which develop in the heat capacity is observed as well as the 
onset of negative values for this quantity. The second singularity 
is not seen because the excitation range does not cover the whole interval 
of energy corresponding to the coexistence of the phases.

\newpage

\section{Summary and conclusions}
\indent

  In the present report we tried to summarize the present knowledge about 
the properties of excited nuclear matter which can be gained through the 
confrontation of simple classical models describing finite systems and 
experimental facts obtained from energetic collisions between finite nuclei.

  As an introduction we developed in section 2 some historical facts 
concerning the beginning of the activities in the field of nuclear 
fragmentation, both on the side of the experiment with the observation of 
power laws in the fragment size distributions, the introduction of the 
droplet model and the theoretical developments based on  mean field 
approaches to nuclear matter and microscopic transport descriptions in the 
mean field approximation and beyond.

  We recalled that 
there are two main difficulties in the interpretation of experimental 
data obtained by means of energetic nuclear collisions. The first concerns 
the finite size of nuclei and the second the fact that the colliding 
objects evolve with time, from an initial compact system to an expanded 
system at the nuclear freeze-out where the nuclear interaction between 
fragments does no longer act, and finally to the asymptotic stage where the 
remnants of the reaction are identified by detectors. The second point 
leads to the question of the setting up of chemical and thermodynamic 
equilibrium. At lower energies there are signs which show that nuclear 
systems are able to absorb and spread sizable amounts of energy more or 
less uniformely over phase space.

  Thermodynamic equilibration is discussed in section 3 in which
some of the efforts which have been made in order to get information about 
this point are developed. For the moment, equilibration seems to be a 
sensible concept since experimental results can be consistently interpreted 
in this framework, even though no strong and clear-cut proof for it exists 
up tp now.

  One of the first and more spectacular successes in the field concerns the 
multiplicities of fragments of different sizes which are amazingly well 
reproduced by means of simple minded percolation models. This point is 
presented and discussed at length in section 4. Universality properties are 
clearly identified and the intriguing power law behaviour which was observed 
and was of central interest in the early days appears again in a new context. 
Since percolation concepts are 
minimal information concepts it may appear that there is not much 
information to be gained from observables like fragment size distributions. 
The reasons for this may be related to the fact that highly excited 
nuclei behave like disordered systems, and as it is the case for many other 
systems of this type in other fields of physics, the details of their 
dynamical properties are of very restricted importance. Universality 
properties of physical systems are remarkable because they work as a link 
between objects and phenomena which are seemingly very different from each 
other and at the same time are insensitive to the detailed information about 
the system.

  The success of percolation drew the attention towards other simple 
classical models which have a strong kinship with percolation but are 
defined in terms of Hamiltonians. Some of them, so called lattice and 
cellular models, are described and analysed in section 5. Their simplicity 
and at the same time their universal character leads to the observation 
that they work as good tools for the investigation of the properties of 
fragmenting nuclei. Most of them are able to reproduce the properties 
related to fragment multiplicities. They show very useful for the 
investigations of effects related to finiteness in search for and analysis 
of phase transitions. They predict the possible existence of two types of 
transitions which are not necessarily correlated in an evident way, a
thermodynamic transition and a 
continuous percolation type transition which for systems with a density 
between nuclear density and half nuclear density occurs at temperatures 
which are higher than those which correspond to the thermodynamic 
transition. This point is not yet clearly understood, although it does not 
lead, a priori, to any contradiction. It opens up the old question 
concerning the link with Fisher's droplet model. Numerical simulations on 
different models lead to very similar thermodynamic properties. 

  The final part, section 6, deals more specifically with the problem of 
the existence and characterization of phase transitions and signs which 
show their existence in finite systems. For such types of analyses specific 
tools are needed. Some of them have been proposed very recently, at a time 
where similar considerations are developed in other fields of physics 
dealing with small systems. Some of these tools are explicitly described. 
Theoretical methods relying on a microcanonical description and aimed to 
detect transitions which have been developed in the framework of lattice 
models have been presented. Their application in the analysis of 
experimental data is described at the end of the section.

  A great wealth of information has been accumulated about the 
characteristics and behaviour of highly excited nuclei, their fragmentation 
properties and the existence of a genuine phase transition of a type which 
may be similar to the classical liquid-gas transition. The last point 
remains however to be confirmed. Many pieces of the puzzle have been 
brought together, however the whole picture is not yet totally clear. 
Experimental confirmation of the transition has to be brought, scaling 
properties further investigated, the question of the possible coexistence 
of two types of transitions elucidated. Quantum mechanical aspects which 
have been quoted and the importance of quantum effects have to be further 
worked out and analysed. Last, there remains the difficult problem of the 
possible 
off-equilibrium character of the fragmentation process of a finite critical 
system. The study of critical off-equilibrium dynamics is a relatively 
recent field of research, still in its infancy. It is a field to which the 
nuclear physics community interested in the behaviour of highly excited and 
decaying systems of particles could certainly bring its contribution.

\section*{\protect Acknowledgements}

  The present report would not have been written without informative and 
clarifying discussions with many colleagues who provided advice and insight. 
We would like to thank A.~Bonasera, M.~Ploszajczak, J.~Perdang for 
interesting exchanges of ideas and informations. We are grateful to 
C.~Barbagallo, T.~Bir\'o, J.M.~Carmona, B.~Elattari, M.~Henkel, J.~Knoll,
A.~Lejeune,
S.K.~Samaddar, A.~Taranc\'on, Y.M.~Zheng for their active collaboration in 
those parts which concern our own contribution. Special thanks go 
to X.~Campi, Ph.~Chomaz, D.H.E.~Gross, F.~Gulminelli, H.~Krivine and
W.~Trautmann for the direct and indirect supply of good ideas and efficient 
information, illuminating explanations, critical and constructive remarks.
We acknowledge clarifying remarks made by R.~Bougault and L.~Phair about 
the interpretation of experimental figures.

\newpage

\clearpage

\end{document}